\documentclass[runningheads,envcountsame]{llncs}

\makeatletter
\RequirePackage[bookmarks,unicode,colorlinks=true]{hyperref}%
   \def\@citecolor{blue}%
   \def\@urlcolor{blue}%
   \def\@linkcolor{blue}%

\def\orcidID#1{\href{http://orcid.org/#1}{\smash{\protect\raisebox{-1.25pt}{\protect\includegraphics{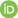}}}}}
\makeatother

\usepackage{xcolor,colortbl}
\usepackage{mathtools}
\usepackage{multirow}
\usepackage{stmaryrd}
\usepackage{cancel}
\usepackage{amsmath,amssymb}
\usepackage{thmtools, thm-restate}
\usepackage[shortlabels]{enumitem}
\usepackage{microtype}
\usepackage{wrapfig}
\usepackage{pbox}
\usepackage{marvosym}
\usepackage{listings}
\usepackage[ruled, vlined, linesnumbered]{algorithm2e}
\hypersetup{
    pdftitle={A Complete Dependency Pair Framework for Almost-Sure Innermost Termination of
    Probabilistic Term Rewriting}, colorlinks=true, linkcolor=blue, citecolor=olive, filecolor=magenta, urlcolor=cyan
}
\usepackage{todonotes}
\usepackage{placeins}
\usepackage{float}
\usepackage{tikz}
\usepackage{caption}
\usepackage{subcaption}
\usepackage{mymatrix}
\usepackage{array,longtable}
\usepackage{nicefrac,xfrac}
\usetikzlibrary{shapes,calc,arrows,automata,decorations.pathmorphing,backgrounds,arrows.meta,shapes.geometric,shapes.multipart,shapes.misc}
\pgfdeclarelayer{edgelayer}
\pgfdeclarelayer{nodelayer}
\pgfsetlayers{background,edgelayer,nodelayer,main}
\tikzstyle{none}=[inner sep=0mm]
\tikzstyle{moveBlock}=[fill=white, draw=black, shape=rectangle]
\tikzstyle{target}=[fill=white, draw=black, shape=circle]

\tikzstyle{dotHead}=[dotted, ->]
\tikzstyle{dotWithoutHead}=[dotted, -]
\tikzstyle{dashHead}=[dashed,->]
\tikzstyle{dashWithoutHead}=[dashed,-]
\tikzstyle{arrow}=[->]

\RequirePackage{makecell}

\usepackage[capitalize,nameinlink]{cleveref}
\usepackage{IEEEtrantools}

\newenvironment{myproof}{
	\noindent{\it Proof.}
}{\qed
	\medskip
}
\def\namedlabel#1#2{\begingroup
    #2%
    \def\@currentlabel{#2}%
    \phantomsection\label{#1}\endgroup
}

\newcommand{\disabledcomment}[1]{}
\newcommand{\oldcomment}[1]{}

\newcommand{\aprove}{\textsf{AProVE}\xspace}

\renewcommand{\emph}[1]{\index{#1}\textit{#1}}

\renewcommand{\emptyset}{\varnothing}

\newcommand{\IN}{\mathbb{N}}
\newcommand{\IR}{\mathbb{R}}

\newcommand{\F}[1]{\mathfrak{#1}}

\makeatletter
\def\moverlay{\mathpalette\mov@rlay}
\def\mov@rlay#1#2{\leavevmode\vtop{%
   \baselineskip\z@skip \lineskiplimit-\maxdimen
   \ialign{\hfil$\m@th#1##$\hfil\cr#2\crcr}}}
\newcommand{\charfusion}[3][\mathord]{
    #1{\ifx#1\mathop\vphantom{#2}\fi
        \mathpalette\mov@rlay{#2\cr#3}
      }
    \ifx#1\mathop\expandafter\displaylimits\fi}
\makeatother

\newcommand{\Proc}{\operatorname{Proc}}
\newcommand{\projOne}{\mathrm{proj}_1}
\newcommand{\projTwo}{\mathrm{proj}_2}

\newcommand{\PosDPoss}{\pos_{\text{Poss}}}

\newcommand{\Junk}{\mathrm{Junk}}

\newcommand{\TSet}[2]{\mathcal{T}\left(#1,#2\right)}

\newcommand{\VSet}{\mathcal{V}}

\newcommand{\R}{\mathcal{R}}

\newcommand{\DTuple}[1]{\mathcal{DT}(#1)}

\newcommand{\DPair}[1]{\mathcal{DP}(#1)}

\newcommand{\FDist}{\operatorname{FDist}}
\newcommand{\Supp}{\operatorname{Supp}}

\newcommand{\rootsym}{\operatorname{root}}
\newcommand{\capterm}{\operatorname{Cap}}

\newcommand{\rules}{\operatorname{Rules}}
\newcommand{\urules}{\mathcal{U}}

\newcommand{\vr}{\operatorname{vr}}
\newcommand{\capt}{\operatorname{capt}}

\newcommand{\Pol}{\operatorname{Pol}}

\newcommand{\Com}[1]{\tcom_{#1}}

\newcommand{\cont}{cont}

\newcommand{\nonprob}{\normalfont{\text{np}}}
\newcommand{\nonprobDP}{\normalfont{\text{dp}}}
\newcommand{\PP}{\mathcal{P}}
\newcommand{\QQ}{\mathcal{Q}}

\newcommand{\tex}{\mathsf{ex}}
\newcommand{\textwo}{\mathsf{ex2}}

\renewcommand{\ts}{\mathsf{s}}
\newcommand{\tz}{\mathsf{0}}

\newcommand{\tf}{\mathsf{f}}
\newcommand{\tg}{\mathsf{g}}
\renewcommand{\th}{\mathsf{h}}
\newcommand{\ta}{\mathsf{a}}
\newcommand{\tb}{\mathsf{b}}
\newcommand{\tc}{\mathsf{c}}
\newcommand{\td}{\mathsf{d}}
\newcommand{\te}{\mathsf{e}}

\newcommand{\tq}{\mathsf{q}}

\newcommand{\tffg}{\mathsf{ffg}}

\newcommand{\trw}{\mathsf{rw}}
\newcommand{\tic}{\mathsf{incpl}}
\newcommand{\tcons}{\mathsf{cons}}
\newcommand{\tnil}{\mathsf{nil}}

\newcommand{\tmoveelements}{\mathsf{moveElements}}
\newcommand{\tor}{\mathsf{or}}

\newcommand{\tif}{\mathsf{if}}
\newcommand{\ttrue}{\mathsf{true}}
\newcommand{\tfalse}{\mathsf{false}}

\newcommand{\tgt}{\mathsf{gt}}
\newcommand{\tp}{\mathsf{p}}

\newcommand{\tF}{\mathsf{F}}
\newcommand{\tG}{\mathsf{G}}
\newcommand{\tH}{\mathsf{H}}
\newcommand{\tA}{\mathsf{A}}

\newcommand{\trotate}{\mathsf{rotate}}
\newcommand{\tapp}{\mathsf{app}}
\newcommand{\tqs}{\mathsf{qsrt}}

\newcommand{\tifhigh}{\mathsf{ifHigh}}
\newcommand{\thigh}{\mathsf{high}}
\newcommand{\tiflow}{\mathsf{ifLow}}
\newcommand{\tlow}{\mathsf{low}}
\newcommand{\tleq}{\mathsf{leq}}

\newcommand{\thd}{\mathsf{hd}}
\newcommand{\ttail}{\mathsf{tl}}
\newcommand{\tisempty}{\mathsf{empty}}

\newcommand{\teven}{\mathsf{even}}
\newcommand{\tloop}{\mathsf{loop}}
\newcommand{\tevenif}{\mathsf{if}}
\newcommand{\tstop}{\mathsf{stop}}
\newcommand{\xs}{\mathit{xs}}
\newcommand{\ys}{\mathit{ys}}

\newcommand{\tcom}{\mathsf{com}}

\newcommand{\ctleaf}{\mathtt{Leaf}}

\newcommand{\ctdepth}{\operatorname{d}}

\newcommand{\ctlevelTwo}{\mathcal{L}}
\newcommand{\ctlevelTwowithborder}{\F{L}}

\crefname{definition}{Def.}{Def.}
\crefname{example}{Ex.}{Ex.}
\crefname{counterexample}{Counterex.}{Counterex.}
\crefname{appendix}{App.}{App.}
\crefname{ex}{Ex.}{Ex.}
\crefname{theorem}{Thm.}{Thm.}
\crefname{lemma}{Lemma}{Lemmas}
\crefname{remark}{Rem.}{Rem.}
\crefname{section}{Sect.}{Sect.}
\crefname{subsection}{Sect.}{Sect.}
\crefname{subsubsection}{Sect.}{Sect.}
\crefname{line}{Line}{Lines}
\crefname{corollary}{Cor.}{Cor.}
\crefname{figure}{Fig.}{Fig.}
\crefname{enumi}{}{}
\crefname{algorithm}{Alg.}{Alg.}

\makeatletter
\NewDocumentCommand{\dparrow}{+O{} +O{0.5cm}}{%
\begin{tikzpicture}[baseline=-0.5ex] {
\node[inner sep=0](@1) at (0,0) {};
\node[inner sep=0](@2) at (#2,0) {};
\draw [arrows={-Triangle[open]},shorten >= 1pt,shorten <= 1pt](@1)--(@2) node[pos=.5,above,inner sep=1pt] {\ensuremath{#1}};}
\end{tikzpicture}\xspace
}

\NewDocumentCommand{\myto}{+O{} +O{0.5cm}}{%
\begin{tikzpicture}[baseline=-0.5ex] {
\node[inner sep=0](@1) at (0,0) {};
\node[inner sep=0](@2) at (#2,0) {};
\draw [arrows={-to}](@1)--(@2) node[pos=.5,above,inner sep=1pt] {\ensuremath{#1}};}
\end{tikzpicture}\xspace
}

\NewDocumentCommand{\paraarrow}{+O{} +O{0.4cm}}{%
\begin{tikzpicture}[baseline=-0.5ex] {
\node[inner sep=0](@1) at (0,0) {};
\node[inner sep=0](@2) at (#2,0) {};
\node[inner sep=0](@3) at (0.07,0) {};
\draw [arrows={-to}](@1)--(@2) node[pos=.5,above,inner sep=1pt] {\ensuremath{#1}};
\draw [arrows={-to}](@1)--(@3);}
\end{tikzpicture}\xspace
}

\NewDocumentCommand{\paradparrow}{+O{} +O{0.4cm}}{%
\begin{tikzpicture}[baseline=-0.5ex] {
\node[inner sep=0](@1) at (0,0) {};
\node[inner sep=0](@2) at (#2,0) {};
\node[inner sep=0](@3) at (0.07,0) {};
\draw [arrows={-Triangle[open]}](@1)--(@2) node[pos=.5,above,inner sep=1pt] {\ensuremath{#1}};
\draw [arrows={-to}](@1)--(@3);}
\end{tikzpicture}\xspace
}

\newcommand{\oset}[2]{%
  {\mathop{#2}\limits^{\vbox to 1\ex@{\kern-\tw@\ex@
   \hbox{\scriptsize #1}\vss}}}}

\newcommand{\osetthree}[2]{%
  {\mathop{#2}\limits^{\vbox to 3\ex@{\kern-\tw@\ex@
   \hbox{\scriptsize #1}\vss}}}}

\newcommand{\osetfive}[2]{%
  {\mathop{#2}\limits^{\vbox to 5\ex@{\kern-\tw@\ex@
   \hbox{\scriptsize #1}\vss}}}}

\newcommand{\osetminus}[2]{%
  {\mathop{#2}\limits^{\vbox to -2\ex@{\kern-\tw@\ex@
   \hbox{\scriptsize #1}\vss}}}}
\makeatother

\newcommand{\itodrEX}{\mathrel{\oset{\scriptsize $\mathsf{i}$\qquad\qquad\qquad\,}{{\to}_{\DPair{\R_{\tex}},\R_{\tex}}}}}
\newcommand{\itorstarEX}{\mathrel{\smash{\stackrel{\raisebox{2pt}{\scriptsize $\mathsf{i}$\,}}%
{\smash{\rightarrow}}}_{\R_{\tex}}^*}}

\newcommand{\ito}{\mathrel{\smash{\stackrel{\raisebox{3.4pt}{\tiny $\mathsf{i}\:$}}{\smash{\rightarrow}}}}}

\newcommand{\idparrow}{\mathrel{\smash{\stackrel{\raisebox{3.4pt}{\tiny $\mathsf{i}\:$}}{\smash{\dparrow}}}}}

\newcommand{\itorstar}{\mathrel{\ito_{\R}^{*}}}

\newcommand{\itor}{\mathrel{\ito_{\R}}}

\newcommand{\itodr}{\mathrel{\ito_{\mathcal{P},\R}}}

\newcommand{\fs}[1]{\mathsf{#1}}
\newcommand{\fun}[1]{\mathrm{#1}}

\renewcommand{\phi}{\varphi}
\renewcommand{\emptyset}{\varnothing}
\newcommand{\true}{\fs{true}}

\newcommand{\pos}{\fun{pos}}

\newcommand{\posT}{\fun{pos}_{\SignatureA}}

\newcommand{\posD}{\fun{pos}_{\SignatureD}}
\newcommand{\posDT}{\fun{pos}_{\SignatureD \cup \SignatureA}}

\newcommand{\anno}{\#} 

\newcommand{\annoEps}{\anno_{\varepsilon}}
\newcommand{\annoD}{\anno_{\SignatureD}}
\newcommand{\disannoPos}[1]{\flat_{#1}^{\uparrow}}

\newcommand{\SignatureDC}{\Sigma}
\newcommand{\SignatureADC}{\Sigma^\#}

\newcommand{\SignatureC}{\mathcal{C}}
\newcommand{\SignatureD}{\mathcal{D}}
\newcommand{\SignatureA}{\mathcal{D}^\#}

\newcommand{\NF}{\mathtt{NF}}
\newcommand{\ANF}{\mathtt{ANF}}

\newcommand{\NN}{\mathbb{N}}

\newcommand{\ruleArr}[3]{
  \mathrel{
    \xrightarrow{{}_{\scriptstyle #1}}
    \!\!{}^{#2}_{#3}
  }
}
\newcommand{\tored}[3]{
  \mathrel{
    \xhookrightarrow{{}_{\scriptstyle #1}}
    \!\!{}^{#2}_{#3}
  }
}
\newcommand{\itored}[3]{
  \mathrel{
    \smash{\stackrel{\raisebox{3.4pt}{\tiny $\mathsf{i}\:$}}{\smash{\hookrightarrow}}}^{#2}_{#3}
  }
}

\newcommand{\defemph}[1]{{\rm #1}}

\SetKwInOut{Input}{input}
\SetKwInOut{Output}{output}

\definecolor{Gray}{gray}{0.85}
\definecolor{LightCyan}{rgb}{0.88,1,1}

\newcolumntype{a}{>{\columncolor{Gray}}c}
\newcolumntype{b}{>{\columncolor{white}}c}

\newcommand{\makeproof}[2]{}
\newcommand{\accepted}[1]{}
\newcommand{\paper}[1]{}
\newcommand{\report}[1]{#1}

\report{
    \setlength{\textwidth}{125mm}
    \setlength{\textheight}{198mm}
}

\report{\usepackage{cite}}

 \newcommand{\notnew}[1]{}

\newcounter{counter-completeness-ctr}
\newcounter{counter-completeness2-ctr}
\newcounter{counter-completeness-adp-ctr}
\newcounter{auxctr}
\newcounter{soundnessInstantiationCtr}
\newcounter{soundnessRewriteCtr}

\paper{\usepackage[ hyperref=true, backend=bibtex, firstinits=true, maxbibnames=99, sortcites, style=numeric-comp ]{biblatex}
\addbibresource{biblioPaper.bib}}

\title{A Complete Dependency Pair Framework for\\Almost-Sure Innermost Termination of\\
  Probabilistic Term Rewriting\thanks{funded by the
    Deutsche Forschungsgemeinschaft (DFG, German Research Foundation) - 235950644 (Project
    GI 274/6-2) and DFG Research Training Group 2236 UnRAVeL}}
\titlerunning{A Complete DP Framework for iAST of PTRSs}
\paper{
    \author{Jan-Christoph Kassing\accepted{$^{(\href{mailto:kassing@cs.rwth-aachen.de}{\mbox{\Letter}})}$}\orcidID{0009-0001-9972-2470}
    \and Stefan Dollase\orcidID{0009-0006-2150-2554}
    \and Jürgen Giesl\accepted{$^{(\href{mailto:giesl@informatik.rwth-aachen.de}{\mbox{\Letter}})}$}\orcidID{0000-0003-0283-8520}}
    \institute{LuFG Informatik 2, RWTH Aachen University, Aachen, Germany}
    \authorrunning{J.-C.\ Kassing, S.\ Dollase, J.\ Giesl}
}
\report{\author{Jan-Christoph Kassing\orcidID{0009-0001-9972-2470}
    \and Stefan Dollase\orcidID{0009-0006-2150-2554}
    \and Jürgen Giesl\orcidID{0000-0003-0283-8520}}
    \institute{LuFG Informatik 2, RWTH Aachen University, Aachen, Germany}
    \authorrunning{J.-C.\ Kassing, S.\ Dollase, J.\ Giesl}
}

\begin{document}
\allowdisplaybreaks

\maketitle
\vspace*{-.6cm}
\begin{abstract}
    Recently, we adapted the well-known dependency pair (DP) framework to a
    \emph{dependency tuple} framework in order to prove
    almost-sure innermost termination (iAST) of probabilistic term rewrite systems.
    While this approach was \emph{incomplete},
    in this paper, we improve it into a \emph{complete} criterion for iAST by
    presenting a new, more elegant definition of DPs for probabilistic term rewriting. 
    Based on this, we extend the probabilistic DP framework by new \emph{transformations}. 
    Our implementation in the tool \aprove{} shows that they increase its power considerably.
\end{abstract}

\section{Introduction}\label{sec-introduction}

Termination of term rewrite systems (TRSs) has been studied
for decades and TRSs are used for automated termination analysis of many 
programming langua\-ges.
One of  the most powerful techniques integrated in essentially all current
termi\-nation tools for
TRSs is  the \emph{dependency pair} (DP) framework
\cite{arts2000termination,gieslLPAR04dpframework,hirokawa2005automating,giesl2006mechanizing} which
allows modular proofs that apply different
techniques in different sub-proofs.

In \cite{BournezRTA02,bournez2005proving}, term rewriting was extended to the probabilistic setting.
Probabilistic programs describe randomized algorithms and probability
distributions, with applications in many areas.
In the probabilistic setting, there are several notions of ``termination''.
A program is \emph{almost-surely terminating} (AST) if the probability of termination is $1$. 
A strictly stronger notion is \emph{positive AST} (PAST), which requires that
the expected runtime is finite. 
While numerous techniques exist to prove (P)AST of imperative programs on numbers (e.g.,
\cite{kaminski2018weakest,mciver2017new,TACAS21,lexrsm,FoundationsTerminationMartingale2020,FoundationsExpectedRuntime2020,rsm,cade19,dblp:journals/pacmpl/huang0cg19,amber,ecoimp,absynth}),
there are only few automatic approaches for programs with complex non-tail recursive
structure
\cite{beutner2021probabilistic,Dallago2017ProbSizedTyping,lago_intersection_2021}. The
approaches that are also suitable for algorithms on recursive data structures
\cite{wang2020autoexpcost,LeutgebCAV2022amor,KatoenPOPL23}
are mostly specialized for specific data structures and cannot easily be adjusted to
other (possibly user-defined) ones, or are not yet fully automated.
In contrast, our goal is
a fully automatic termination analysis for (arbitrary) probabilistic TRSs (PTRSs).

Up to now, only two approaches for automatic termination analysis of PTRSs were
developed
\cite{avanzini2020probabilistic,kassinggiesl2023iAST}.
In \cite{avanzini2020probabilistic},  orderings based on interpretations were adapted to prove
PAST. However, already for non-probabilistic TRSs such
a direct application of orderings is limited in power. 
To obtain a powerful approach, one should
combine such orderings in a modular way, as in the DP framework. 

Indeed, in \cite{kassinggiesl2023iAST}, we adapted
the DP framework to the probabilistic setting in order to prove
innermost AST (iAST), i.e., AST for rewrite sequences which follow
the innermost evaluation strategy.
However, in contrast to the DP framework for ordinary TRSs, 
the probabilistic \emph{dependency tuple} (DT) framework in \cite{kassinggiesl2023iAST} is
\emph{incomplete},  i.e., there are PTRSs which are iAST but where this cannot be proved
with DTs. In this paper, we introduce a new concept of probabilistic DPs and
a corresponding new re\-write relation. In this way, we obtain a novel \emph{complete} 
criterion for iAST via DPs while maintaining soundness for all processors that were developed 
in the probabilistic DT framework of \cite{kassinggiesl2023iAST}.
Moreover, our improvement allows us to introduce additional more powerful
``transformational'' probabilistic DP processors which were not possible in the 
framework of \cite{kassinggiesl2023iAST}.

We recapitulate the DP framework for non-probabilistic TRSs
in \cref{Preliminaries}.
Then, we present our novel ADPs (\emph{annotated dependency pairs}) for probabilistic TRSs in \cref{Probabilistic
  Annotated Dependency Pairs}.
In \cref{The ADP Framework}, we show how to adapt the processors from the framework of
\cite{kassinggiesl2023iAST} to our probabilistic ADP framework.
In addition, our framework allows for the definition of new processors which \emph{transform} ADPs.
As an example, in \cref{ADP Transformations} we adapt the \emph{rewriting processor} to the probabilistic setting, 
which benefits from our new, more precise rewrite relation.
The implementation of our approach in the tool \aprove{} is evaluated in \cref{Evaluation}.
We refer to\paper{ \cite{report} for all proofs.}\report{ App.\ \ref{appendix} for all proofs. In App.\ \ref{moreTrans} we show how the
other transformational processors of the DP framework can also be adapted to the
probabilistic setting. Finally, in App.\ \ref{Examples} we present selected
examples from our new set of benchmarks.}

\section{The DP Framework}\label{Preliminaries}\label{DP Framework}

We assume familiarity with term rewriting \cite{baader_nipkow_1999}
and recapitulate the DP framework with its core processors (see e.g.,
\cite{arts2000termination,gieslLPAR04dpframework,giesl2006mechanizing,hirokawa2005automating}
for details).
We regard finite TRSs $\R$ over a finite signature $\Sigma$
and let $\TSet{\Sigma}{ \VSet}$ denote the set of terms over $\Sigma$ and a
set of variables $\VSet$.
We decompose $\SignatureDC = \SignatureD \uplus \SignatureC$ such that 
$f \in \SignatureD$ if $f = \rootsym(\ell)$ for some $\ell \to r \in \R$.
The symbols in $\SignatureD$ are called \emph{defined symbols}.
For every $f \in \SignatureD$, we introduce a fresh \emph{annotated symbol} $f^{\#}$ of the same 
arity.\footnote{The symbols $f^{\#}$ were called \emph{tuple symbols} in the original DP framework \cite{giesl2006mechanizing} 
and also in \cite{kassinggiesl2023iAST}, 
as they represent the tuple of arguments of the original defined symbol $f$.} 
Let $\SignatureA$ be the set of all annotated symbols and $\SignatureADC = \SignatureA \uplus \SignatureDC$.
For any $t = f(t_1,\linebreak[2] \ldots,t_n) \in \TSet{\SignatureDC}{\VSet}$ with $f \in \SignatureD$, let $t^{\#} = f^{\#}(t_1,\ldots,t_n)$.
For every rule $\ell \to r$ and every (not necessarily proper) subterm $t$ of $r$ with defined root symbol, 
one obtains a \emph{dependency pair} (DP) $\ell^\# \to t^\#$.
$\DPair{\R}$ denotes the set of all dependency pairs of $\R$.
As an example, consider $\R_{\tex} = \{ \eqref{R-ex-1}, \eqref{R-ex-2} \}$ with its
dependency pairs $\DPair{\R_\tex}=\{\eqref{R-ex-3}, \eqref{R-ex-4}\}$.
To ease readability, we often write $\tF$ instead of $\tf^\#$, etc.

\vspace*{-.3cm}
\noindent
\begin{minipage}{0.5\textwidth}
  \begin{align}
		\label{R-ex-1} \tf(\ts(x)) &\!\to\! \tc(\tf(\tg(x)))\\
		\label{R-ex-2} \tg(x)      &\!\to\! \ts(x)
  \end{align}
\end{minipage}
\begin{minipage}{0.4\textwidth}
  \begin{align}
		\label{R-ex-3} \tF(\ts(x)) &\!\to\! \tF(\tg(x))\\
		\label{R-ex-4} \tF(\ts(x)) &\!\to\! \tG(x)
  \end{align}
\end{minipage}

\vspace*{.2cm}

The DP framework uses \emph{DP problems} $(\mathcal{P}, \R)$ where $\mathcal{P}$ is a (finite) set of DPs and $\R$ is a TRS\@.
A (possibly infinite) sequence $t_0, t_1, t_2, \ldots$
with $t_i \itodr \circ \itorstar t_{i+1}$
for all $i$ is an (innermost) $(\mathcal{P}, \R)$-\emph{chain} which represents subsequent ``function calls'' in evaluations.
Here, ``$\circ$'' denotes composition and
steps with $\itodr$ are called \emph{$\mathbf{p}$-steps}, where $\itodr$ is the restriction of $\to_{\mathcal{P}}$ to rewrite steps
where the used redex is in $\NF_{\R}$ (the set of
normal forms w.r.t.\ $\R$). 
Steps with $\itorstar$ are called \emph{$\mathbf{r}$-steps}  
and are used to evaluate the arguments of an annotated function symbol.
So an infinite chain
consists of an infinite number of $\mathbf{p}$-steps with a finite number of $\mathbf{r}$-steps between consecutive $\mathbf{p}$-steps.
For example, $\tF(\ts(x)), \tF(\ts(x)), \ldots$
is an infinite $(\DPair{\R_{\tex}}, \R_{\tex})$-chain,
as $\tF(\ts(x)) \itodrEX \tF(\tg(x)) \itorstarEX \tF(\ts(x))$.
Throughout the paper, we restrict ourselves to innermost rewriting (``$\itor$''), because our adaption
of DPs to the probabilistic setting relies on this evaluation
strategy.\footnote{Moreover, already in the non-probabilistic setting, the restriction to
innermost rewrit\-ing makes termination analysis with DPs substantially more
powerful, e.g., by allowing the application of additional techniques like \emph{usable
rules} and \emph{rewriting} of DPs \cite{gieslLPAR04dpframework,giesl2006mechanizing}. Indeed, we also adapt these techniques in our
novel ADP framework for probabilistic rewriting. Nevertheless, we conjecture that ADPs are
also suitable for an adaption to analyze full  instead of innermost AST, and we will investigate
that in future work.}

A DP problem $(\mathcal{P}, \R)$ is called \emph{innermost terminating} (iTerm) if there is no infinite innermost $(\mathcal{P}, \R)$-chain.
The main result on DPs is the \emph{chain criterion} which states
that there is no infinite sequence $t_1 \itor t_2 \itor \ldots$, i.e., $\R$ is iTerm, iff $(\DPair{\R},\R)$ is iTerm.
The  DP framework is  a \emph{divide-and-conquer}
approach, which
applies \emph{DP processors} to
transform DP problems into simpler  sub-problems.
A \emph{DP processor} $\Proc$ has the form $\Proc(\mathcal{P}, \R) = \{(\mathcal{P}_1,\R_1), \ldots,
(\mathcal{P}_n,\R_n)\}$, where
$\mathcal{P}, \mathcal{P}_1, \ldots, \mathcal{P}_n$ are sets of DPs and  $\R, \R_1, \ldots,
\R_n$ are TRSs. 
A processor $\Proc$ is \emph{sound} if $(\mathcal{P}, \R)$ is iTerm whenever 
$(\mathcal{P}_i,\R_i)$ is iTerm for all 
$1 \leq i \leq n$. 
It is \emph{complete} if $(\mathcal{P}_i,\R_i)$ is iTerm for all 
$1 \leq i \leq n$ whenever  $(\mathcal{P}, \R)$ is iTerm.

So given a TRS $\R$, one starts with the initial
DP problem $(\DPair{\R}, \R)$ and applies sound 
(and preferably complete) DP processors repeatedly until all sub-problems are ``solved'' 
(i.e., sound processors transform them to the empty set). This yields a modular framework for termination
proofs, as different techniques can be used for different sub-problems $(\mathcal{P}_i,\R_i)$.
The following three theorems recapitulate the three most important processors of the DP framework.

The (innermost) \emph{$(\mathcal{P}, \R)$-dependency graph} is a control flow graph that indicates
which DPs can be used after each other in a chain.
Its set of nodes is $\mathcal{P}$
and there is an edge from $\ell_1^\# \to t_1^\#$ to $\ell_2^\# \to t_2^\#$ if there exist
substitutions $\sigma_1, \sigma_2$ such that $t_1^\# \sigma_1 \itorstar \ell_2^\# \sigma_2$ and $\ell_1^\# \sigma_1, \ell_2^\# \sigma_2 \in \NF_{\R}$.
Any infinite $(\mathcal{P}, \R)$-chain corresponds to
an infinite path in the dependency graph, and since the graph is finite, this infinite
path must end in some strongly connected component (SCC).\footnote{Here, a
set $\mathcal{P}'$ of DPs is  an \emph{SCC} if it is a maximal cycle,
i.e., it is a maximal set such that for any $\ell_1^\# \to t_1^\#$ and $\ell_2^\# \to
t_2^\#$ in $\mathcal{P}'$ there is
a non-empty path from $\ell_1^\# \to t_1^\#$ to $\ell_2^\# \to
t_2^\#$ which only traverses nodes from $\mathcal{P}'$.}
Hence, it suffices to consider the SCCs of this graph independently.

   \begin{restatable}[Dependency Graph Processor]{theorem}{depgraph}\label{DGP}
        For the SCCs $\mathcal{P}_1, \ldots, \mathcal{P}_n$ of the $(\mathcal{P}, \R)$-dependency graph,  
        $\Proc_{\mathtt{DG}}(\mathcal{P},\R) = \{(\mathcal{P}_1,\R), \ldots, (\mathcal{P}_n,\R)\}$ is sound and complete. 
    \end{restatable}

    \begin{example}[Dependency Graph]
        Consider the TRS $\R_{\tffg} \!=\! \{ \eqref{R-fgf-1}\}$
        with $\DPair{\R_{\tffg}} \!=\! \{ \eqref{R-fgf-2}, \eqref{R-fgf-3}, \eqref{R-fgf-4} \}$.
        The $(\DPair{\R_{\tffg}}, \R_{\tffg})$-dependency graph is on the right.
        \pagebreak

        \vspace*{-.7cm}
        \noindent
        \begin{minipage}{0.42\textwidth}
        \begin{align}
                \label{R-fgf-1} \tf(\tf(\tg(x))) &\!\to\! \tf(\tg(\tf(\tg(\tf(x)))))
        \end{align}
        \end{minipage}\hfill%
        \begin{minipage}{0.42\textwidth}
        \begin{align}
                \label{R-fgf-2} \tF(\tf(\tg(x))) &\!\to\! \tF(\tg(\tf(\tg(\tf(x)))))\\
                \label{R-fgf-3} \tF(\tf(\tg(x))) &\!\to\! \tF(\tg(\tf(x)))\\
                \label{R-fgf-4} \tF(\tf(\tg(x))) &\!\to\! \tF(x)
        \end{align}
        \end{minipage}\hfill%
        \begin{minipage}{0.15\textwidth}
        \scriptsize
        \vspace*{.5cm}
        \hspace*{.5cm}
        \begin{tikzpicture}
            \node[shape=rectangle,draw=black!100] (A) at (0,1.4)   {\eqref{R-fgf-2}};
            \node[shape=rectangle,draw=black!100] (B) at (0,.7)    {\eqref{R-fgf-3}};
            \node[shape=rectangle,draw=black!100] (C) at (.8,1.4)  {\eqref{R-fgf-4}};

            \path [->,in=290,out=250,looseness=5] (C) edge (C);
            \path [->] (C) edge (A);
            \path [->] (C) edge (B);            
        \end{tikzpicture}
        \end{minipage}
    \end{example}

While the exact dependency graph is not computable in general, there exist sev\-eral
techniques to over-approximate it automatically, see, e.g.,
\cite{arts2000termination,giesl2006mechanizing,hirokawa2005automating}.
In our example, $\Proc_{\mathtt{DG}}(\DPair{\R_{\tffg}}, \R_{\tffg})$
yields the DP problem $(\{ \eqref{R-fgf-4} \}, \R_\tffg)$.

The next processor removes rules that cannot be used for
right-hand sides of dependency pairs when their variables are instantiated with normal forms.

\begin{restatable}[Usable Rules Processor]{theorem}{usablerules}\label{URP}
    Let $\R$ be a TRS\@.
    For every $f \in \SignatureADC$ let $\rules_\R(f) = \{\ell \to r \in \R \mid \rootsym(\ell) = f\}$.
    For any  $t \in  \TSet{\SignatureADC}{ \VSet}$,  
    its \defemph{usable rules} $\urules_\R(t)$ are
    the smallest set
    such that $\urules_\R(x) = \emptyset$ for all $x \in \VSet$ and $\urules_\R(f(t_1, \ldots, t_n)) = \rules_\R(f) \cup \bigcup_{i = 1}^n \urules_\R(t_i) \; \cup \; \bigcup_{\ell \to r \in \rules_\R(f)} \urules_\R(r)$.
    The \defemph{usable rules} for the DP problem $(\mathcal{P}, \R)$ are $\urules(\mathcal{P},\R) =
    \bigcup_{\ell^\# \to t^\# \in \mathcal{P}} \urules_\R(t^\#)$.
    Then $\Proc_{\mathtt{UR}}(\mathcal{P},\R) = \{(\mathcal{P},\urules(\mathcal{P},\R))\}$ is sound but not
    complete.\footnote{\label{CompletenessUsableRules}See \cite{gieslLPAR04dpframework} for a complete version of this processor.
    It extends DP problems by an additional set to store the left-hand sides of all rules
    (including the non-usable ones) 
    to determine whether a rewrite step is innermost.
    We omit this here for readability.}
\end{restatable}

$\Proc_{\mathtt{UR}}\bigl(\{ \eqref{R-fgf-4} \}, \R_{\tffg}\bigr)$
yields the problem $(\{ \eqref{R-fgf-4} \}, \emptyset)$, i.e., 
it removes all rules,
because the right-hand side of $\eqref{R-fgf-4}$ does not contain the defined symbol $\tf$.

A \emph{polynomial interpretation} $\Pol$ is a $\SignatureDC$-algebra which maps every
function symbol $f \in \SignatureDC$ to a polynomial $f_{\Pol} \in \IN[\VSet]$ over the
variables $\VSet$ with coefficients from $\IN$, see \cite{lankford1979proving}.
$\Pol(t)$ denotes the \emph{interpretation} of a term $t$ by the $\Sigma$-algebra $\Pol$.
An arithmetic inequation like $\Pol(t_1) > \Pol(t_2)$ \emph{holds} if it is true for all
instantiations of its variables by natural numbers.
The reduction pair processor\footnote{In this paper, 
we only regard the reduction pair processor with
polynomial interpretations, because for most other classical orderings it
is not clear how to extend them to probabilistic TRSs, where one has to
consider ``expected values of terms''.} allows us to use \emph{weakly monotonic} polynomial interpretations
that do not have to depend on all of their arguments,
i.e., $x \geq y$ implies $f_{\Pol}(\ldots, x, \ldots) \geq f_{\Pol}(\ldots, y, \ldots)$ for all $f \in \SignatureADC$.
The processor requires that all rules and DPs are weakly decreasing
and it removes those DPs that are strictly decreasing.

\begin{restatable}[Reduction Pair Processor]{theorem}{rpp}\label{RPP}
    Let $\Pol:\TSet{\SignatureADC}{\VSet} \to \IN[\VSet]$ be a weakly monotonic polynomial interpretation.
    Let $\mathcal{P} = \mathcal{P}_{\geq} \uplus \mathcal{P}_{>}$  with $\mathcal{P}_{>} \neq
    \emptyset$ such that:
    \begin{itemize}
        \item[(1)] For every $\ell \to r \in \R$, we have $\Pol(\ell) \geq \Pol(r)$.
        \item[(2)] For every $\ell^\# \to t^\# \in \mathcal{P}$, we have $\Pol(\ell^\#) \geq \Pol(t^\#)$.
        \item[(3)] For every $\ell^\# \to t^\# \in \mathcal{P}_{>}$, we have $\Pol(\ell^\#) > \Pol(t^\#)$.
        \end{itemize}
    Then $\Proc_{\mathtt{RP}}(\mathcal{P},\R) = \{(\mathcal{P}_{\geq},\R)\}$ is sound and complete.
\end{restatable}

For $(\{ \eqref{R-fgf-4} \}, \emptyset)$, one can use the reduction pair processor with the poly\-nomial interpretation
that maps $\tf(x)$ to $x+1$
and both $\tF(x)$ and $\tg(x)$ to $x$. Then,
$\Proc_{\mathtt{RP}}\bigl(\{ \eqref{R-fgf-4} \}, \emptyset\bigr) = \{\bigl(\emptyset, \emptyset\bigr)\}$.
As $\Proc_{\mathtt{DG}}(\emptyset, \ldots) = \emptyset$ and all processors used are sound,
this means that there is no infinite innermost chain
for the initial DP problem $(\DPair{\R_{\tffg}}, \R_{\tffg})$ and thus,
$\R_{\tffg}$ is innermost terminating.

\section{Probabilistic Annotated Dependency Pairs}\label{Probabilistic Annotated Dependency Pairs}

In this section we present our novel adaption of DPs to the probabilistic setting.
As in \cite{avanzini2020probabilistic,Faggian2019ProbabilisticRN,bournez2005proving,kassinggiesl2023iAST},
the rules of a probabilistic TRS
have finite multi-distributions on the right-hand sides.
A finite \emph{multi-distribution} $\mu$ on a set $A \neq \emptyset$ is a finite multiset
of pairs $(p:a)$, where $0 < p \leq 1$ is a probability and $a \in A$, with $\sum
_{(p:a) \in \mu} \, p = 1$.  
$\FDist(A)$ is the set of all finite multi-distributions on $A$. 
For $\mu\in\FDist(A)$, its \emph{support} is the multiset $\Supp(\mu)\!=\!\{a
\mid (p\!:\!a)\!\in\!\mu$ for some $p\}$.

A pair  $\ell \to \mu \in \TSet{\Sigma}{ \VSet} \times \FDist(\TSet{\Sigma}{\VSet})$ such
that $\ell \not\in \VSet$ and $\VSet(r) \subseteq \VSet(\ell)$ for every $r \in
\Supp(\mu)$ is a \emph{probabilistic rewrite rule}. 
A \emph{probabilistic TRS} (PTRS) is a finite set of probabilistic rewrite rules.
As an example, consider the PTRS $\R_{\trw}$  with the rule $\tg(x) \!\to\!
\{\nicefrac{1}{2}:\tg(\tg(x)), \, \nicefrac{1}{2}:x \}$, which corresponds to a symmetric
random walk. Let $\tg^2(x)$ abbreviate $\tg(\tg(x))$, etc.

A PTRS $\R$ induces a \emph{rewrite relation} ${\to_{\R}} \subseteq
\TSet{\Sigma}{\VSet} \times \FDist(\TSet{\Sigma}{\VSet})$ where $s \to_{\R} \{p_1:t_1,
\ldots, p_k:t_k\}$  if there is a position $\pi$ of $s$, a rule $\ell \to \{p_1:r_1,
\ldots, p_k:r_k\} \in \R$, and a substitution $\sigma$ such that $s|_{\pi}=\ell\sigma$ and
$t_j = s[r_j\sigma]_{\pi}$ for all $1 \leq j \leq k$.
We call $s \to_{\R} \mu$ an \emph{innermost} rewrite step (denoted $s \itor \mu$) if
$\ell\sigma \in \ANF_{\R}$, where
$\ANF_\R$ is the set of all \emph{terms in argument
normal form w.r.t.\ $\R$}, i.e., 
$t \in \ANF_\R$ iff $t' \in \NF_\R$ for all proper subterms $t'$ of $t$.

To track all possible rewrite sequences (up to non-determinism) with their 
probabilities, we \emph{lift} $\itor$ to \emph{(innermost) rewrite sequence trees} (RSTs). 
An (innermost) $\R$-RST is a tree whose nodes $v$ are labeled by pairs $(p_v, t_v)$ of a
probability $p_v$ and a term $t_v$ such that the edge relation represents a
probabilistic innermost rewrite step. 
More precisely,
$\F{T}\!=\!(V,E,L)$ {\normalsize is an (innermost)}
$\R${\normalsize\emph{-RST}} if (1)
$(V, E)$ is a (possibly infinite) directed tree with nodes
$V \neq \emptyset$ and directed edges
$E \subseteq V \times V$ 
where $vE = \{ w \mid (v,w) \in E \}$ is
finite for every $v \in V$, (2) $L : V \rightarrow (0,1] \times \TSet{\Sigma}{\VSet}$
labels every node $v$ by a probability $p_v$ and a term $t_v$
where $p_v = 1$ for the root $v \in V$
of the tree, and (3) for all  $v \in V$: if
$vE = \{w_1, \ldots,
w_k\} \neq \emptyset$, then $t_v \itor \{\tfrac{p_{w_1}}{p_v}:t_{w_1}, \ldots,
\tfrac{p_{w_k}}{p_v}:t_{w_k}\}$. 
For any innermost $\R$-RST $\F{T}$ we define $|\F{T}|_{\ctleaf} =
\sum_{v \in \ctleaf} \, p_v$, where $\ctleaf$ is the set of $\F{T}$'s leaves. 
An RST $\F{T}$ is \emph{innermost almost-surely terminating}
(iAST) if $|\F{T}|_{\ctleaf} = 1$.    
Similarly,  a PTRS $\R$ is \emph{iAST}
if all innermost $\R$-RSTs are iAST. 
While $|\F{T}|_{\ctleaf} = 1$ holds for every finite RST $\F{T}$, for infinite
RSTs $\F{T}$ we may have $|\F{T}|_{\ctleaf} <1$, and even $|\F{T}|_{\ctleaf} = 0$
if $\F{T}$ has no leaf at all.
This notion is equivalent to the notions of AST in
\cite{avanzini2020probabilistic,kassinggiesl2023iAST}, where one uses a lifting to
multisets instead of trees. 
For example, the infinite
\begin{wrapfigure}[6]{r}{0.37\textwidth}
    \scriptsize
    \vspace*{-0.7cm}
    \begin{tikzpicture}
        \tikzstyle{adam}=[thick,draw=black!100,fill=white!100,minimum
          size=4mm,shape=rectangle split, rectangle split parts=2,rectangle split
          horizontal] 
        \tikzstyle{empty}=[rectangle,thick,minimum size=4mm]
        \node[adam] at (-4, 0)  (a) {$1$ \nodepart{two} $\tg(x)$};
        \node[adam] at (-5, -0.65)  (b) {$\nicefrac{1}{2}$ \nodepart{two} $\tg^2(x)$};
        \node[adam] at (-3, -0.65)  (c) {$\nicefrac{1}{2}$ \nodepart{two} $x$};
        \node[adam] at (-6, -1.3)  (d) {$\nicefrac{1}{4}$ \nodepart{two} $\tg^3(x)$};
        \node[adam] at (-4, -1.3)  (e) {$\nicefrac{1}{4}$ \nodepart{two} $\tg(x)$};
        \node[empty] at (-6.5, -1.9)  (f) {$\ldots$};
        \node[empty] at (-5.5, -1.9)  (g) {$\ldots$};
        \node[empty] at (-4.5, -1.9)  (h) {$\ldots$};
        \node[empty] at (-3.5, -1.9)  (i) {$\ldots$};
        \draw (a) edge[->] (b);
        \draw (a) edge[->] (c);
        \draw (b) edge[->] (d);
        \draw (b) edge[->] (e);
        \draw (d) edge[->] (f);
        \draw (d) edge[->] (g);
        \draw (e) edge[->] (h);
        \draw (e) edge[->] (i);
    \end{tikzpicture}
\end{wrapfigure}
$\R_{\trw}$-RST $\F{T}$ on the side has $|\F{T}|_{\ctleaf} = 1$. 
In fact, $\R_{\trw}$ is iAST, because $|\F{T}|_{\ctleaf} = 1$
holds for all innermost $\R_{\trw}$-RSTs $\F{T}$. 

As shown in \cite{kassinggiesl2023iAST}, to adapt the DP framework in order to prove iAST
of PTRSs, one has to regard all DPs resulting from the same rule \emph{at
once}. Otherwise, one would not be able to distinguish between the DPs of the TRS with the
rule $\ta \to \{\nicefrac{1}{2}: \tb, \nicefrac{1}{2}: \tc(\ta,\ta) \}$ which is iAST and
the rule  $\ta \to \{\nicefrac{1}{2}: \tb, \nicefrac{1}{2}: \tc(\ta,\ta,\ta) \}$, which is
not iAST.  For that reason, in the adaption of
the DP framework to PTRSs in \cite{kassinggiesl2023iAST},
one constructs \emph{dependency tuples} (DTs) whose right-hand sides combine
the right-hand sides of all dependency pairs
resulting from one  rule. However, 
a drawback of this approach is that the resulting
chain criterion is not complete, i.e.,
it allows for chains that do
not correspond to any rewrite sequence of the original PTRS $\R$.

\addtocounter{theorem}{1}
\setcounter{counter-completeness-ctr}{\value{theorem}}

\noindent \emph{Example \arabic{theorem}}. 
    Consider the PTRS $\R_{\tic}$ with the rules 

    \vspace*{-0.7cm}
    \begin{minipage}[t]{5cm}
        \begin{align}
            \ta &\to \{1 : \tf(\th(\tg),\tg)\} \label{Rtic-rule-1}\\
            \tg &\to\{\nicefrac{1}{2} : \tb_1, \nicefrac{1}{2} : \tb_2\} \label{Rtic-rule-2}\!
        \end{align}
    \end{minipage}
    \begin{minipage}[t]{5cm}
        \begin{align}
            \th(\tb_1) &\to \{1 : \ta\} \label{Rtic-rule-3}\\
            \tf(x,\tb_2) &\to \{1 : \ta\} \label{Rtic-rule-4}\!
        \end{align}
    \end{minipage}

    \vspace*{.1cm}
    
    \noindent
    and the  $\R_{\tic}$-RST below.  So $\ta$ can be rewritten to the normal form 
    $\tf(\th(\tb_2),\tb_1)$

    \begin{wrapfigure}[8]{r}{0.72\textwidth}
    \scriptsize

\vspace*{-.8cm}

    \begin{tikzpicture}
            \tikzstyle{adam}=[thick,draw=black!100,fill=white!100,minimum size=4mm,shape=rectangle split, rectangle split parts=2,rectangle split horizontal]
            \tikzstyle{adam2}=[thick,draw=red!100,fill=white!100,minimum size=4mm,shape=rectangle split, rectangle split parts=2,rectangle split horizontal]
            \tikzstyle{empty}=[rectangle,thick,minimum size=4mm]
            
            \node[adam] at (0, 0)  (a) {$1$ \nodepart{two} $\ta$};
            \node[adam] at (0, -0.6)  (b) {$1$ \nodepart{two} $\tf(\th(\tg),\tg)$};
            \node[adam] at (-2.2, -1.2)  (c) {$\nicefrac{1}{2}$ \nodepart{two} $\tf(\th(\tg),\tb_1)$};
            \node[adam] at (2.2, -1.2)  (d) {$\nicefrac{1}{2}$ \nodepart{two} $\tf(\th(\tg),\tb_2)$};
            \node[adam] at (-3.3, -1.8) (e1) {$\nicefrac{1}{4}$ \nodepart{two} $\tf(\th(\tb_1),\tb_1)$};
            \node[adam2,label=below:{normal form}] at (-1.1, -1.8)  (e2) {$\nicefrac{1}{4}$ \nodepart{two} $\tf(\th(\tb_2),\tb_1)$};
            \node[adam] at (1.1, -1.8)  (e3) {$\nicefrac{1}{4}$ \nodepart{two} $\tf(\th(\tb_1),\tb_2)$};
            \node[adam] at (3.3, -1.8)  (e4) {$\nicefrac{1}{4}$ \nodepart{two} $\tf(\th(\tb_2),\tb_2)$};
            \node[adam] at (-3.3, -2.4)  (e11) {$\nicefrac{1}{4}$ \nodepart{two} $\tf(\ta,\tb_1)$};
            \node[adam] at (1.1, -2.4)  (e33) {$\nicefrac{1}{4}$ \nodepart{two} $\tf(\ta,\tb_2)$};
            \node[adam] at (3.3, -2.4)  (e44) {$\nicefrac{1}{4}$ \nodepart{two} $\ta$};
            \node[empty] at (-3.3, -3)  (e111) {$\ldots$};
            \node[empty] at (1.1, -3)  (e333) {$\ldots$};
            \node[empty] at (3.3, -3)  (e444) {$\ldots$};
            
            \draw (a) edge[->] (b);
            \draw (b) edge[->] (c);
            \draw (b) edge[->] (d);
            \draw (c) edge[->] (e1);
            \draw (c) edge[->] (e2);
            \draw (d) edge[->] (e3);
            \draw (d) edge[->] (e4);
            \draw (e1) edge[->] (e11);
            \draw (e3) edge[->] (e33);
            \draw (e4) edge[->] (e44);
            \draw (e11) edge[->] (e111);
            \draw (e33) edge[->] (e333);
            \draw (e44) edge[->] (e444);
        \end{tikzpicture}
    \end{wrapfigure}    

    \noindent
    with probability $\nicefrac{1}{4}$ and to
    the terms $\tf(\ta,\tb_1)$ and $\ta$    that contain the redex $\ta$ with  a
    probability of $\nicefrac{1}{4} + \nicefrac{1}{4}
    = \nicefrac{1}{2}$. In the term $\tf(\ta,\tb_2)$, one can rewrite the subterm
    $\ta$, and if that ends in a normal form, one can still rewrite the outer $\tf$ which will
    yield $\ta$ again.  So to over-approximate
    the probability of non-termination, one
    could consider the term $\tf(\ta,\tb_2)$ as if one had two occurrences of $\ta$. Then this
    would correspond to a random walk where the number of $\ta$ symbols is decreased by 1 with
    probability $\nicefrac{1}{4}$, increased by 1 with probability $\nicefrac{1}{4}$, and
    kept the same with probability $\nicefrac{1}{2}$. Such a random walk is
    AST, and since a similar observation holds for all $\R_{\tic}$-RSTs, $\R_{\tic}$ is iAST
    (we will prove iAST of $\R_{\tic}$ with our new ADP framework in \Cref{The ADP Framework,ADP Transformations}).

    In contrast, the DT framework from \cite{kassinggiesl2023iAST}
    fails on this example. As mentioned, the right-hand sides of DTs
    combine
    the right-hand sides of all dependency pairs
    resulting from one  rule. So the right-hand side of the DT for \eqref{Rtic-rule-1} 
    contains the term $\tcom_4(\tF(\th(\tg),\tg), \tH(\tg), \tG, \tG)$, where $\tcom_4$ is a special
    compound symbol of arity 4.
    However, here it is no longer clear which occurrence of the annotated symbol
    $\tG$ corresponds to which occurrences of $\tg$. Therefore, when rewriting an occurrence of
    $\tG$, in the ``chains'' of \cite{kassinggiesl2023iAST} one may also rewrite arbitrary
    occurrences of $\tg$ simultaneously. (For that reason, in
    \cite{kassinggiesl2023iAST} one also couples the DT together with its
    original rule.)
    In particular, \cite{kassinggiesl2023iAST} also allows a simultaneous
    rewrite step of all underlined symbols in $\tcom(\tF(\th(\tg),\underline{\tg}),
    \tH(\underline{\tg}), \underline{\tG}, \tG)$ even though the underlined $\tG$ cannot
    correspond to both underlined $\tg$ symbols. As shown in\report{ \Cref{Incompleteness of Chain Criterion for Example} in App.\ \ref{appendix}}\paper{
    \cite{report}},
    this leads to a chain that is not iAST and
    that does not correspond to any $\R_{\tic}$-rewrite sequence. To avoid this problem,
    one would have to keep track of the connections between annotated symbols and the
    corresponding original subterms. However, such an improvement would become very complicated in the
    formalization of \cite{kassinggiesl2023iAST}.

\setcounter{theorem}{\value{counter-completeness-ctr}}
\addtocounter{theorem}{1}

Therefore, in contrast to \cite{kassinggiesl2023iAST}, in our new notion of DPs,
we annotate defined symbols directly in the original rewrite rule
instead of extracting annotated subterms from its right-hand side.
This makes the definition easier, more elegant, and more readable, and 
allows us to solve the incompleteness problem of \cite{kassinggiesl2023iAST}.
  
\begin{definition}[Annotations]
    Let $t \in \TSet{\SignatureADC}{\VSet}$ be an \defemph{annotated term}
    and for $\Sigma' \subseteq \SignatureADC$, let $\pos_{\Sigma'}(t)$ be all positions of $t$
    with symbols from $\Sigma'$.
    For a set of positions $\Phi \subseteq \posDT(t)$, let
    $\anno_\Phi(t)$ be the variant of $t$ where the symbols at positions from $\Phi$ in $t$
    are annotated and all other annotations are removed.
    Thus, $\posT(\anno_\Phi(t)) = \Phi$, and
    $\anno_\emptyset(t)$ removes all annotations from $t$, where we often write
    $\flat(t)$ instead of $\anno_\emptyset(t)$. We extend $\flat$ to multi-distributions,
    rules, and sets of rules by removing the annotations of all occurring terms.
  We write $\annoD(t)$ instead of $\anno_{\pos_{\SignatureD}(t)}(t)$ to annotate all
    defined symbols in $t$, and $\anno_{\varepsilon}(t)$ instead of
    $\anno_{\{\varepsilon\}}(t)$ to annotate just the root symbol of $t$.
   Moreover, let  $\disannoPos{\pi}(t)$ result from
    removing all annotations from $t$ that are strictly above the position $\pi$.
   Finally, we write $t \trianglelefteq_{\#} s$ if
there is a $\pi \in \posT(s)$ and $t = \flat(s|_\pi)$, i.e., $t$ results from a subterm of
$s$ with annotated root symbol by removing its annotation.
\end{definition}

\begin{example}
    So if  $\tg \in \SignatureD$, then we have $\anno_{\{1\}}(\tg(\tg(x))) =
    \anno_{\{1\}}(\tG(\tG(x))) = \tg(\tG(x))$, 
    $\anno_\SignatureD(\tg(\tg(x))) =
    \anno_{\{\varepsilon, 1 \}}(\tg(\tg(x))) =
    \tG(\tG(x))$, and $\flat(\tG(\tG(x))) = \tg(\tg(x))$. 
    Moreover, $\disannoPos{1}(\tG(\tG(x))) = \tg(\tG(x))$ and $\tg(x)  \trianglelefteq_{\#} \tg(\tG(x))$.
\end{example}

Next, we define the \emph{canonical annotated dependency pairs} for a given PTRS.

\begin{definition}[Canonical Annotated Dependency Pairs]\label{def:canonical-ADPs}
    For a rule $\ell \to \mu = \{ p_1 : r_1, \ldots, p_k : r_k \}$, its canonical
    \defemph{annotated dependency pair} (ADP)
    is 
    \[
        \DPair{\ell \to \mu} \;\; = \;\; \ell \to \{ p_1 : \annoD(r_1), \ldots, p_k : \annoD(r_k)\}^{\ttrue}
    \]
    The canonical ADPs of a PTRS $\R$ are $\DPair{\R} = \{\DPair{\ell \to \mu} \mid \ell \to \mu \in \R\}$.
\end{definition}

\begin{example}
    For $\R_{\trw}$,
    the canonical ADP for $\tg(x) \to \{\nicefrac{1}{2}: \tg(\tg(x))), \nicefrac{1}{2}: x\}$ is
    $\tg(x) \to \{\nicefrac{1}{2}: \tG(\tG(x)), \nicefrac{1}{2}: x\}^{\ttrue}$
      instead of  the (complicated)  DT from \cite{kassinggiesl2023iAST}:
    \[
        \DTuple{\R_{\trw}} = \{ \langle \tG(x),\tg(x) \rangle \to \{
        \nicefrac{1}{2} : \langle \Com{2}(\tG(\tg(x)),\tG(x)),\tg^2(x) \rangle, \nicefrac{1}{2}
        : \langle \Com{0},x \rangle\} \}
    \]  
\end{example}

So the left-hand side of an ADP is just
the left-hand side of the original rule.
The right-hand side of the ADP results from the right-hand side of the original\linebreak rule 
by replacing all  $f\in\SignatureD$ with $f^{\#}$.
Moreover, every ADP has a flag $m \in \{\ttrue,\linebreak \tfalse\}$
to indicate whether this ADP may be used for an $\mathbf{r}$-step at a position below the next $\mathbf{p}$-step.
(This flag will later be modified by our usable rules processor.)
In general, we work with the following rewrite systems in our new framework.

\begin{definition}[Annotated Dependency Pairs, $\itored{}{}{\PP}$]\label{def:ADPs-and-Rewriting}
    An \defemph{ADP} has the form $\ell \ruleArr{}{}{} \{ p_1:r_{1}, \ldots, p_k: r_k\}^{m}$,
    where $\ell \in \TSet{\Sigma}{\VSet}$ with $\ell \notin \VSet$, $m \in \{\ttrue, \tfalse\}$,
    and for all $1 \leq j \leq k$ we have $r_{j} \!\in\! \TSet{\SignatureADC}{\VSet}$ with $\VSet(r_j) \subseteq \VSet(\ell)$.

    Let $\PP$ be a finite set of ADPs (a so-called \defemph{ADP problem}).
    An annotated term $s \in \TSet{\SignatureADC}{\VSet}$ rewrites with $\PP$ to
    $\mu = \{p_1:t_1,\ldots,p_k:t_k\}$ (denoted  $s \itored{}{}{\PP} \mu$)
    if there is a rule $\ell \ruleArr{}{}{} \{ p_1:r_{1}, \ldots, p_k: r_k\}^{m} \in
    \PP$, a substitution $\sigma$, and a position $\pi \in \pos_{\SignatureD \cup
    \SignatureA}(s)$ such that $\flat(s|_\pi)=\ell\sigma \in \ANF_\PP$, and  
    for all $1 \leq j \leq k$ we have
    \begin{equation*}
        \begin{array}{rllllll@{\hspace*{1cm}}l}
        t_j &=                  &s[r_j\sigma]_{\pi}         & \text{if} & \pi \in \posT(s)    & \text{and} & m = \ttrue   & (\mathbf{pr})\\
        t_j &= \disannoPos{\pi}(&s[r_j\sigma]_{\pi})        & \text{if} & \pi \in \posT(s)    & \text{and} & m = \tfalse  & (\mathbf{p})\\
        t_j &=                  &s[\flat(r_j)\sigma]_{\pi}  & \text{if} & \pi \not\in\posT(s) & \text{and} & m = \ttrue   & (\mathbf{r})\\
        t_j &= \disannoPos{\pi}(&s[\flat(r_j)\sigma]_{\pi}) & \text{if} & \pi
        \not\in\posT(s) & \text{and} & m = \tfalse  & (\mathbf{irr}) \!
        \end{array}
    \end{equation*}
    To highlight the position $\pi$ of the redex, we also write $s \itored{}{}{\PP,\pi} t$.
    Again, $\ANF_\PP$ is the set of all terms
    in argument normal form w.r.t.\ $\PP$.
\end{definition}

Rewriting with $\PP$ can be seen as ordinary term rewriting while considering and  modifying annotations.
In the ADP framework, we represent all DPs resulting from a rule as well as the original rule by just one ADP. 
So for example, the ADP  $\tg(x) \to \{\nicefrac{1}{2}: \tG(\tG(x)), \nicefrac{1}{2}: x\}^{\ttrue}$ for the rule
$\tg(x) \to \{\nicefrac{1}{2}: \tg(\tg(x)), \nicefrac{1}{2}: x\}$
represents both DPs resulting from the two occurrences of $\tg$ on the right-hand
side, and the rule itself (by simply disregarding all annotations of the ADP).

As in the classical DP framework, our goal is to track specific reduction sequences where
(1) there are $\mathbf{p}$-steps where the root symbol of the redex is annotated and a DP is applied, and
(2) between two $\mathbf{p}$-steps there can be several $\mathbf{r}$-steps where rules are applied below the position of the next $\mathbf{p}$-step.

A step of the form $(\mathbf{pr})$  in \Cref{def:ADPs-and-Rewriting} can
represent both
$\mathbf{p}$- and $\mathbf{r}$-steps.
All annotations are kept during this step except for annotations of the subterms that
correspond to variables of the applied rule. These subterms are always in normal form due to the
innermost evaluation strategy and we erase their annotations
in order to handle rewriting with non-left-linear rules correctly.
A $(\mathbf{pr})$-step at posi\-tion $\pi$ plays the
role of an $\mathbf{r}$-step
for terms in multi-distributions where one later  rewrites an annotated symbol at a position above $\pi$,
and for all other terms it  plays the role of a $\mathbf{p}$-step.
As an example,
for a PTRS $\R_{\textwo}$ with the rules $\tg(x,x) \to \{1:\tf(x)\}$ and $\tf(\ta) \to \{1:\tf(\tb)\}$,  
we have the canonical ADPs $\tg(x,x) \to \{1:\tF(x)\}^{\ttrue}$ and $\tf(\ta) 
\to \{1:\tF(\tb)\}^{\ttrue}$, and we can rewrite
$\tG(\tF(\tb),\tf(\tb))  \itored{}{}{\DPair{\R_{\textwo}}} \{1: \tF(\tf(\tb))\}$
using the first ADP. Here, we have $\pi =
\varepsilon$, $\flat(s|_\varepsilon) = \tg(\tf(\tb),\tf(\tb)) = \ell \sigma$ where $\sigma$
instantiates $x$ with the normal form $\tf(\tb)$, and $r_1 = \tF(x)$.

A step of the form $(\mathbf{r})$
rewrites at the position of a non-annotated defined symbol. So
this represents an $\mathbf{r}$-step and thus, we remove all annotations
from the right-hand side $r_j$.
As an example, we have $\tG(\tF(\tb),\tf(\ta)) \itored{}{}{\DPair{\R_{\textwo}}}
\{1:\tG(\tF(\tb),\tf(\tb)) \}$  using the ADP $\tf(\ta) 
\to \{1:\tF(\tb)\}^{\ttrue}$.

A step of the form  $(\mathbf{p})$ represents a $\mathbf{p}$-step. Thus,
we remove all annotations above the position $\pi$, 
because no $\mathbf{p}$-steps are possible above $\pi$.
So if $\PP$ contains  $\tf(\ta) 
\to \{1:\tF(\tb)\}^{\tfalse}$, then
$\tG(\tF(\tb),\tF(\ta)) \itored{}{}{\PP}
\{1:\tg(\tF(\tb),\tF(\tb)) \}$.

Finally, a step of the form $(\mathbf{irr})$ is an $\mathbf{r}$-step that is irrelevant for proving iAST, because due to $m = \tfalse$, afterwards there cannot be a $\mathbf{p}$-step at a position above.
For example, if $\PP$ again contains  $\tf(\ta) 
\to \{1:\tF(\tb)\}^{\tfalse}$, then
$\tG(\tF(\tb),\tf(\ta)) \itored{}{}{\PP}
\{1:\tg(\tF(\tb),\tf(\tb)) \}$.
Such $(\mathbf{irr})$-steps are needed to ensure that all rewrite steps with $\R$ are also
possible with the ADP problems $\PP$ that result from $\DPair{\R}$ when applying ADP
processors. These processors only modify annotations, but keep the rest of the rules
unchanged. So for all these
ADP problems $\PP$, we have $\R=\flat(\PP)$ and $\flat(t) \in \ANF_{\R}$ iff $t \in
\ANF_\PP$ for all $t \in \TSet{\SignatureADC}{\VSet}$, i.e., the innermost evaluation strategy is not affected by the
application of ADP processors.
This is different from the classical DP framework, where the usable rules
processor reduces the number of rules. This may result in new
redexes that are allowed for innermost rewriting.
Thus, the usable rules processor in our new ADP framework is \emph{complete},
whereas in \cite{gieslLPAR04dpframework},
one has to extend DP problems by
an additional component to achieve
completeness of this processor (see \Cref{CompletenessUsableRules}).

Now, $s \itor \{p_1:t_1, \ldots, p_k:t_k\}$ essentially\footnote{We have
$\annoD(s) \itored{}{}{\DPair{\R}} \{p_1:t_1', \ldots, p_k:t_k'\}$
where $t_j'$ and $\annoD(t_j)$ are the same up to some annotations of subterms that are
$\DPair{\R}$-normal forms. 
The reason is that as mentioned above, annotations of the subterms (in normal form) that
correspond to variables of the rule are erased. So for example, rewriting
$\tG(\tF(\tb),\tF(\tb))$ with $\DPair{\R_{\textwo}}$ yields $\{1: \tF(\tf(\tb))\}$ and
not $\{1: \tF(\tF(\tb))\}$.
}
implies $\annoD(s) \itored{}{}{\DPair{\R}} \{p_1:\annoD(t_1), \ldots, p_k:\annoD(t_k)\}$, 
and we got rid of any ambiguities in the rewrite relation that led to incompleteness in
\cite{kassinggiesl2023iAST}.
While our ADPs are much simpler than the DTs of \cite{kassinggiesl2023iAST}, due to their
annotations they still
contain all information that is needed to define the required DP processors.

Instead of chains of DPs, in the probabilistic setting one works with \emph{chain trees}
\cite{kassinggiesl2023iAST}, 
where $\mathbf{p}$- and $\mathbf{r}$-steps are indicated by $P$- and $R$-nodes in the tree.
Chain trees are defined analogously to RSTs, but the crucial requirement is that every
infinite path of the tree must contain infinitely many steps of the forms $(\mathbf{pr})$ or
$(\mathbf{p})$. Thus, in our setting
$\F{T}=(V,E,L,P)$ is a $\PP$\emph{-chain tree} (CT) if
\begin{enumerate}
\item $(V, E)$ is a (possibly infinite) directed tree with nodes
  $V \neq \emptyset$ and directed edges $E \subseteq V \times V$  where $vE = \{ w \mid (v,w) \in E \}$ is  finite for every $v \in V$.
    \item $L:V\rightarrow(0,1]\times\TSet{\SignatureADC}{\VSet}$ labels every node $v$ by a probability $p_v$ and a term $t_v$.
    For the root $v \in V$ of the tree, we have $p_v = 1$.
    \item $P \subseteq V \setminus \ctleaf$ (where $\ctleaf$ are all leaves) is a subset
    of the inner nodes to indicate whether we use $(\mathbf{pr})$ or $(\mathbf{p})$ for the next rewrite step.
    $R = V \setminus (\ctleaf \cup P)$ are all inner nodes that are not in $P$, i.e.,
    where we rewrite using $(\mathbf{r})$ or $(\mathbf{irr})$.
    \item For all $v \in P$:
    if $vE = \{w_1, \ldots, w_k\}$, then $t_v  \itored{}{}{\PP} \{\tfrac{p_{w_1}}{p_v}:t_{w_1}, \ldots, \tfrac{p_{w_k}}{p_v}:t_{w_k}\}$ using Case $(\mathbf{pr})$ or $(\mathbf{p})$. 
    \item For all $v \in R$:
    if $vE = \{w_1, \ldots, w_k\}$, then $t_v \itored{}{}{\PP}
    \{\tfrac{p_{w_1}}{p_v}:t_{w_1}, \ldots, \tfrac{p_{w_k}}{p_v}:t_{w_k}\}$ using  Case $(\mathbf{r})$ or $(\mathbf{irr})$. 
    \item Every infinite path in $\F{T}$ contains infinitely many nodes from $P$. 
\end{enumerate}
Let $|\F{T}|_{\ctleaf} = \sum_{v \in \ctleaf} \, p_v$.
We define that $\PP$ is iAST if $|\F{T}|_{\ctleaf} = 1$ for all $\PP$-CTs $\F{T}$.
So Conditions 1--5 ensure that the chain tree corresponds to an RST and 
Condition 6 requires that one may only use finitely many
$\mathbf{r}$-steps before the next $\mathbf{p}$-step.
This yields a chain criterion as in the non-probabilistic setting, where (in contrast to
the chain criterion of \cite{kassinggiesl2023iAST}) we again have ``iff'' instead of ``if''. 

\begin{restatable}[Chain Criterion]{theorem}{ProbChainCriterion}\label{theorem:prob-chain-criterion-new}
  $\R$ is iAST iff $\DPair{\R}$ is iAST.
\end{restatable}

Our chain criterion is complete (``only if''), because
ADPs only add annotations to rules.
Hence, every $\DPair{\R}$-CT
can be turned into an $\R$-RST by omitting all annotations.
So in contrast to \cite{kassinggiesl2023iAST}, the step from the original PTRS to ADPs
does not affect the ``potential power'' of the approach. Moreover, in the future this\linebreak may
also allow the development of techniques to \emph{disprove} iAST within the ADP framework.
To prove soundness (``if''), one has to show that
every $\R$-RST  can be simulated by a 
$\DPair{\R}$-CT. As mentioned, all proofs can be 
found in\paper{ \cite{report}.}\report{ App.\ \ref{appendix}.}

\section{The ADP Framework}\label{The ADP Framework}

Our new (probabilistic) ADP framework again applies processors to
transform an ADP problem into simpler sub-problems.
An \emph{ADP processor} $\Proc$ has the form $\Proc(\PP) = \{\PP_1, \ldots,\PP_n\}$, where
$\PP, \PP_1, \ldots, \PP_n$ are ADP problems. 
$\Proc$ is \emph{sound} if $\PP$ is iAST whenever 
$\PP_i$ is iAST for all $1 \leq i \leq n$. 
It is \emph{complete} if $\PP_i$ is iAST for all 
$1 \leq i \leq n$ whenever $\PP$ is iAST.
For a PTRS $\R$, one starts with the canonical
ADP problem $\DPair{\R}$ and applies sound 
(and preferably complete) ADP processors repeatedly until the
ADPs contain no annotations anymore.
Such an ADP problem is trivially iAST. 
The framework again allows for modular termination proofs, since different techniques can be applied on each sub-problem $\PP_i$.

We now adapt the processors from \cite{kassinggiesl2023iAST} to our new framework.
The (innermost)  $\PP$-\emph{dependency graph} is a control flow graph between ADPs from $\PP$, 
indicating whether an ADP $\alpha$ may lead to an application of another 
ADP $\alpha'$ on an annotated subterm introduced by $\alpha$. 
This possibility is not related to the probabilities. Hence, we can use the \emph{non-probabilistic variant}
$\nonprob(\PP) = \{\ell \to \flat(r_j) \mid \ell \to \{p_1:r_1, \ldots, p_k:r_k\}^{\ttrue} \in \PP, 1
\leq j \leq k\}$, which is an ordinary TRS over the signature $\Sigma$.
Note that for $\nonprob(\PP)$ we only need to consider rules
with the flag $\ttrue$, since only such rules
can be used at a position below the next $\mathbf{p}$-step.

\begin{definition}[Dependency Graph]
    The \defemph{$\PP$-dependency graph} has the nodes $\PP$ and 
    there is an edge from $\ell_1 \ruleArr{}{}{} \{ p_1:r_{1}, \ldots, p_k: r_k\}^{m}$ to $\ell_2 \to \ldots$ 
    if there are substitutions
    $\sigma_1, \sigma_2$ and a $t \trianglelefteq_{\#} r_{j}$ for some $1 \leq j \leq k$ 
    such that $t^\# \sigma_1 \ito_{\nonprob(\PP)}^*
    \ell_2^\# \sigma_2$ and both $\ell_1 \sigma_1$ and $\ell_2 \sigma_2$ are in $\ANF_{\PP}$.
\end{definition}

So there is an edge from an ADP $\alpha$ to an ADP $\alpha'$ if after a step of the form
$(\mathbf{pr})$ or $(\mathbf{p})$ with $\alpha$ at position $\pi$
there may eventually come another
step of the form $(\mathbf{pr})$ or $(\mathbf{p})$ with $\alpha'$ on or below $\pi$.
Hence, for every path in a $\PP$-CT from a $P$-node where an
annotated subterm $f^\#(\ldots)$ is introduced
to the next $P$-node where the subterm $f^\#(\ldots)$ at this position is rewritten, 
there is a corresponding  edge in the
$\PP$-dependency graph. Since every infinite path in a CT contains infinitely
many nodes from $P$, every such path traverses a
cycle of the dependency graph infinitely often.
Thus, it suffices to consider the SCCs of the dependency graph separately.
In our framework, this means that we remove the annotations from all rules except those
that are in the SCC that we want to analyze.
As in \cite{kassinggiesl2023iAST}, to automate the following two processors, the same over-approximation
techniques as for the non-probabilistic dependency graph can be used.

\begin{restatable}[Probabilistic Dependency Graph Processor]{theorem}{ProbDepGraphProc}\label{theorem:prob-DGP}
    For the SCCs $\PP_1, \ldots, \PP_n$ of the
    $\PP$-dependency graph,
    \mbox{\small $\Proc_{\mathtt{DG}}(\PP)\!=\!\{\PP_1 \cup \flat(\PP \setminus \PP_1), ...,\PP_n \cup \flat(\PP \setminus \PP_n)\}$} is sound and complete.
\end{restatable}

\begin{example}
Consider the PTRS $\R_{\tic}$ from Ex.\ \arabic{counter-completeness-ctr}
with the canonical ADPs

\vspace*{-0.6cm}
\hspace*{-.7cm}\begin{minipage}[t]{6cm}
    \begin{align}
        \ta &\to \{1 : \tF(\tH(\tG),\tG)\}^{\ttrue} \label{Rtic-adp-1}\\
        \tg &\to\{\nicefrac{1}{2} : \tb_1, \nicefrac{1}{2} : \tb_2\}^{\ttrue} \label{Rtic-adp-2}\!
    \end{align}
\end{minipage}
\hspace*{.7cm}
\begin{minipage}[t]{5cm}
    \begin{align}
        \th(\tb_1) &\to \{1 : \tA\}^{\ttrue} \label{Rtic-adp-3}\\
        \tf(x,\tb_2) &\to \{1 : \tA\}^{\ttrue} \label{Rtic-adp-4}\!
    \end{align}
\end{minipage}
\end{example}

\begin{wrapfigure}[3]{r}{0.13\textwidth}
    \scriptsize
    \vspace*{-0.6cm}
    \hspace*{-.15cm}\begin{tikzpicture}
            \node[shape=rectangle,draw=black!100, minimum size=3mm] (A) at (0,0) {\eqref{Rtic-adp-1}};
            \node[shape=rectangle,draw=black!100, minimum size=3mm] (B) at (1,0) {\eqref{Rtic-adp-2}};
            \node[shape=rectangle,draw=black!100, minimum size=3mm] (C) at (0,0.7) {\eqref{Rtic-adp-3}};
            \node[shape=rectangle,draw=black!100, minimum size=3mm] (D) at (1,0.7) {\eqref{Rtic-adp-4}};
        
            \path [->] (A) edge (B);
            \path [->] (A) edge (C);
            \path [->] (A) edge (D);
            \path [->] (C) edge (A);    
            \path [->] (D) edge (A);              
        \end{tikzpicture}      
\end{wrapfigure}
\noindent The $\DPair{\R_{\tic}}$-dependency graph can be seen on the right.
As \eqref{Rtic-adp-2} is not contained in the only SCC, we can remove all annotations from
\eqref{Rtic-adp-2}.
However, since \eqref{Rtic-adp-2} already does not contain any annotations, here
the dependency graph processor does not change $\DPair{\R_{\tic}}$.

\smallskip

To remove the annotations of \emph{non-usable}
terms like $\tG$ in \eqref{Rtic-adp-1} that lead out of
the SCCs of the dependency graph, one can
apply the \emph{usable terms processor}. 

\begin{restatable}[Usable Terms Processor]{theorem}{UsableTermsProc}\label{theorem:prob-UPP}
    Let $\ell_1 \in \TSet{\Sigma}{\VSet}$ and $\PP$ be an ADP problem.
    We call $t \in \TSet{\SignatureADC}{\VSet}$ with
    $\rootsym(t) \in \SignatureD^\#$ \defemph{usable} w.r.t.\ $\ell_1$ and
    $\PP$ if there
    are substitutions $\sigma_1, \sigma_2$ and
    an $\ell_2 \ruleArr{}{}{} \mu_2 \in \PP$ where $\mu_2$ contains an
    annotated symbol,
    such that $\anno_{\varepsilon}(t) \sigma_1 \ito_{\nonprob(\PP)}^*
    \ell_2^\# \sigma_2$ and both $\ell_1 \sigma_1$ and $\ell_2
    \sigma_2$ are in $\ANF_{\PP}$.
    Let $\flat_{\ell,\PP}(s)$
    result from $s$ by removing the annotations from the roots of all its
        subterms  that 
    are 
    not usable w.r.t.\ $\ell$ and
    $\PP$, i.e., $\posT(\flat_{\ell,\PP}(s)) = \{ \pi \in \posT(s) \mid s|_\pi$ is usable
    w.r.t.\  $\ell_1$ and
    $\PP\,\}$. 
    The transformation that removes the annotations from the roots of all
    non-usable terms in the
    right-hand sides of ADPs is $\mathcal{T}_\mathtt{UT}(\PP) \!=\! \{ \ell \!\to\! 
        \{ p_1: \flat_{\ell,\PP}(r_1), \ldots, p_k:\flat_{\ell,\PP}(r_k)\}^{m}
        \mid  \ell \!\to\! \{ p_1: r_1, \ldots, p_k:r_k\}^{m} \!\in\! \PP \}$.
    Then $\Proc_{\mathtt{UT}}(\PP) = \{\mathcal{T}_\mathtt{UT}(\PP)\}$ is sound and complete. 
\end{restatable} 

\noindent
So for $\DPair{\R_{\tic}}$, $\Proc_{\mathtt{UT}}$ replaces $\eqref{Rtic-adp-1}$ 
by $\;\ta \to \{1 : \tF(\tH(\tg),\tg)\}^{\ttrue} \quad (\ref{Rtic-adp-1}')$.

As in \Cref{URP} of the ordinary DP framework, 
the idea of the \emph{usable rules processor} remains to find rules
that cannot be used below steps
at annotations in right-hand sides of ADPs when their variables are instantiated with normal
forms.

\begin{restatable}[Prob.\ Usable Rules Processor]{theorem}{ProbUsRulesProc}\label{def:prob-usable-rules}
   For an ADP problem $\PP$
and $f\!\in\!\SignatureADC$, let $\rules_\PP(f) = \{\ell \to \mu^{\true} \in \PP \mid \rootsym(\ell) = f\}$.
    For any  $t\!\in\!\TSet{\SignatureADC}{\VSet}$,
    its\linebreak \defemph{usable rules} $\urules_\PP(t)$ are the smallest set
   with $\urules_\PP(x)=\emptyset$ for all $x\in\VSet$
    and $\urules_\PP(f(t_1,$\linebreak $\ldots, t_n)) = \rules_\PP(f) \cup \bigcup_{i = 1}^n
     \urules_\PP(t_i) \; \cup \; \bigcup_{\ell \to \mu^{\true} \in \rules_\PP(f), r \in \Supp(\mu)} \urules_\PP(\flat(r))$, otherwise.
    The \defemph{usable rules} for $\PP$ are
    $\urules(\PP) =    
    \bigcup_{\ell \to \mu^{m} \in \PP, r \in \Supp(\mu), t \trianglelefteq_{\#} r}
    \urules_{\PP}(t^\#)$.
    Then $\Proc_{\mathtt{UR}}(\PP) = \{
    \urules(\PP) \cup
    \{\ell \to \mu^{\tfalse} \mid \ell
    \to \mu^{m} \in \PP \setminus \urules(\PP)\} \}$ is sound and complete,
    i.e., we turn the  flag of all non-usable rules to $\tfalse$.
\end{restatable}

\begin{example}
    For our ADP problem
    $\{ (\ref{Rtic-adp-1}'), \eqref{Rtic-adp-2}, \eqref{Rtic-adp-3}, 
    \eqref{Rtic-adp-4} \}$,  $\eqref{Rtic-adp-4}$ is not
    usable because neither $\tf$ nor $\tF$ occur below annotated symbols on right-hand sides. 
    Hence, $\Proc_{\mathtt{UR}}$ replaces  $\eqref{Rtic-adp-4}$
    by $\tf(x,\tb_2) \to \{1 : \tA\}^{\tfalse} \quad
    (\ref{Rtic-adp-4}')$. 
    As discussed after \Cref{def:ADPs-and-Rewriting}, in contrast to the processor of \Cref{URP},
    our usable rules processor is complete
    since we do not remove non-usable rules but only set their flag to $\tfalse$.
\end{example}

Finally, we adapt the reduction pair processor.
Here, (1) for every rule with the flag $\ttrue$ (which can therefore be used for $\mathbf{r}$-steps), 
the expected value must be weakly decreasing when removing the annotations.
Since rules can also be used for $\mathbf{p}$-steps, (2) we also require a weak decrease
when comparing the annotated left-hand side with the expected value of all annotated subterms in the right-hand side. 
Since we sum up the values of the annotated subterms of each right-hand side, we can again
use \emph{weakly monotonic} interpretations.
As in \cite{avanzini2020probabilistic,kassinggiesl2023iAST}, to ensure 
``monotonicity'' w.r.t.\ expected values
we have to restrict
ourselves to interpretations with multilinear polynomials, where all monomials have the
form $c \cdot x_1^{e_1} \cdot \ldots \cdot x_n^{e_n}$ with $c \in \NN$ and $e_1,\ldots,e_n \in \{0,1\}$.
The processor then removes the annotations from those ADPs where (3) in addition there is at
least one right-hand side $r_j$ whose annotated subterms are strictly
decreasing.\footnote{In addition, the corresponding non-annotated
right-hand side $\flat(r_j)$ must be at least weakly decreasing.
The reason is that in contrast to the original DP
framework, we may now have nested annotated symbols and thus, we have to ensure that they
behave ``monotonically''.
So we have to ensure that  $\Pol(A) > \Pol(B)$
also implies that the measure of $F(A)$ is greater than $F(B)$.
Every term $r$ is ``measured'' as 
$\sum_{t \trianglelefteq_{\#} r} \Pol(t^\#)$, i.e., $F(A)$ is measured as $\Pol(F(a)) + \Pol(A)$.
Hence, in this example  we must ensure
that $\Pol(A) > \Pol(B)$ implies
$\Pol(F(a)) + \Pol(A) > \Pol(F(b)) + \Pol(B)$.
For that reason, we also have to require $\Pol(a) \geq 
\Pol(b)$.}

\begin{restatable}[Probabilistic Reduction Pair Processor]{theorem}{ProbRPP}\label{theorem:prob-RPP}
    Let $\Pol: \mathcal{T}(\SignatureADC, \linebreak \VSet) \to \IN[\VSet]$ be a
    weakly monotonic, multilinear polynomial interpretation. 
    Let $\PP = \PP_{\geq} \uplus \PP_{>}$ with $\PP_{>}\neq\emptyset$ such that:
	\begin{itemize}
		\item[(1)] For every $\ell \ruleArr{}{}{} \{ p_1:r_{1}, \ldots, p_k: r_k\}^{\ttrue} \in \PP$, we have\\ 
		    $\Pol(\ell) \geq \sum_{1 \leq j \leq k} p_j \cdot \Pol(\flat(r_j))$.
		\item[(2)] For every $\ell \ruleArr{}{}{} \{ p_1:r_{1}, \ldots, p_k: r_k\}^{m} \in \PP$, we have\\ 
            $\Pol(\ell^\#) \geq \sum_{1 \leq j \leq k} p_j \cdot \sum_{t \trianglelefteq_{\#} r_j} \Pol(t^\#)$. 
		\item[(3)] For every $\ell \ruleArr{}{}{} \{ p_1:r_{1}, \ldots, p_k: r_k\}^{m} \in \PP_{>}$, 
            there exists a $1 \leq j \leq k$ with $\Pol(\ell^\#) > \sum_{t \trianglelefteq_{\#} r_j} \Pol(t^\#)$.\\
		    If $m = \ttrue$, then we additionally have $\Pol(\ell) \geq \Pol(\flat(r_j))$. 
	\end{itemize}
	Then $\Proc_{\mathtt{RP}}(\PP) = \{\PP_{\geq} \cup \flat(\PP_{>}) \}$ is sound and complete.
\end{restatable}

\begin{example}\label{RPPexample}
    In \Cref{ADP Transformations}, we will present a new
    \emph{rewriting processor} and show how the ADP $(\ref{Rtic-adp-1}')$ can be transformed
    into
    \[
    \mbox{\small $\ta \to \{\nicefrac{1}{4}:\tf(\tH(\tb_1),\tb_1), \nicefrac{1}{4}:\tf(\th(\tb_2),\tb_1), 
        \nicefrac{1}{4}:\tF(\tH(\tb_1),\tb_2),
        \nicefrac{1}{4}:\tF(\th(\tb_2),\tb_2)\}^{\ttrue} \tag{\ref{Rtic-adp-1}$''$}$}
    \]
    For the resulting  ADP problem $\{(\ref{Rtic-adp-1}''), \eqref{Rtic-adp-2},
    \eqref{Rtic-adp-3}, (\ref{Rtic-adp-4}')\}$ with
    \[ 
    \mbox{\small $\tg \to\{\nicefrac{1}{2} : \tb_1, \nicefrac{1}{2} : \tb_2\}^{\ttrue} \; \eqref{Rtic-adp-2}
    \quad
    \th(\tb_1) \to \{1 : \tA\}^{\ttrue}\; \eqref{Rtic-adp-3}
    \quad
    \tf(x,\tb_2) \to \{1 : \tA\}^{\tfalse}\; (\ref{Rtic-adp-4}')$}
    \]
    we use the reduction pair processor with the polynomial interpretation that maps $\tA$, $\tF$,
    and $\tH$ to $1$ and all other symbols  to $0$, to remove all annotations from the
    $\ta$-ADP $(\ref{Rtic-adp-1}'')$,
    because it contains the right-hand side
    $\tf(\th(\tb_2),\tb_1)$ without annotations and thus,
    $\Pol(\tA) = 1 > \sum_{t \trianglelefteq_{\#}\tf(\th(\tb_2),\tb_1)}
    \Pol(t^\#) = 0$.    
    Another application of the usable terms processor removes the remaining
    $\tA$-annotations from \eqref{Rtic-adp-3} and $(\ref{Rtic-adp-4}')$. Since there are no more annotations left, this
    proves iAST of $\R_{\tic}$.
\end{example}

Finally, in proofs with the ADP framework, one may obtain
ADP problems $\PP$ that have a non-probabilistic structure, 
i.e., every ADP has the form $\ell \to \{1:r\}^{m}$.
Then the \emph{probability removal processor} allows us to switch
to ordinary DPs.

\begin{restatable}[Probability Removal Processor]{theorem}{NPP}\label{theorem:prob-NPP}
	Let $\PP$ be an ADP problem where every ADP in $\PP$ has the
    form $\ell \to \{1:r\}^{m}$.
    Let $\nonprobDP(\PP) = \{\ell^\# \to t^\# \mid \ell \to \{1\!:r\}^{m}
    \in \PP, t \trianglelefteq_{\#} r\}$.
    Then $\PP$ is iAST iff the
    non-probabilistic DP problem $(\nonprobDP(\PP),\nonprob(\PP))$ is iTerm.  
    So the processor $\Proc_{\mathtt{PR}}(\PP) = \emptyset$
    is sound and complete iff $(\nonprobDP(\PP), \nonprob(\PP))$ is iTerm.
\end{restatable}

\section{Transforming ADPs}\label{ADP Transformations}

Compared to the DT framework for PTRSs in \cite{kassinggiesl2023iAST}, our
new ADP framework is not only easier, more elegant, and yields a
complete chain criterion, but it also has important practical advantages, because
every processor that performs a rewrite step benefits from our novel
definition of rewriting with ADPs (whereas the rewrite relation with DTs in
\cite{kassinggiesl2023iAST} was an ``incomplete over-approximation'' of the rewrite
relation of the original TRS).
To illustrate this, we
adapt the \emph{rewriting} processor
from the original DP framework 
\cite{giesl2006mechanizing} to the probabilistic setting, which
allows us to
prove iAST of $\R_{\tic}$ from Ex.\ \arabic{counter-completeness-ctr}.
Such transformational processors
had not been adapted in the probabilistic DT framework of \cite{kassinggiesl2023iAST}.
While one could also adapt the rewriting processor to the setting of
\cite{kassinggiesl2023iAST},
then it would be substantially weaker, and we would
fail in proving iAST of $\R_{\tic}$.
We refer to\report{ App.\ \ref{moreTrans}}\paper{ \cite{report}} for our adaption
of the remaining transformational processors from
\cite{giesl2006mechanizing} (based on \emph{instantiation}, \emph{forward instantiation}, and \emph{narrowing})
to the probabilistic setting.

In the non-probabilistic setting,
the rewriting processor may rewrite a redex 
in the right-hand side of a DP if this does not affect the construction of
chains. 
To ensure that, the usable rules for this redex must be non-overlapping (NO).
If the DP occurs in a chain, then
this redex is weakly innermost terminating, hence by NO also
terminating and confluent, and thus, it has a unique normal form \cite{Gramlich1995AbstractRB}.

In the probabilistic setting,
to ensure that the probabilities for the normal forms stay the same, 
in addition to NO we require that the rule used for the rewrite step
is linear (L) 
(i.e., every variable occurs at most once in the left-hand side 
and in each term of the multi-distribution $\mu$ on the right-hand side) and non-erasing (NE)
(i.e., each variable of the
left-hand side occurs in each term of $\Supp(\mu)$).

\begin{definition}[Rewriting Processor] \label{def:RewritingProcessor}
    Let $\PP$ be an ADP problem with $\PP = \PP' \uplus \{
    \ell \to \{p_1:r_1, \ldots, p_k:r_k\}^{m}\}$. 
    Let $\tau\!\in\!\posD(r_j)$ for some $1 \leq j \leq k$ such that $r_j|_{\tau} \in \TSet{\Sigma}{\VSet}$, i.e., 
    there is no annotation below or at the position $\tau$.
    If $r_j \tored{}{\true}{\PP,\tau} \{q_1\!:\!e_{1}, \ldots,
    q_h\!:\!e_{h}\}$, where $\tored{}{\true}{\PP,\tau}$ is defined like $\itored{}{}{\PP,\tau}$ 
    but the used redex $r_j|_{\tau}$ does not have to be in $\ANF_\PP$ and the applied
    rule from $\PP$ must have the flag $m = \true$,
    then we define \[
        \begin{array}{rlll}
            \Proc_{\mathtt{r}}(\PP) = &\Bigl\{  \PP' \cup \{&\ell \to &\{
            p_1:\flat(r_{1}), \ldots, p_k: \flat(r_{k})\}^{m},\\ 
            &&\ell \to &
\begin{array}[t]{l}
            \{p_1:r_1, \ldots, p_k:r_k\} \setminus \{p_j:r_j\} \\
                   \cup \; \{p_j \cdot q_1:e_1, \ldots, p_j \cdot q_h:e_h\}^{m} \;\}\;\Bigr\}
\end{array} 
\end{array}
    \]
\end{definition}

In the non-probabilistic DP framework, one only transforms the DPs by rewriting, but the
rules are left unchanged. But since our ADPs represent both DPs and rules, when rewriting an ADP, we
add a copy of the original ADP without any annotations (i.e., this corresponds to the
original rule which can now only be used for $(\mathbf{r})$-steps).
Another change to the rewriting processor in the classic DP framework is the requirement
that there exists no annotation below $\tau$. 
Otherwise, rewriting would potentially remove annotations from $r_j$.
For the soundness of the processor, we have to ensure
that this cannot happen.

\setcounter{soundnessRewriteCtr}{\value{theorem}}

\begin{restatable}[Soundness\footnote{\label{RewritingComplete}For completeness in the
    non-probabilistic setting \cite{giesl2006mechanizing}, one uses a different definition
    of ``non-terminating'' (or ``infinite'') DP problems. 
    In future work, we will examine if such a definition would
    also yield completeness of
$\Proc_{\mathtt{r}}$
    in the
    probabilistic case.} of the Rewriting Processor]{theorem}{RewritingProc}\label{theorem:ptrs-rewriting-proc}$\Proc_{\mathtt{r}}$
    as in \Cref{def:RewritingProcessor} is \defemph{sound} if one of the following cases
    holds:
    \begin{enumerate}
        \item $\urules_{\PP}(r_j|_{\tau})$ is NO, and the rule used for rewriting 
        $r_j|_{\tau}$ is L and NE.
        \item $\urules_{\PP}(r_j|_{\tau})$ is NO, and all its rules
        have the form $\ell' \to \{1:r'\}^{\true}$.
        \item $\urules_{\PP}(r_j|_{\tau})$ is NO, $r_j|_\tau$
        is a ground term, and $r_j \itored{}{}{\PP,\tau} \{q_1:e_{1}, \ldots, q_h:e_{h}\}$ is an innermost step.
    \end{enumerate}
\end{restatable}

We refer to\report{ App.\ \ref{appendix}}\paper{ \cite{report}}
for a discussion on the requirements L and NE in the first case\report{ (see the
counterexamples in \Cref{example:rew-proc-ll-for-soundness,example:rew-proc-rl-for-soundness})}.
The second case corresponds to the original rewrite processor where all usable rules of
$r_j|_{\tau}$ are non-probabilistic.
In the last case, for any instantiation
only a single innermost rewrite step is possible for $r_j|_{\tau}$.
The restriction to innermost rewrite steps is only useful if $r_j|_\tau$
is ground. Otherwise, an innermost step on $r_j|_\tau$ might 
become a non-innermost step when instantiating $r_j|_\tau$'s variables.

The rewriting processor
benefits from our ADP framework, because 
it applies the rewrite relation $\tored{}{}{\PP}$. 
In contrast, a rewriting processor in the DT framework of
\cite{kassinggiesl2023iAST} would have to replace a DT by \emph{multiple} new DTs, 
due to the ambiguities in their rewrite relation.
Such a rewriting processor would fail for 
$\R_{\tic}$ whereas with the processor of
\Cref{theorem:ptrs-rewriting-proc} we can now prove that $\R_{\tic}$ is iAST.

\begin{example}\label{leading ex rewriting}
    After applying the usable terms and the
    usable rules processor to $\DPair{\R_{\tic}}$, we obtained:

    \vspace*{-0.7cm}
    \begin{minipage}[t]{6cm}
        \begin{align}
            \ta &\to \{1 : \tF(\tH(\tg),\tg)\}^{\ttrue} \tag{\ref{Rtic-adp-1}$'$}\\
            \tg &\to\{\nicefrac{1}{2} : \tb_1, \nicefrac{1}{2} :
            \tb_2\}^{\ttrue} \tag{\ref{Rtic-adp-2}}
        \end{align}
    \end{minipage}
    \hspace*{.5cm}
    \begin{minipage}[t]{5cm}
        \begin{align}
            \th(\tb_1) &\to \{1 : \tA\}^{\ttrue} \tag{\ref{Rtic-adp-3}}\\
            \tf(x,\tb_2) &\to \{1 : \tA\}^{\tfalse} \tag{\ref{Rtic-adp-4}$'$}
        \end{align}
    \end{minipage}

    \vspace*{.1cm}

    Now we can apply the rewriting processor on $(\ref{Rtic-adp-1}')$ repeatedly until all $\tg$s are rewritten
    and replace it by the ADP {\small$\ta \to \{\nicefrac{1}{4}:\tF(\tH(\tb_1),\tb_1), \nicefrac{1}{4}:\tF(\tH(\tb_2),\tb_1), 
    \nicefrac{1}{4}:\tF(\tH(\tb_1),\tb_2),
    \nicefrac{1}{4}:\tF(\tH(\tb_2),\tb_2)\}^{\ttrue}$} as well as several resulting ADPs
    $\ta \to \ldots$ without annotations.
    Now in the subterms $\tF(\ldots,\tb_1)$ and $\tH(\tb_2)$, the annotations
    are removed from the roots by the usable
    terms processor, as these subterms cannot rewrite to
    annotated instances of left-hand sides of ADPs. 
    So the $\ta$-ADP is changed to
    {\small$\ta \to \{\nicefrac{1}{4}:\tf(\tH(\tb_1),\tb_1), \nicefrac{1}{4}:\tf(\th(\tb_2),\tb_1), 
    \nicefrac{1}{4}:\tF(\tH(\tb_1),\tb_2),
    \nicefrac{1}{4}:\tF(\th(\tb_2),\tb_2)\}^{\ttrue}\; (\ref{Rtic-adp-1}'')$}.\report{
    This ADP corresponds to the observations that explain why $\R_{\tic}$ is iAST 
    in Ex.\ \arabic{counter-completeness-ctr}: We have two terms in the right-hand 
    side that correspond to one $\tA$ each, both with
    probability $\nicefrac{1}{4}$, one term that corresponds to a normal form with
    probability $\nicefrac{1}{4}$, and one that corresponds to two $\tA$s with
    probability $\nicefrac{1}{4}$. So again this corresponds to a random walk where the number
    of $\tA$s is decreased by 1 with
    probability $\nicefrac{1}{4}$, increased by 1 with probability $\nicefrac{1}{4}$, and
    kept the same with probability $\nicefrac{1}{2}$.}
    Then we use the reduction pair processor as in \Cref{RPPexample}
    to prove iAST for $\R_{\tic}$.
\end{example}

\section{Conclusion and Evaluation}\label{Evaluation}

We developed a new ADP framework, which advances our work in \cite{kassinggiesl2023iAST} into a
\emph{com\-plete} criterion for almost-sure innermost termination by using annotated
DPs in\-stead of dependency tuples, which also simplifies the framework substantially. 
More\-over, we adapted the \emph{rewriting} processor
of the classic DP frame\-work to the probabilistic setting.
We also adapted the other trans\-formational processors of the non-probabilistic DP
framework, see\report{ App.\ \ref{moreTrans}}\paper{ \cite{report}}.
The soundness proofs for 
the adap\-ted processors are much more
involved than in the non-probabilistic setting, due to the
more complex structure of chain trees. 
However, the
processors themselves are analogous to their non-probabilistic counterparts, and
thus, existing imple\-mentations of the 
processors can easily be adapted to their probabilistic versions.

We implemented our new contributions in our termination prover 
\textsf{AProVE} \cite{JAR-AProVE2017}
and compared the new probabilistic ADP framework 
with transfor\-mational processors (\textsf{ADP}) to the DT framework
from \cite{kassinggiesl2023iAST} (\textsf{DT}) and to \textsf{AProVE}'s tech\-niques for
ordinary non-probabilistic TRSs (\textsf{AProVE-NP}),
which include many addi\-tio\-nal processors and which benefit from using separate 
dependency pairs instead of ADPs or DTs.
For the processors in \Cref{The ADP Framework}, we could
re-use the existing implementation of \cite{kassinggiesl2023iAST} for our ADP framework.
The main goal for probabilistic termination 
analysis is to become as powerful as termination analysis in the non-pro\-ba\-bilistic setting.
Therefore, in our first experiment, we  
considered the non-probabilistic TRSs of the \emph{TPDB} \cite{TPDB} (the
benchmark set used in the annual \emph{Termination and Com\-plexity Competition (TermComp)}
\cite{TermComp}) and compared \textsf{ADP} and \textsf{DT} with
\textsf{AProVE-NP}, because at the current \emph{TermComp},
\textsf{AProVE-NP} was the
most powerful tool for termination of ordinary
non-probabilistic TRSs. 
Clearly, a TRS can be represented as a PTRS with trivial probabilities, 
and then (innermost) AST  is the same as (innermost) termination.
While
both \textsf{ADP} and \textsf{DT} have a probability removal processor to
switch to  the classical DP framework for such problems, we disabled that
processor in this experiment. 
Since \textsf{ADP} and \textsf{DT} can only deal with innermost evaluation, 
we used the benchmarks from the ``TRS Innermost'' and ``TRS Standard'' categories of
the \emph{TPDB},
but only considered innermost evaluation for all examples.
We used a timeout of 300 seconds for each example.
The ``TRS Innermost'' category contains 366 benchmarks, where \textsf{AProVE-NP} proves innermost termination for 293,
\textsf{DT} is able to prove it for 133 (45\% of \textsf{AProVE-NP}),
and for \textsf{ADP} this number rises to 159 (54\%).
For the 1512 benchmarks from the ``TRS Standard'' category, \textsf{AProVE-NP} can prove innermost termination for 1114,
\textsf{DT} for 611 (55\% of \textsf{AProVE-NP}),
and \textsf{ADP} for 723 (65\%).
This shows that the transformations are very important for automatic termination proofs 
as we get around 10\% closer to \textsf{AProVE-NP}'s results in both categories.

As a second experiment, we extended the PTRS benchmark set from \cite{kassinggiesl2023iAST}
by 33\linebreak new PTRSs for typical probabilistic programs, 
including some examples with complicated probabilistic structure.
For instance, we added the following  PTRS $\R_{\tqs}$ for
probabilistic quicksort. Here, we write $r$
instead of $\{1:r\}$ for readability.

\vspace*{-.3cm}

{\small
\[
    \begin{array}{rcl}
    \trotate(\tcons(x,\xs)) & \to & \{ \nicefrac{1}{2} : \tcons(x,\xs), \; \nicefrac{1}{2} :
    \trotate(\tapp(\xs, \tcons(x,\tnil)))\}\\
    \tqs(\xs) & \to &  \tif(\tisempty(\xs),\;
    \tlow(\thd(\xs), \ttail(\xs)), \; \thd(\xs), \;
    \thigh(\thd(\xs), \ttail(\xs)))\\
    \tif(\ttrue, \xs, x, \ys) & \to&  \tnil \qquad \tisempty(\tnil)  \to  \ttrue \qquad \tisempty(\tcons(x,\xs))  \to  \tfalse\\
    \tif(\tfalse, \xs, x, \ys) & \to& \tapp(\tqs(\trotate(\xs)),\;  \tcons(x, \,
    \tqs(\trotate(\ys))))\\
    \thd(\tcons(x, \xs))   &\to&   x \qquad \quad
    \ttail(\tcons(x, \xs))  \to   \xs \!
    \end{array}
\]}

The $\trotate$-rules rotate
a list  randomly often (they are 
AST, but not termina\-ting). Thus, by choosing the first element
of the resulting list, one obtains random\linebreak pivot elements for the recursive calls of
$\tqs$ in the second $\tif$-rule.
In addition to the  rules above, $\R_{\tqs}$ contains rules for
list concatenation ($\tapp$), and rules such that
$\tlow(x,\xs)$ ($\thigh(x,\xs)$)
returns all elements of the list $\xs$ that are smaller (greater or equal) than
$x$, see\report{ \Cref{Probabilistic Quicksort}}\paper{ \cite{report}}. In contrast to the quicksort example in \cite{kassinggiesl2023iAST},
proving iAST of the 
above rules requires transformational processors 
to instantiate and rewrite the $\tisempty$-,
$\thd$-,
and $\ttail$-subterms in the
right-hand side of the $\tqs$-rule.
So while \textsf{DT}  fails for this example,
\textsf{ADP} can prove iAST of $\R_{\tqs}$.

90 of the 100 PTRSs in our
set are iAST, and
\textsf{DT} succeeds for 54 of them (60~\%) with the tech\-nique of
\cite{kassinggiesl2023iAST} that does not use transformational processors.
Adding the new processors in \textsf{ADP} increases this number to
77 (86~\%), which dem\-on\-strates their power for PTRSs with non-trivial probabilities.
For details on our experiments and for instructions on how to run our implementation
in \textsf{AProVE} via its \emph{web interface} or locally, see:\report{
  \[\mbox{\url{https://aprove-developers.github.io/ProbabilisticADPs/}}\]}\paper{
 \mbox{\url{https://aprove-developers.github.io/ProbabilisticADPs/}}}
  
On this website, we also performed experiments where we disabled individual transformational processors
of the ADP framework, which shows the usefulness of each new processor.
In addition to the ADP and DT framework, an alternative
technique to analyze PTRSs via a direct application of interpretations was presented in
\cite{avanzini2020probabilistic}.
However, \cite{avanzini2020probabilistic} analyzes PAST (or rather \emph{strong} AST),
and a comparison between the DT framework and their technique can
be found in \cite{kassinggiesl2023iAST}.
In future work, we will adapt more processors of the DP framework to 
the probabilistic setting. Moreover, we work on analyzing AST also for full instead of
innermost rewriting and already developed criteria when iAST implies
full AST  \cite{FOSSACS24}.

\paper{\printbibliography}
  \report{
      \bibliographystyle{splncs04}
       \providecommand{\noopsort}[1]{}

  }
\report{
    \clearpage
    \appendix    
    
\vspace*{-.4cm}

\section*{Appendix}

\vspace*{-.1cm}

In \Cref{appendix} we give all proofs for our new results and observations.
Then in \Cref{moreTrans} we adapt further transformational DP processors to the probabilistic setting.
More precisely, we consider the \emph{instantiation}, \emph{forward instantiation},
and the \emph{narrowing} processor. 
Because the original \emph{narrowing} processor turns out to be unsound in the probabilistic
setting, the probabilistic narrowing processor only
instantiates variables according to the
narrowing substitutions. Hence, we call it
the \emph{rule overlap instantiation} processor.
Finally, in \Cref{Examples}, we present some examples from our collection of benchmarks that demonstrate certain aspects of
our new contributions.

\section{Proofs}\label{appendix}

We start by showing that $\R_{\tic}$ is a counterexample for completeness of the DT framework from \cite{kassinggiesl2023iAST}.
First, we introduce the framework and all needed notation.

For any term $t \in \TSet{\Sigma}{\VSet}$ with $f \in \SignatureD$, we say that $t$ is the \emph{flattened copy} of $t^\#$.
In the probabilistic adaption of the DP framework from \cite{kassinggiesl2023iAST}, for
any term $r$ in the right-hand side of a rule, one has
to consider all subterms  of $r$ with defined root symbol
\emph{at once}.
In order to deal just with terms instead of multisets, one defines
$r$'s \emph{dt transformation}
$dt(r) = \Com{n}(t^\#_1, \ldots, t^\#_n)$,
if $\{t_{1}, \dots, t_{n}\}$ are all subterms of $r$ with defined root symbol.
Here, $\SignatureC$ is extended by fresh \emph{compound} constructor symbols $\Com{n}$ of
arity $n$ for every $n \in \mathbb{N}$.
To make $dt(r)$ unique, one uses the lexicographic ordering $<$ on positions where $t_{i} =
r|_{\pi_{i}}$ and $\pi_{1} < \ldots < \pi_{n}$. 
As an example, $dt(\tg^2(x)) = \Com{2}(\tG(\tg(x)),\tG(x))$.

To abstract from nested compound symbols and from the order of their arguments, 
the following normalization is introduced.
For any term $t$, its \emph{content} $\cont(t)$ is the multiset defined by
$\cont(\Com{n}(t_1,\ldots,t_n)) = \cont(t_1) \cup \ldots \cup  \cont(t_n)$ and $\cont(t) =
\{t\}$ for other terms $t$.
For any term $t$ with $\cont(t) = \{ t_1, \ldots, t_n \}$, the term
$\Com{n}(t_1,\ldots,t_n)$ is a \emph{normalization} of $t$. 
For two terms $t, t'$, one defines $t \approx t'$ if $\cont(t) = \cont(t')$. 
So for example, $\Com{3}(x, x, y)$ is a normalization of $\Com{2}(\Com{1}(x), \Com{2}(y,x))$.
One does not distinguish between terms that are equal
w.r.t.\ $\approx$ and writes $\Com{n}(t_1,\ldots,t_n)$ for any term $t$ with a compound
root symbol where $\cont(t) = \{t_1,\ldots,t_n\}$, i.e., all such $t$ are considered to be
normalized. 

For every rule $\ell \to \mu = \{p_1:r_1, \ldots, p_k:r_k\}$,
the corresponding \emph{dependency tuple} (DT) relates
$\ell^\#$ with $\{p_1:dt(r_1), \ldots, p_k:dt(r_k)\}$.
However, in addition, DTs also store the original rule $\ell \to \mu$. As
shown in \cite{kassinggiesl2023iAST}, this is needed for the soundness of the approach,
because otherwise, one cannot simulate every possible rewrite sequence with dependency tuples. 
So one defines $\DTuple{\ell \to \mu}
= \langle \ell^\#,\ell \rangle \to \{ p_1 : \langle dt(r_1),r_1 \rangle, \ldots, p_k :
\langle dt(r_k),r_k \rangle\}$. 
$\DTuple{\R}$ denotes the set of all dependency tuples of a PTRS $\R$.
For example, $\DTuple{\R_{\trw}} = \{ \langle \tG(x),\tg(x) \rangle \to \{
\nicefrac{1}{2} : \langle \Com{2}(\tG(\tg(x)),\tG(x)),\tg^2(x) \rangle, \nicefrac{1}{2}
: \langle \Com{0},x \rangle\} \}$. 
This type of rewrite system is called a \emph{probabilistic pair term rewrite system
(PPTRS)}.

\begin{definition}[PPTRS,$\idparrow_{\PP, \R}$] \label{def:PPTRS-rewriting-on-terms}
  Let $\PP$ be a finite set of rules of the form $\langle \ell^\#,\ell \rangle\linebreak \to  \{
p_1:\langle d_1,r_1 \rangle, \ldots, p_k:\langle d_k,r_k \rangle\}$. 
For every such rule, let $\projOne(\PP)$ contain $\ell^\# \to\linebreak \{ p_1:d_1, \ldots,
p_k:d_k \}$ and $\projTwo(\PP)$ contain $\ell \to \{ p_1:r_1, \ldots, p_k:r_k \}$. \pagebreak[2]
If $\projTwo(\PP)$ is a PTRS and $\cont(d_j) \subseteq \cont(dt(r_j))$ for all
$1 \leq j \leq k$,  then $\PP$ is a PPTRS.

A normalized term $\Com{n}(s_1,\ldots,s_n)$ \defemph{rewrites} with the
\mbox{PPTRS} $\PP$ to $\{p_1:b_1, \ldots,\linebreak p_k:b_k\}$ w.r.t.\ a PTRS $\R$ (denoted
$\idparrow_{\PP, \R}$) if there are an 
$\langle \ell^\#,\ell \rangle \to \{ p_1:\langle d_1,r_1 \rangle, \ldots, p_k:\langle d_k,r_k \rangle\}
\in \PP$, a substitution $\sigma$, and an $1 \leq i \leq n$ with $s_i = \ell^\# \sigma \in \ANF_{\R}$, and for all $1 \leq j \leq k$ we have $b_j =
\Com{n}(t_1^j,\ldots,t_n^j)$ where:
    \begin{itemize}
	    \item[(a)] $t_i^j = d_j \sigma$ for all $1 \leq j \leq k$, i.e., we
              rewrite the term $s_i$ using $\projOne(\PP)$. 
		\item[(b)] For every $1 \leq i' \leq n$ with $i' \neq i$ we have
        \begin{itemize}
            \item[\normalfont{(i)}] $t_{i'}^j = s_{i'}$ for all $1 \leq j \leq k$ \quad or
            \item[\normalfont{(ii)}] $t_{i'}^j =   s_{i'}[r_j \sigma]_{\tau}$    for all
              $1 \leq j \leq k$,\\              
            if $s_{i'}|_\tau = \ell \sigma$ for some position $\tau$ and if $\ell \to  \{
            p_1:r_1, \ldots, p_k:r_k\} \in \R$.  
        \end{itemize}
        So $s_{i'}$ stays the same in all $b_j$ or one can apply the rule from
        $\projTwo(\PP)$ to rewrite $s_{i'}$ in all $b_j$, provided that this rule is also
        contained in $\R$.  
        Note that even if the rule is applicable, the term $s_{i'}$ can still stay the
        same in all $b_j$.  
	\end{itemize}
\end{definition}

Now one can simulate the rewrite step $\tg^2(x) \ito_{\R_{\trw}}
\{\nicefrac{1}{2}: \tg^3(x), \nicefrac{1}{2}: \tg(x)\}$ by\linebreak
\vspace*{-0.35cm}
{\small
\[\Com{2}(\tG(\tg(x)),\tG(x)) \!\idparrow_{\DTuple{\R_{\trw}},\R_{\trw}}\!
\{\nicefrac{1}{2}:\Com{3}(\tG(\tg^2(x)),\tG(\tg(x)), \tG(x)), \nicefrac{1}{2}:\Com{1}(\tG(x))\}\]
}using $\DTuple{\R_{\trw}}$. 
In $\Com{2}(\tG(\textcolor{blue}{\underline{\underline{\tg(x)}}}),\textcolor{red}{\underline{\tG(x)}})$, 
due to (a), the (underlined) second argument $s_i = s_2 = \textcolor{red}{\underline{\tG(x)}}$
is rewritten with
$\projOne(\DTuple{\R_{\trw}})$
to $\Com{2}(\tG(\tg(x)),\tG(x))$ or $\Com{0}$, both with probability $\nicefrac{1}{2}$.
At the same time, due to (b)(ii),
the (twice underlined) sub\-term $\textcolor{blue}{\underline{\underline{\tg(x)}}}$ of the first argument $s_{i'} = s_1 =
\tG(\textcolor{blue}{\underline{\underline{\tg(x)}}})$ is rewritten using the original rule $\tg(x) \to
\{\nicefrac{1}{2}:\tg^2(x), \; \nicefrac{1}{2}:x \}$ to
$\tg^2(x)$ or $x$, both  with probability
$\nicefrac{1}{2}$. So when rewriting $s_i = s_2 = \textcolor{red}{\underline{\tG(x)}}$ one can also
perform the corresponding rewrite step on its flattened copy inside $s_{i'} = s_1 =
\tG(\textcolor{blue}{\underline{\underline{\tg(x)}}})$.
But the ambiguity in \cref{def:PPTRS-rewriting-on-terms} also allows the step
$\Com{2}(\tG(\tg(x)),\textcolor{red}{\underline{\tG(x)}}) \idparrow_{\DTuple{\R_{\trw}},\R_{\trw}}
\{\nicefrac{1}{2}:\Com{3}(\tG(\tg(x)),\tG(\tg(x)), \tG(x)),
\nicefrac{1}{2}\!:\!\Com{1}(\tG(\tg(x)))\}$ that does not simulate any original rewrite step
with $\R_{\trw}$. Therefore, the approach of \cite{kassinggiesl2023iAST} is not complete in the
probabilistic setting.

In \cite{kassinggiesl2023iAST}, there is also
an analogous rewrite relation for PTRSs, where one can apply the same
rule simultaneously to the same subterms in a single rewrite step.

\begin{definition}[$\idparrow_{\R}$] \label{def:PTRS-rewriting-on-terms}
    For a normalized term $\Com{n}(s_1,\ldots,s_n)$ and a PTRS $\R$, let
    $\Com{n}(s_1,...,s_n) \idparrow_{\R}  \{p_1\!:\!b_1,..., p_k\!:\!b_k\}$ if there are
     $\ell\!\to\!\{p_1\!:\!r_1, \ldots, p_k\!:\!r_k\} \in \R$,
    a position $\pi$, a substitution $\sigma$, and an
    $1\!\leq\!i\!\leq n$ with $s_i|_{\pi}\!=\!\ell \sigma \in \ANF_{\R}$, and for all $1\!\leq
    j\!\leq k$ we have $b_j = \Com{n}(t_1^j,\ldots,t_n^j)$ where 
	\begin{itemize}
		\item[(a)] $t_i^j = s_i[r_j \sigma]_{\pi}$ for all $1 \leq j \leq
                  k$, i.e., we rewrite the term $s_i$ using $\R$. 
        \item[(b)] For every $1 \leq i' \leq n$ with $i' \neq i$ we have
        \begin{itemize}
            \item[\normalfont{(i)}]  $t_{i'}^j = s_{i'}$  for all $1 \leq j \leq k$ \quad or
            \item[\normalfont{(ii)}]   $t_{i'}^j = s_{i'}[r_j \sigma]_{\tau}$ for all $1
              \leq j \leq k$, if $s_{i'}|_\tau = \ell \sigma$ for some position $\tau$. 
        \end{itemize}
    \end{itemize}
\end{definition}

The DT framework of \cite{kassinggiesl2023iAST}
works on \emph{(probabilistic) DT problems} $(\PP,\R)$, where $\PP$ is a
PPTRS and $\R$ is a PTRS.
To analyze a PTRS $\R$, one starts with the DT problem $(\DTuple{\R}, \R)$. 
For the \emph{chain criterion} for DTs,
one uses a notion of \emph{chain trees} that is defined as follows. 
$\F{T}=(V,E,L,P)$ is a $(\PP,\R)$\emph{-chain tree (CT)} if
\begin{enumerate}
\item $(V, E)$ is a (possibly infinite) directed tree with nodes
  $V \neq \emptyset$ and directed edges $E \subseteq V \times V$  where $vE = \{ w \mid (v,w) \in E \}$ is \pagebreak[3] finite for every $v \in V$.
    \item $L:V\rightarrow(0,1]\times\TSet{\SignatureADC}{\VSet}$ labels every node $v$ by a probability $p_v$ and a term $t_v$.
    For the root $v \in V$ of the tree, we have $p_v = 1$.
    \item $P \subseteq V \setminus \ctleaf$ (where $\ctleaf$ are all leaves) is a subset
      of the inner nodes to indicate whether we use $\PP$ or $\R$ for the next rewrite step.
    $R = V \setminus (\ctleaf \cup P)$ are all inner nodes that are not in $P$. 
    Thus, $V = P \uplus R \uplus \ctleaf$.
    \item For all  $v\!\in\!P$: If $vE\!=\!\{w_1, \ldots, w_k\}$, then $t_v \idparrow_{\PP, \R} \{\tfrac{p_{w_1}}{p_v}:t_{w_1}, \ldots, \tfrac{p_{w_k}}{p_v}:t_{w_k}\}$.
    \item For all  $v \in R$: If $vE = \{w_1, \ldots, w_k\}$, then $t_v \idparrow_{\R} \{\tfrac{p_{w_1}}{p_v}:t_{w_1}, \ldots, \tfrac{p_{w_k}}{p_v}:t_{w_k}\}$.
    \item Every infinite path in $\F{T}$ contains infinitely many nodes from $P$. 
\end{enumerate}
Again, one defines $|\F{T}|_{\ctleaf} =
\sum_{v \in \ctleaf} \, p_v$ and says that $(\PP,\R)$ is iAST if $|\F{T}|_{\ctleaf} = 1$
for all
$(\PP,\R)$-CTs $\F{T}$.     
This yields a chain criterion which is only \emph{incomplete}, i.e., here we only have
``if'' instead of ``iff''.

\begin{restatable}[Chain Crit.\ of \cite{kassinggiesl2023iAST}]{theorem}{ProbChainCriterionOld}\label{theorem:prob-chain-criterion}
 \hspace*{-.3cm} A PTRS $\R$ is iAST if $(\DTuple{\R},\!\R)$ is iAST.
\end{restatable}

\begin{lemma}[Incompleteness of Chain Criterion for $\R_{\tic}$]\label{Incompleteness of Chain Criterion for Example}
    The DT Problem $(\DTuple{\R_{\tic}},\R_{\tic})$ is not iAST.
\end{lemma}

\begin{myproof}
    Consider $\R_{\tic}$ from Ex.\ \arabic{counter-completeness-ctr} with the rules 
    
    \vspace*{-0.5cm}
    \begin{minipage}[t]{5cm}
        \begin{align*}
            \ta &\to \{1 : \tf(\th(\tg),\tg)\} \\
            \tg &\to\{\nicefrac{1}{2} : \tb_1, \nicefrac{1}{2} : \tb_2\} \!
        \end{align*}
    \end{minipage}
    \begin{minipage}[t]{5cm}
        \begin{align*}
            \th(\tb_1) &\to \{1 : \ta\} \\
            \tf(x,\tb_2) &\to \{1 : \ta\} \!
        \end{align*}
    \end{minipage}
    
    \vspace*{.4cm}
    \noindent The set $\DTuple{\R_{\tic}}$ consists of the dependency tuples 
    \begin{align*}
        \langle \tA, \ta \rangle &\to \{1:\langle \Com{4}(\tF(\th(\tg),\tg), \tH(\tg), \tG, \tG),\tf(\th(\tg),\tg)\rangle\}\\
        \langle \tG, \tg \rangle &\to \{\nicefrac{1}{2} : \langle \Com{0}, \tb_1 \rangle, \nicefrac{1}{2} : \langle \Com{0}, \tb_2 \rangle \}\\
        \langle \tH(\tb_1), \th(\tb_1) \rangle &\to \{1 : \langle \Com{1}(\tA), \ta \rangle\}\\
        \langle \tF(x,\tb_2), \tf(x,\tb_2) \rangle &\to \{1 : \langle \Com{1}(\tA), \ta \rangle\}
    \end{align*}
    Now we obtain the following $(\DTuple{\R_{\tic}},\R_{\tic})$-chain tree $\F{T}$
    with $|\F{T}|_{\ctleaf} = 0$
     that uses both $\idparrow_{\DTuple{\R_{\tic}},\R_{\tic}}$ and $\idparrow_{\R_{\tic}}$ for the edge relation:
    \begin{center}
        \tiny
        \begin{tikzpicture}
            \tikzstyle{adam}=[thick,draw=black!100,fill=white!100,minimum
              size=4mm,shape=rectangle split, rectangle split parts=2,rectangle split
              horizontal]
            \tikzstyle{adamgray}=[thick,draw=black!100,fill=lightgray!100,minimum size=4mm,shape=rectangle split, rectangle split parts=2,rectangle split horizontal]
            \tikzstyle{empty}=[rectangle,thick,minimum size=4mm]
            \tikzstyle{mycircle}=[circle,draw=black!100,fill=white!100,thick,minimum size=4mm]
            
            \node[adam] at (0, 0)  (a) {$1$ \nodepart{two} $\Com{1}(\textcolor{red}{\underline{\tA}})$};
            \node[adam] at (0, -0.7)  (b) {$1$ \nodepart{two} $\Com{4}(\tF(\th(\tg),\textcolor{blue}{\underline{\underline{\tg}}}), \tH(\textcolor{blue}{\underline{\underline{\tg}}}), \textcolor{red}{\underline{\tG}}, \tG)$};
            \node[adam] at (-3.2, -1.4)  (c) {$\nicefrac{1}{2}$ \nodepart{two} $\Com{3}(\tF(\th(\textcolor{blue}{\underline{\underline{\tg}}}),\tb_1), \tH(\tb_1), \textcolor{red}{\underline{\tG}})$};
            \node[adam] at (3.2, -1.4)  (d) {$\nicefrac{1}{2}$ \nodepart{two} $\Com{3}(\tF(\th(\textcolor{blue}{\underline{\underline{\tg}}}),\tb_2), \tH(\tb_2), \textcolor{red}{\underline{\tG}})$};
            \node[adam] at (-4.9, -2.1)  (e1) {$\nicefrac{1}{4}$ \nodepart{two} $\Com{2}(\tF(\th(\tb_1),\tb_1), \textcolor{red}{\underline{\tH(\tb_1)}})$};
            \node[adam] at (-1.6, -2.1)  (e2) {$\nicefrac{1}{4}$ \nodepart{two} $\Com{2}(\tF(\th(\tb_2),\tb_1), \textcolor{red}{\underline{\tH(\tb_1)}})$};
            \node[adam] at (1.6, -2.1)  (e3) {$\nicefrac{1}{4}$ \nodepart{two} $\Com{2}(\tF(\textcolor{red}{\underline{\th(\tb_1)}},\tb_2), \tH(\tb_2))$};
            \node[adam] at (4.9, -2.1)  (e4) {$\nicefrac{1}{4}$ \nodepart{two} $\Com{2}(\textcolor{red}{\underline{\tF(\th(\tb_2),\tb_2)}}, \tH(\tb_2))$};
            \node[adam] at (-4.9, -2.8)  (e11) {$\nicefrac{1}{4}$ \nodepart{two} $\Com{2}(\tF(\th(\tb_1),\tb_1), \tA)$};
            \node[adam] at (-1.6, -2.8)  (e22) {$\nicefrac{1}{4}$ \nodepart{two} $\Com{2}(\tF(\th(\tb_2),\tb_1), \tA)$};
            \node[mycircle] at (1.6, -2.8)  (e33) {$\ldots$};
            \node[adam] at (4.9, -2.8)  (e44) {$\nicefrac{1}{4}$ \nodepart{two} $\Com{2}(\tA, \tH(\tb_2))$};
            \node[empty] at (-4.9, -3.5)  (e111) {$\ldots$};
            \node[empty] at (-1.6, -3.5)  (e222) {$\ldots$};
            \node[adamgray] at (1.6, -3.5)  (e333) {$\nicefrac{1}{4}$ \nodepart{two} $\Com{2}(\tA, \tH(\tb_2))$};
            \node[empty] at (4.9, -3.5)  (e444) {$\ldots$};
            \node[empty] at (1.6, -4.2)  (e3333) {$\ldots$};
            
            \draw (a) edge[->] (b);
            \draw (b) edge[->, in=70, out=250, looseness=0.50] (c);
            \draw (b) edge[->, in=110, out=290, looseness=0.50] (d);
            \draw (c) edge[->, in=70, out=250, looseness=0.50] (e1);
            \draw (c) edge[->, in=100, out=290, looseness=0.50] (e2);
            \draw (d) edge[->, in=70, out=250, looseness=0.50] (e3);
            \draw (d) edge[->, in=110, out=290, looseness=0.50] (e4);
            \draw (e1) edge[->] (e11);
            \draw (e2) edge[->] (e22);
            \draw (e3) edge[->] (e33);
            \draw (e4) edge[->] (e44);
            \draw (e11) edge[->] (e111);
            \draw (e22) edge[->] (e222);
            \draw (e33) edge[->] (e333);
            \draw (e44) edge[->] (e444);
            \draw (e333) edge[->] (e3333);
        \end{tikzpicture}
    \end{center}
    The part marked with a circle has the following form:
    \begin{center}
        \tiny
        \begin{tikzpicture}
            \tikzstyle{adam}=[thick,draw=black!100,fill=white!100,minimum size=4mm,shape=rectangle split, rectangle split parts=2,rectangle split horizontal]
            \tikzstyle{adam2}=[thick,draw=red!100,fill=white!100,minimum size=4mm,shape=rectangle split, rectangle split parts=2,rectangle split horizontal]
            \tikzstyle{empty}=[rectangle,thick,minimum size=4mm]
            
            \node[adam] at (0, 0)  (a) {$1$ \nodepart{two} $\Com{2}(\tF(\textcolor{blue}{\underline{\underline{\ta}}},\tb_2), \tH(\tb_2))$};
            \node[adam] at (0, -0.7)  (b) {$1$ \nodepart{two} $\Com{2}(\tF(\tf(\th(\tg),\textcolor{blue}{\underline{\underline{\tg}}}),\tb_2), \tH(\tb_2))$};
            \node[adam] at (-3, -1.4)  (c) {$\nicefrac{1}{2}$ \nodepart{two} $\Com{2}(\tF(\tf(\th(\textcolor{blue}{\underline{\underline{\tg}}}),\tb_1),\tb_2), \tH(\tb_2))$};
            \node[adam] at (3, -1.4)  (d) {$\nicefrac{1}{2}$ \nodepart{two} $\Com{2}(\tF(\tf(\th(\textcolor{blue}{\underline{\underline{\tg}}}),\tb_2),\tb_2), \tH(\tb_2))$};
            \node[adam] at (-4.5, -2.1)  (e1) {$\nicefrac{1}{4}$ \nodepart{two} $\Com{2}(\tF(\tf(\textcolor{blue}{\underline{\underline{\th(\tb_1)}}},\tb_1),\tb_2), \tH(\tb_2))$};
            \node[adam2,label=below:{$\NF_{\R_{\tic}}$}] at (-1.5, -2.8)  (e2) {$\nicefrac{1}{4}$ \nodepart{two} $\Com{2}(\tF(\tf(\th(\tb_2),\tb_1),\tb_2), \tH(\tb_2))$};
            \node[adam] at (1.5, -2.1)  (e3) {$\nicefrac{1}{4}$ \nodepart{two} $\Com{2}(\tF(\tf(\textcolor{blue}{\underline{\underline{\th(\tb_1)}}},\tb_2),\tb_2), \tH(\tb_2))$};
            \node[adam] at (4.5, -2.8)  (e4) {$\nicefrac{1}{4}$ \nodepart{two} $\Com{2}(\tF(\textcolor{blue}{\underline{\underline{\tf(\th(\tb_2),\tb_2)}}},\tb_2), \tH(\tb_2))$};
            \node[adam] at (-4.5, -3.5)  (e11) {$\nicefrac{1}{4}$ \nodepart{two} $\Com{2}(\tF(\textcolor{green}{\tf(\ta,\tb_1)},\tb_2), \tH(\tb_2))$};
            \node[adam] at (1.5, -3.5)  (e33) {$\nicefrac{1}{4}$ \nodepart{two} $\Com{2}(\tF(\textcolor{green}{\tf(\ta,\tb_2)},\tb_2), \tH(\tb_2))$};
            \node[adam] at (4.5, -4.2)  (e44) {$\nicefrac{1}{4}$ \nodepart{two} $\Com{2}(\tF(\textcolor{green}{\ta},\tb_2), \tH(\tb_2))$};
            \node[empty] at (-4.5, -4.9)  (e111) {$\ldots$};
            \node[empty] at (1.5, -4.9)  (e333) {$\ldots$};
            \node[empty] at (4.5, -4.9)  (e444) {$\ldots$};
            
            \draw (a) edge[->] (b);
            \draw (b) edge[->, in=70, out=250, looseness=0.50] (c);
            \draw (b) edge[->, in=110, out=290, looseness=0.50] (d);
            \draw (c) edge[->, in=70, out=250, looseness=0.50] (e1);
            \draw (c) edge[->, in=110, out=290, looseness=0.50] (e2);
            \draw (d) edge[->, in=70, out=250, looseness=0.50] (e3);
            \draw (d) edge[->, in=110, out=290, looseness=0.50] (e4);
            \draw (e1) edge[->] (e11);
            \draw (e3) edge[->] (e33);
            \draw (e4) edge[->] (e44);
            \draw (e11) edge[->] (e111);
            \draw (e33) edge[->] (e333);
            \draw (e44) edge[->] (e444);
        \end{tikzpicture}
    \end{center}
    This tree has the form of the RST from Ex.\ \arabic{counter-completeness-ctr}
    with the additional context $\Com{2}(\tF(\square,\linebreak \tb_2), \tH(\tb_2))$
    around it.
    Again, the part inside this context will reach a normal form with probability $1$ and
    afterwards, e.g., at the node labeled with
    $\Com{2}(\tF(\tf(\th(\tb_2), \tb_1),\linebreak \tb_2), \tH(\tb_2))$, we
    can rewrite $\tF(\ldots)$ with the DT
    to $\tA$, which yields
    $\Com{2}(\tA, \tH(\tb_2))$ in the gray node. 
    In this way, one obtains infinitely many nodes labeled with
    $\Com{2}(\tA, \tH(\tb_2))$ whose probabilities add up to
    $\nicefrac{1}{4}$ (in the tree we depicted this by a gray node instead).

    Note that this tree satisfies the conditions (1)-(5) of a CT, but it does not satisfy
    condition (6) as
    the part corresponding to the circle contains infinite paths without $P$ nodes.
    We have to additionally cut these infinite paths to create a valid CT $\F{T}$
    and show that we still have $|\F{T}|_{\ctleaf} < 1$.
    Since these infinite subtrees of the parts that do not use any DTs are all iAST, we can
    use the same idea as in the proof of the P-Partition Lemma of \cite[Lemma
    50]{reportkg2023iAST} (see also \cref{lemma:p-partition}).
    The variant of the P-Partition Lemma that we apply here
    is called \emph{Cutting Lemma} (\Cref{lemma:cutting})
    below. Then we obtain
    the desired $(\DTuple{\R_{\tic}},\R_{\tic})$-CT $\F{T}$
    with $|\F{T}|_{\ctleaf} < 1$.

The problem with this CT is that when applying a DT to rewrite the underlined 
$\textcolor{red}{\underline{\tG}}$ in the child of the root, then one should not rewrite
both flattened copies 
$\textcolor{blue}{\underline{\underline{\tg}}}$ in
$\Com{4}(\tF(\th(\tg),\textcolor{blue}{\underline{\underline{\tg}}}),
\tH(\textcolor{blue}{\underline{\underline{\tg}}}), \textcolor{red}{\underline{\tG}},
\tG)$. The reason is that $\textcolor{red}{\underline{\tG}}$ either corresponds to the two
flattened copies
$\tg$ in the arguments of $\th$ and $\tH$ or to the flattened copy $\tg$ in the second argument of
$\tF$. This ``wrong'' rewriting yields terms like
$\Com{2}(\tF(\th(\tb_2),\tb_1), \textcolor{red}{\underline{\tH(\tb_1)}})$ 
which do not correspond to any
terms in the original RST
since the arguments of $\th$ and $\tH$ are different.
Indeed, while the corresponding term
$\tf(\th(\tb_2),\tb_1)$ in the RST is a normal form, 
$\Com{2}(\tF(\th(\tb_2),\tb_1), \textcolor{red}{\underline{\tH(\tb_1)}})$ contains the
subterm $\textcolor{red}{\underline{\tH(\tb_1)}}$ which is not a normal form.  
\end{myproof}

\medskip

We first recapitulate the notion of a \emph{sub-chain tree} from \cite{reportkg2023iAST}.

\begin{definition}[Subtree, Sub-CT] \label{def:chain-tree-induced-sub}
    Let $(\PP, \R)$ be a DT problem and let $\F{T} = (V,E,L,P)$ be a tree that satisfies Conditions (1)-(5) of a $(\PP, \R)$-CT.
	Let $W \subseteq V$ be non-empty, weakly connected, and for all $x \in W$ we have $xE \cap W = \emptyset$ or $xE \cap W = xE$.
	Then, we define the \defemph{subtree} (or \defemph{sub-CT} if it satisfies Condition (6) as well) $\F{T}[W]$ by $\F{T}[W] = (W,E \cap (W \times W),L^W,P \cap (W \setminus W_{\ctleaf}))$.
	Here, $W_{\ctleaf}$ denotes the leaves of the tree $G^{\F{T}[W]} = (W, E \cap (W \times W))$ so that the new set $P \cap (W \setminus W_{\ctleaf})$ only contains inner nodes.
	Let $w \in W$ be the root of $G^{\F{T}[W]}$.
	To ensure that the root of our subtree has the probability $1$ again,
	we use the labeling $L^W(x) = (\frac{p_{x}^{\F{T}}}{p_w^{\F{T}}}: t_{x}^{\F{T}})$
	for all nodes $x \in W$.
        If $W$ contains the root of $(V, E)$, then we call the
sub-chain tree \defemph{grounded}.
\end{definition}

The property of being non-empty and weakly connected ensures that the resulting 
graph $G^{\F{T}[W]}$ is a tree again.
The property that we either have $xE \cap W = \emptyset$ or $xE \cap W = xE$ ensures that the sum of 
probabilities for the successors of a node $x$ is equal to the probability for the node $x$ itself.

We say that a CT (or RST) $\F{T}$ \emph{converges (or terminates) with probability}
$p \in \IR$ if we have $|\F{T}|_{\ctleaf} = p$.
Now we can prove the cutting lemma that is needed for the proof of \Cref{Incompleteness of Chain Criterion for Example}. 

\begin{lemma}[Cutting Lemma]\label{lemma:cutting}
    Let $(\PP, \R)$ be a DT problem, let $\F{T} = (V,E,L,P)$ be a tree that only 
    satisfies Conditions (1)-(5) of a $(\PP,\R)$-CT, and let $\F{T}$ converge with probability $<1$.
    Assume that every subtree that only contains nodes from $R$ converges with probability $1$.
    Then there exists a subtree $\F{T}'$ that converges with probability $<1$ such that every infinite path has an infinite number of nodes from $P$, i.e., $\F{T}'$ is a valid $(\PP, \R)$-CT as it now also satisfies Condition (6).
\end{lemma}
  
\begin{myproof}
    Let $\F{T} = (V,E,L,P)$ be a tree that only satisfies Conditions (1)-(5) of a $(\PP, \R)$-CT with $|\F{T}|_{\ctleaf} = c < 1$ for some $c \in \IR$.
    Since we have $0 \leq c < 1$, there is an $\varepsilon > 0$ such that $c + \varepsilon < 1$.
    Remember that the formula for the geometric series is:
    \[
        \sum_{n = 1}^{\infty} \left(\frac{1}{d}\right)^n = \frac{1}{d-1}, \text{ for all } d \in \IR \text{ such that } \frac{1}{|d|} < 1
    \]
    Let $d = \frac{1}{\varepsilon} + 2$. Now, we have $\frac{1}{d} = \frac{1}{\frac{1}{\varepsilon} + 2} < 1$ and:
    \begin{equation}\label{eq:setting-d}
        \frac{1}{\varepsilon} + 1 < \frac{1}{\varepsilon} + 2 \Leftrightarrow \frac{1}{\varepsilon} + 1 < d \Leftrightarrow \frac{1}{\varepsilon} < d-1 \Leftrightarrow \frac{1}{d-1} < \varepsilon \Leftrightarrow \sum_{n = 1}^{\infty} \left(\frac{1}{d}\right)^n < \varepsilon
    \end{equation}
    We will now construct a subtree $\F{T}' = (V',E',L',P')$ such that every infinite path has an infinite number of $P$ nodes and such that 
    \begin{equation}\label{eq:sum-after-all-cuts}
        |\F{T}'|_{\ctleaf} \leq |\F{T}|_{\ctleaf} + \sum_{n = 1}^{\infty} \left(\frac{1}{d}\right)^n
    \end{equation}
    and then, we finally have
    \[
        |\F{T}'|_{\ctleaf} \stackrel{\eqref{eq:sum-after-all-cuts}}{\leq} |\F{T}|_{\ctleaf} + \sum_{n = 1}^{\infty} \left(\frac{1}{d}\right)^n = c + \sum_{n = 1}^{\infty} \left(\frac{1}{d}\right)^n \stackrel{\eqref{eq:setting-d}}{<} c + \varepsilon < 1
    \]
    
    The idea of this construction is that we cut infinite subtrees of pure $R$ nodes as soon as the probability for normal forms is high enough.
    In this way, one obtains paths where after finitely many $R$ nodes, there is a $P$ node, or we reach a leaf.

    The construction works as follows.
    For any node $x \in V$, let $\ctlevelTwo(x)$ be the number of $P$ nodes in the path from the root to $x$.
    Furthermore, for any set $W \subseteq V$ and $k \in \IN$, let
    $\ctlevelTwowithborder(W,k) = \{x \in W \mid \ctlevelTwo(x) \leq k \lor (x
    \in P \land \ctlevelTwo(x) \leq k+1)\}$ be the set of all nodes in $W$ that have at most $k$ nodes from $P$ in the path from the root to its predecessor.
    So if $x \in \ctlevelTwowithborder(W,k)$ is not in $P$, then we have at most $k$ nodes from $P$ in the
    path from the root to $x$
    and if $x \in \ctlevelTwowithborder(W,k)$ is in $P$, 
    then we have at most $k+1$ nodes from $P$ in the path from the root to $x$.
    We will inductively define a set $U_k \subseteq V$ such that $U_k \subseteq
    \ctlevelTwowithborder(V,k)$ and then define the subtree as
    $\F{T}' = \F{T}[\bigcup_{k \in \IN} U_k]$.

    We start by considering the subtree
    $\F{T}_0 = \F{T}[\ctlevelTwowithborder(V,0)]$.
    This tree only contains nodes from $R$.
    While the node set $\ctlevelTwowithborder(V,0)$ itself may contain nodes from $P$, they can only occur at the leaves of $\F{T}_0$, and by definition of a subtree, we remove every leaf from $P$ in the creation of $\F{T}_0$.
    Using the prerequisite of the lemma, we get $|\F{T}_0|_{\ctleaf}=1$.
    In \cref{Possibilities for Te} one can see the different possibilities for $\F{T}_0$.
    Either $\F{T}_0$ is finite or $\F{T}_0$ is infinite.
    In the first case, we can add all the nodes to $U_0$ since there is no infinite path of pure $R$ nodes.
    Hence, we define $U_0 = \ctlevelTwowithborder(V,0)$.
    In the second case, we have to cut the tree at a specific depth once the probability of leaves is high enough.
    Let $\ctdepth_{0}(y)$ be the depth of the node $y$ in the tree $\F{T}_0$.
    Moreover, let $D_{0}(k) = \{x \in \ctlevelTwowithborder(V,0) \mid \ctdepth_{0}(y) \leq k\}$ be the set of nodes in $\F{T}_0$ that have a depth of at most $k$.
    Since $|\F{T}_0|_{\ctleaf}=1$ and $|\cdot|_{\ctleaf}$ is monotonic w.r.t.\ the depth of
    the tree $\F{T}_0$, we can find an $N_{0} \in \IN$ such that
    \[
        \sum_{x \in \ctleaf^{\F{T}_0}, d_{0}(x) \leq N_{0}} p_x^{\F{T}_0} \geq  1 - \frac{1}{d}
    \]
    Here, $\ctleaf^{\F{T}}$ and $p_x^{\F{T}}$ denote the set of leaves and the probability of the
    node $x$ in the tree $\F{T}$, resp.
            
    We include all nodes from $D_{0}(N_{0})$ in $U_0$ and delete every other node of $\F{T}_0$.
    In other words, we cut the tree after depth $N_{0}$.
    This cut can be seen in \cref{Possibilities for Te}, indicated by the dotted line.
    We now know that this cut may increase the probability of leaves by at most $\frac{1}{d}$.
    Therefore, we define $U_0 = D_{0}(N_{0})$ in this case.

    \begin{figure}
        \centering
        \begin{subfigure}[b]{0.4\textwidth}
            \centering
            \begin{tikzpicture}
                \tikzstyle{adam}=[circle,thick,draw=black!100,fill=white!100,minimum size=3mm]
                \tikzstyle{empty}=[circle,thick,minimum size=3mm]
                
                \node[adam] at (0, 0)  (a) {};
                \node[adam, label=center:{\tiny $R$}] at (2, -3)  (b) {};
                \node[adam, label=center:{\tiny $P$}] at (-2, -3)  (c) {};
                \node[adam, label=center:{\tiny $P$}] at (0.75, -3)  (d) {};
                \node[adam, label=center:{\tiny $R$}] at (-0.75, -3)  (e) {};
                \node[empty, label=center:{\small $R$}] at (0, -1.5)  (middleA) {};
                
                \node[adam, label=center:{\tiny $\NF$}] at (-1.5, -5)  (nf1) {};
                \node[adam, label=center:{\tiny $\NF$}] at (0, -5)  (nf2) {};
                
                \node[adam, label=center:{\tiny $\NF$}] at (2.75, -5)  (bb) {};
                \node[adam, label=center:{\tiny $P$}] at (1.25, -5)  (bc) {};
                \node[adam, label=center:{\tiny $P$}] at (2, -5)  (bd) {};
                \node[empty, label=center:{\small $R$}] at (2, -4)  (middleA) {};
    
                \node[empty, label=center:{\small $R$}] at (-0.75, -4)  (middleA) {};
                
                \node[empty] at (0, -6)  (stretch) {};
            
                \draw (a) edge[-] (b);
                \draw (b) edge[-] (d);
                \draw (d) edge[-] (e);
                \draw (e) edge[-] (c);
                \draw (a) edge[-] (c);
                
                \draw (b) edge[-] (bb);
                \draw (bb) edge[-] (bd);
                \draw (bd) edge[-] (bc);
                \draw (b) edge[-] (bc);
                
                \draw (e) -- (nf1) -- (nf2) -- (e);
                
                \begin{scope}[on background layer]
                \fill[green!20!white,on background layer] (0, 0) -- (-2, -3) -- (2, -3);
                \fill[green!20!white,on background layer] (2, -3) -- (1.25, -5) -- (2.75, -5);
                \fill[green!20!white,on background layer] (-0.75, -3) -- (-1.5, -5) -- (0, -5);
                \end{scope}
            \end{tikzpicture}
            \caption{$\F{T}_{x}$ finite}
        \end{subfigure}
        \hspace{30px}
        \begin{subfigure}[b]{0.4\textwidth}
            \centering
            \begin{tikzpicture}
                \tikzstyle{adam}=[circle,thick,draw=black!100,fill=white!100,minimum size=3mm]
                \tikzstyle{empty}=[circle,thick,minimum size=3mm]
                
                \node[adam] at (0, 0)  (a) {};
                \node[adam, label=center:{\tiny $R$}] at (2, -3)  (b) {};
                \node[adam, label=center:{\tiny $P$}] at (-2, -3)  (c) {};
                \node[adam, label=center:{\tiny $P$}] at (0.75, -3)  (d) {};
                \node[adam, label=center:{\tiny $R$}] at (-0.75, -3)  (e) {};
                \node[empty, label=center:{\small $R$}] at (0, -1.5)  (middleA) {};
                
                \node[adam, label=center:{\tiny $\NF$}] at (-1.5, -5)  (nf1) {};
                \node[adam, label=center:{\tiny $\NF$}] at (0, -5)  (nf2) {};
                
                \node[adam, label=center:{\tiny $R$}] at (2.75, -5)  (bb) {};
                \node[empty] (bbi) at (2.75,-6)  {};
                \node[adam, label=center:{\tiny $P$}] at (1.25, -5)  (bc) {};
                \node[adam, label=center:{\tiny $R$}] at (2, -5)  (bd) {};
                \node[empty] (bdi) at (2,-6)  {};
                \node[empty, label=center:{\small $R$}] at (2, -4)  (middleA) {};
    
                \node[empty, label=center:{\small $R$}] at (-0.75, -4)  (middleA) {};

                \node[empty, label=center:{\small \textcolor{red}{$N_{x}$}}] at (3, -3.7)  (cut) {};
            
                \draw (a) edge[-] (b);
                \draw (b) edge[-] (d);
                \draw (d) edge[-] (e);
                \draw (e) edge[-] (c);
                \draw (a) edge[-] (c);
                
                \draw (b) edge[-] (bb);
                \draw (bb) edge[-] (bd);
                \draw (bd) edge[-] (bc);
                \draw (b) edge[-] (bc);
                
                \draw (e) -- (nf1) -- (nf2) -- (e);
                
                \draw[] (bb) edge ($(bb)!0.3cm!(bbi)$) edge [dotted] ($(bb)!0.7cm!(bbi)$);
                \draw[] (bd) edge ($(bd)!0.3cm!(bdi)$) edge [dotted] ($(bd)!0.7cm!(bdi)$);

                \draw[] (-2, -3.7) edge [red, dotted] (cut);
                
                \begin{scope}[on background layer]
                \fill[green!20!white,on background layer] (0, 0) -- (-2, -3) -- (2, -3);
                \fill[green!20!white,on background layer] (2, -3) -- (1.25, -5) -- (2.75, -5);
                \fill[green!20!white,on background layer] (-0.75, -3) -- (-1.5, -5) -- (0, -5);
                \end{scope}
            \end{tikzpicture}
            \caption{$\F{T}_{x}$ infinite}
        \end{subfigure}
        \caption{Possibilities for $\F{T}_x$}\label{Possibilities for Te}
    \end{figure}
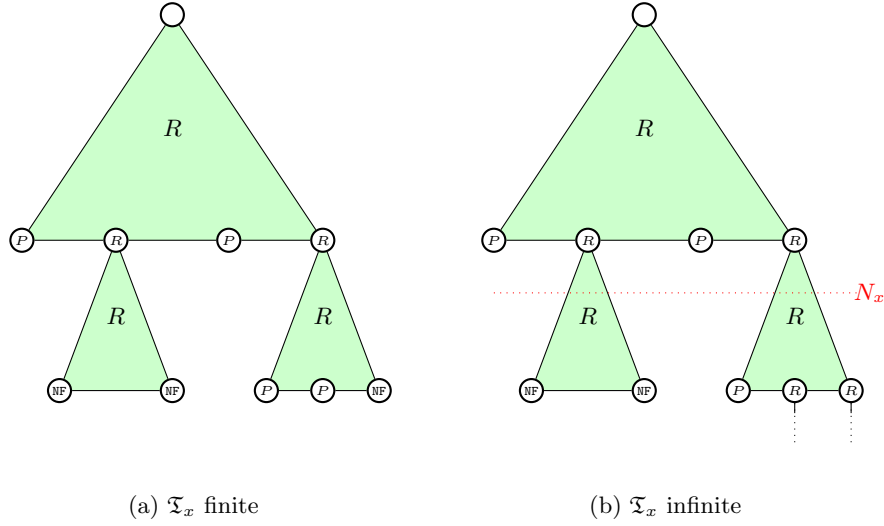

    For the induction step, assume that we have already defined a subset $U_i \subseteq \ctlevelTwowithborder(V,i)$.
    Let $H_i = \{x \in U_i \mid x \in P, \ctlevelTwo(x) = i+1\}$ be the set of leaves in
    $\F{T}[U_i]$ that are in $P$.
    For each $x \in H_i$, we consider the subtree that starts at $x$ until we
    reach the next node from $P$, including the node itself.
    Everything below such a node will be cut.
    To be precise, we regard the tree $\F{T}_{x} = (V_x,E_x,L_x,P_x) =
    \F{T}[\ctlevelTwowithborder(xE^*,i+1)]$. Here, $xE^*$ is the set of all nodes that
    are reachable from $x$ by arbitrary many steps.
    
    First, we show that $|\F{T}_{x}|_{\ctleaf} = 1$.
    For every direct successor $y$ of $x$, the subtree $\F{T}_{y} = \F{T}_{x}[y E_x^*]$ of $\F{T}_{x}$ that starts at $y$ does not contain any nodes from $P$.
    Hence, we have $|\F{T}_{y}|_{\ctleaf} = 1$ by the prerequisite of the lemma, and hence
    \[
        |\F{T}_x|_{\ctleaf} = \sum_{y \in xE} p_y \cdot |\F{T}_{y}|_{\ctleaf} = \sum_{y \in xE} p_y \cdot 1 = \sum_{y \in xE} p_y = 1.
    \]
    For the construction of $U_{i+1}$, we have the same cases as before, see \Cref{Possibilities for Te}.
    Either $\F{T}_x$ is finite or $\F{T}_x$ is infinite.
    Let $Z_{x}$ be the set of nodes that we want to add to our node set $U_{i+1}$ from the tree $\F{T}_x$.
    In the first case we can add all the nodes again and set $Z_{x} = V_x$.
    In the second case, we once again cut the tree at a specific depth once the probability for leaves is high enough.
    Let $\ctdepth_{x}(z)$ be the depth of the node $z$ in the tree $\F{T}_{x}$.
    Moreover, let $D_{x}(k) = \{x \in V_x \mid \ctdepth_{x}(z) \leq k\}$ be the set of nodes in $\F{T}_x$ that have a depth of at most $k$.
    Since $|\F{T}_{x}|_{\ctleaf}=1$ and $|\cdot|_{\ctleaf}$ is monotonic w.r.t.\ the depth of the tree $\F{T}_{x}$, we can find an $N_x \in \IN$ such that
    \[
        \sum_{y \in \ctleaf^{\F{T}_{x}}, d_{x}(y) \leq N_{x}} p_y^{\F{T}_{x}} \geq 1 - \left(\frac{1}{d}\right)^{i+1} \cdot \frac{1}{|H_i|}
    \]
    We include all nodes from $D_{x}(N_{x})$ in $U_{i+1}$ and delete every other node of $\F{T}_{x}$.
    In other words, we cut the tree after depth $N_{x}$.
    We now know that this cut may increase the probability of leaves by at most $\left(\frac{1}{d}\right)^{i+1} \cdot \frac{1}{|H_i|}$.
    Therefore, we set $Z_{x} = D_{x}(N_{x})$.
    
    We do this for each $x \in H_i$ and in the end, we set
        $U_{i+1} = U_i \cup \bigcup_{x \in H_i} Z_{x}$.

    It is straightforward to see that $\bigcup_{k \in \IN} U_k$ satisfies the conditions of \Cref{def:chain-tree-induced-sub}, as we only cut after certain nodes in our construction.
    Hence, $\bigcup_{k \in \IN} U_k$ is non-empty and weakly connected, and for each
    of its nodes, it either contains no or all successors.
    Furthermore, $\F{T}' = \F{T}[\bigcup_{k \in \IN} U_k]$ is a sub-chain
    tree which
    does not contain an infinite path of pure $R$ nodes as we cut every such
    path after a finite depth.
        
    It remains to prove that $|\F{T}'|_{\ctleaf} \leq |\F{T}|_{\ctleaf} + \sum_{n = 1}^{\infty} \left(\frac{1}{d}\right)^n$ holds.
    During the $i$-th iteration of the construction, we may increase the value of $|\F{T}|_{\ctleaf}$ by the sum of all probabilities corresponding to the new leaves resulting from the cuts.
    As we cut at most $|H_i|$ trees in the $i$-th iteration and for each such tree, we added at most a total probability of $\left(\frac{1}{d}\right)^{i+1} \cdot \frac{1}{|H_i|}$ for the new leaves, the value of $|\F{T}|_{\ctleaf}$ might increase by 
    \[
        |H_i| \cdot \left(\frac{1}{d}\right)^{i+1} \cdot \frac{1}{|H_i|} = \left(\frac{1}{d}\right)^{i+1}
    \]
    in the $i$-th iteration, and hence in total, we then get
    \[
        |\F{T}'|_{\ctleaf} \leq |\F{T}|_{\ctleaf} + \sum_{n = 1}^{\infty} \left(\frac{1}{d}\right)^n,
    \]
    as desired (see \eqref{eq:sum-after-all-cuts}).
\end{myproof}

Next, we want to prove our new chain criterion for ADPs (\Cref{theorem:prob-chain-criterion-new}), which is both sound and complete.
Due to our new definition of ADPs, this proof is easier than for the
probabilistic chain criterion that is only sound from \cite{kassinggiesl2023iAST}
(\Cref{theorem:prob-chain-criterion}).
We start by defining an important set of positions that we will use throughout the proof.

\begin{definition}[{\normalfont{$\PosDPoss$}}] \label{def:prop-important-sets}
	Let $\R$ be a PTRS.
    For a term $t \in \TSet{\Sigma}{\VSet}$ we define
    $\normalfont{\PosDPoss}(t,\R) = \{\pi \mid \pi \in \posD(t), t|_\pi \notin \NF_{\R}\}$. 
    Here, $\NF_{\R}$ again denotes the set of all normal forms w.r.t.\ $\R$.
\end{definition}

So $\PosDPoss(t,\R)$ contains all positions of subterms of $t$
that may be used as a redex now or in future rewrite steps, because 
the subterm has a defined root symbol and is not in $\NF_{\R}$.

\begin{example} \label{example:important-sets}
    Consider the following PTRS $\R$ over a signature with $\SignatureD = \{\tf,\tg\}$ and $\SignatureC = \{\ta,\ts\}$ with the rules $\tf(\ta,\ta) \to \{ 1: \ts(\tf(\tg,\tg))\}$ and $\tg \to \{ 1: \ta\}$.
	For the term $t = \ts(\tf(\tg,\tg))$ we have $\PosDPoss(t,\R) = \{1, 1.1, 1.2\}$.
\end{example}

Finally, for two (possibly annotated) terms $s,t$ we define $s \doteq t$ if $\flat(s) = \flat(t)$.

\ProbChainCriterion*

\begin{myproof}
    In the following, we will often implicitly use that for an annotated term $t \in \TSet{\Sigma^\#}{\VSet}$, we have
    $\flat(t) \in \ANF_{\R}$ iff $t \in \ANF_{\DPair{\R}}$ since a rewrite rule 
    and its corresponding canonical annotated dependency pair have the same left-hand side.
    \smallskip
    
    \noindent
    \underline{\emph{Soundness:}} Assume that $\R$ is not iAST.
    Then, there exists an innermost $\R$-RST $\F{T}=(V,E,L)$
    whose root is labeled with $(1:t)$ for some term $t \in \TSet{\Sigma}{\VSet}$
    that converges with probability $<1$.
    We will construct a $\DPair{\R}$-CT $\F{T}' = (V,E,L',V \setminus \ctleaf^{\F{T}})$ with the same underlying tree structure and an adjusted labeling such that $p_x^{\F{T}} = p_x^{\F{T}'}$ for all $x \in V$, where all the inner nodes are in $P$.
    Since the tree structure and the probabilities are the same, we then get $|\F{T}|_{\ctleaf} = |\F{T}'|_{\ctleaf}$.
    To be precise, the set of leaves in $\F{T}$ is equal to the set of leaves in $\F{T}'$, and they have the same probabilities.
    Since $|\F{T}|_{\ctleaf} < 1$, we thus have $|\F{T}'|_{\ctleaf} < 1$.
    Hence, there exists a $\DPair{\R}$-CT $\F{T}'$ that converges with probability $<1$
    and $\DPair{\R}$ is not iAST either.
    \begin{center}
        \scriptsize
        \begin{tikzpicture}
            \tikzstyle{adam}=[thick,draw=black!100,fill=white!100,minimum size=4mm, shape=rectangle split, rectangle split parts=2,rectangle split
            horizontal]
            \tikzstyle{empty}=[rectangle,thick,minimum size=4mm]
            
            \node[adam] at (-3.5, 0)  (a) {$1$ \nodepart{two} $t$};
            \node[adam] at (-5, -0.8)  (b) {$p_1$ \nodepart{two} $t_{1}$};
            \node[adam] at (-2, -0.8)  (c) {$p_2$ \nodepart{two} $t_{2}$};
            \node[adam] at (-6, -1.6)  (d) {$p_3$ \nodepart{two} $t_3$};
            \node[adam] at (-4, -1.6)  (e) {$p_4$ \nodepart{two} $t_4$};
            \node[adam] at (-2, -1.6)  (f) {$p_5$ \nodepart{two} $t_5$};
            \node[empty] at (-6, -2.4)  (g) {$\ldots$};
            \node[empty] at (-4, -2.4)  (h) {$\ldots$};
            \node[empty] at (-2, -2.4)  (i) {$\ldots$};

            \node[empty] at (-0.5, -1.2)  (arrow) {\Huge $\leadsto$};
            
            \node[adam,pin={[pin distance=0.1cm, pin edge={,-}] 140:\tiny \textcolor{blue}{$P$}}] at (3.5, 0)  (a2) {$1$ \nodepart{two} $\annoD(t)$};
            \node[adam,pin={[pin distance=0.1cm, pin edge={,-}] 140:\tiny \textcolor{blue}{$P$}}] at (2, -0.8)  (b2) {$p_1$ \nodepart{two} $t'_1$};
            \node[adam,pin={[pin distance=0.1cm, pin edge={,-}] 45:\tiny \textcolor{blue}{$P$}}] at (5, -0.8)  (c2) {$p_2$ \nodepart{two} $t'_2$};
            \node[adam,pin={[pin distance=0.1cm, pin edge={,-}] 140:\tiny \textcolor{blue}{$P$}}] at (1, -1.6)  (d2) {$p_3$ \nodepart{two} $t'_3$};
            \node[adam,pin={[pin distance=0.1cm, pin edge={,-}] 45:\tiny \textcolor{blue}{$P$}}] at (3, -1.6)  (e2) {$p_4$ \nodepart{two} $t'_4$};
            \node[adam,pin={[pin distance=0.1cm, pin edge={,-}] 45:\tiny \textcolor{blue}{$P$}}] at (5, -1.6)  (f2) {$p_5$ \nodepart{two} $t'_5$};
            \node[empty] at (1, -2.4)  (g2) {$\ldots$};
            \node[empty] at (3, -2.4)  (h2) {$\ldots$};
            \node[empty] at (5, -2.4)  (i2) {$\ldots$};
        
            \draw (a) edge[->] (b);
            \draw (a) edge[->] (c);
            \draw (b) edge[->] (d);
            \draw (b) edge[->] (e);
            \draw (c) edge[->] (f);
            \draw (d) edge[->] (g);
            \draw (e) edge[->] (h);
            \draw (f) edge[->] (i);

            \draw (a2) edge[->] (b2);
            \draw (a2) edge[->] (c2);
            \draw (b2) edge[->] (d2);
            \draw (b2) edge[->] (e2);
            \draw (c2) edge[->] (f2);
            \draw (d2) edge[->] (g2);
            \draw (e2) edge[->] (h2);
            \draw (f2) edge[->] (i2);
        \end{tikzpicture}
    \end{center}
    We construct the new labeling $L'$ for the $\DPair{\R}$-CT inductively such that for all inner nodes $x \in V \setminus \ctleaf$ with children nodes $xE = \{y_1,\ldots,y_k\}$ we have $t_x' \itored{}{}{\DPair{\R}} \{\tfrac{p_{y_1}}{p_x}:t_{y_1}', \ldots, \tfrac{p_{y_k}}{p_x}:t_{y_k}'\}$.
    Let $X \subseteq V$ be the set of nodes $x$ where we have already defined the labeling $L'(x)$.
    During our construction, we ensure that the following property holds:
    \begin{equation} \label{chain-crit-1-soundness-induction-hypothesis}
        \parbox{.9\textwidth}{For every node $x \in X$ we have $t_x \doteq t_x'$ and $\PosDPoss(t_x,\R) \subseteq \posT(t_x')$.}
    \end{equation}
    This means that the corresponding term $t_x$ for the node $x$ in $\F{T}$ has the same structure as the term $t_x'$ in $\F{T}'$,
    and additionally, all the possible redexes in $t_x$ are annotated in $t_x'$.
    The annotations ensure that we rewrite with Case $(PR)$ of \Cref{def:ADPs-and-Rewriting} so that the node $x$ is contained in $P$. 
    We label the root of $\F{T}'$ with $\annoD(t)$.
    Here, we have $t \doteq \annoD(t)$ and $\PosDPoss(t,\R) \subseteq \posD(t) = \posT(\annoD(t))$.
    As long as there is still an inner node $x \in X$ such that its successors are not contained in $X$, we do the following.
    Let $xE = \{y_1, \ldots, y_k\}$ be the set of its successors.
    We need to define the corresponding terms $t_{y_1}', \ldots, t_{y_k}'$ for the nodes $y_1, \ldots, y_k$.
    Since $x$ is not a leaf, we have $t_x \itor \{\tfrac{p_{y_1}}{p_x}:t_{y_1}, \ldots, \tfrac{p_{y_k}}{p_x}:t_{y_k}\}$.
    This means that there is a rule $\ell \to \{p_1:r_1, \ldots, p_k:r_k\} \in \R$, a position $\pi$, and a substitution $\sigma$ such that ${t_x}|_\pi = \ell\sigma \in \ANF_{\R}$.
    Furthermore, we have $t_{y_j} = t_x[r_j \sigma]_{\pi}$ for all $1 \leq j \leq k$.
    So the labeling of the successor $y_j$ in $\F{T}$ is $L(y_j) = (p_x
    \cdot p_j: t_x[r_j\sigma]_\pi)$ for all $1 \leq j \leq k$.
    
    The corresponding ADP for the rule is $\ell \to \{ p_1 : \annoD(r_1), \ldots, p_k : \annoD(r_k) \}^{\ttrue}$.
    Furthermore, $\pi \in \PosDPoss(t_x,\R) \subseteq_{(IH)} \posT(t_x')$ and $t_x \doteq_{(IH)} t_x'$.
    Hence, we can rewrite $t_x'$ with $\ell \to \{ p_1 : \annoD(r_1), \ldots, p_k : \annoD(r_k) \}^{\ttrue}$, 
    using the position $\pi$
    and the substitution $\sigma$, and Case $(PR)$ of \Cref{def:ADPs-and-Rewriting} applies.
    We get $t_x' \itored{}{}{\DPair{\R}} \{p_1: t_{y_1}', \ldots, p_k: t_{y_k}'\}$ with
    $t_{y_j}' = t_x'[\annoD(r_j) \sigma]_{\pi}$ by $(PR)$.
    This means that we have $t_{y_j} \doteq t_{y_j}'$.
    It remains to prove $\PosDPoss(t_{y_j},\R) \subseteq \posT(t_{y_j}')$ for all $1 \leq j \leq k$.
    For all positions $\tau \in \PosDPoss(t_{y_j},\R) = \PosDPoss(t_x[r_j \sigma]_{\pi},\R)$ that are orthogonal or above $\pi$, we have $\tau \in \PosDPoss(t_{x},\R) \subseteq_{(IH)} \posT(t_{x}')$, and all annotations orthogonal or above $\pi$ remain in $t_{y_j}'$ as they were in $t_{x}'$.
    For all positions $\tau \in \PosDPoss(t_{y_j},\R) = \PosDPoss(t_x[r_j \sigma]_{\pi},\R)$ that are below $\pi$, we know that,
    due to innermost evaluation, at least the defined root symbol of a term that is not in normal form must be inside $r_j$, 
    and thus $\tau \in \posT(t_{y_j}')$, as all defined symbols of $r_j$ are annotated in 
    $t_{y_j}' = t_x'[\annoD(r_j) \sigma]_{\pi}$.
    This ends the induction proof for this direction.
    \smallskip
   
    \noindent
    \underline{\emph{Completeness:}} Assume that $\DPair{\R}$ is not iAST.
    Then, there exists a $\DPair{\R}$-CT $\F{T} = (V,E,L,P)$ whose root is labeled with $(1:t)$ for some annotated term $t \in \TSet{\SignatureADC}{\VSet}$
    that converges with probability $<1$.
    We will construct an $\R$-RST $\F{T}' = (V,E,L')$ with the same underlying tree structure and an adjusted labeling such that $p_x^{\F{T}} = p_x^{\F{T}'}$ for all $x \in V$.
    Since the tree structure and the probabilities are the same, we then get $|\F{T}'|_{\ctleaf} = |\F{T}|_{\ctleaf} < 1$.
    Therefore, there exists an $\R$-RST $\F{T}'$ that converges with probability $<1$.
    Hence, $\R$ is not iAST either.

    We construct the new labeling $L'$ for the $\R$-RST inductively such that for all inner nodes $x \in V \setminus \ctleaf$ with children nodes $xE = \{y_1,\ldots,y_k\}$ we have $t_x' \itor \{\tfrac{p_{y_1}}{p_x}:t_{y_1}', \ldots, \tfrac{p_{y_k}}{p_x}:t_{y_k}'\}$.
    Let $X \subseteq V$ be the set of nodes $x$ where we have already defined the labeling $L'(x)$.
    During our construction, we ensure the following property:
    \begin{equation} \label{chain-crit-1-completeness-induction-hypothesis}
        \parbox{.8\textwidth}{For every node $x \in X$ we have $t_x \doteq t_x'$ and $\posT(t_x') = \emptyset$.}
    \end{equation}
    \begin{center}
        \scriptsize
        \begin{tikzpicture}
            \tikzstyle{adam}=[thick,draw=black!100,fill=white!100,minimum size=4mm, shape=rectangle split, rectangle split parts=2,rectangle split
            horizontal]
            \tikzstyle{empty}=[rectangle,thick,minimum size=4mm]
            
            \node[adam,pin={[pin distance=0.1cm, pin edge={,-}] 140:\tiny \textcolor{blue}{$P$}}] at (-3.5, 0)  (a) {$1$ \nodepart{two} $t$};
            \node[adam,pin={[pin distance=0.1cm, pin edge={,-}] 140:\tiny \textcolor{blue}{$P$}}] at (-5, -0.8)  (b) {$p_1$ \nodepart{two} $t_{1}$};
            \node[adam,pin={[pin distance=0.1cm, pin edge={,-}] 45:\tiny \textcolor{blue}{$P$}}] at (-2, -0.8)  (c) {$p_2$ \nodepart{two} $t_{2}$};
            \node[adam,pin={[pin distance=0.1cm, pin edge={,-}] 140:\tiny \textcolor{blue}{$P$}}] at (-6, -1.6)  (d) {$p_3$ \nodepart{two} $t_3$};
            \node[adam,pin={[pin distance=0.1cm, pin edge={,-}] 45:\tiny \textcolor{blue}{$P$}}] at (-4, -1.6)  (e) {$p_4$ \nodepart{two} $t_4$};
            \node[adam,pin={[pin distance=0.1cm, pin edge={,-}] 45:\tiny \textcolor{blue}{$P$}}] at (-2, -1.6)  (f) {$p_5$ \nodepart{two} $t_5$};
            \node[empty] at (-6, -2.4)  (g) {$\ldots$};
            \node[empty] at (-4, -2.4)  (h) {$\ldots$};
            \node[empty] at (-2, -2.4)  (i) {$\ldots$};

            \node[empty] at (-0.5, -1.2)  (arrow) {\Huge $\leadsto$};
            
            \node[adam] at (3.5, 0)  (a2) {$1$ \nodepart{two} $\flat(t)$};
            \node[adam] at (2, -0.8)  (b2) {$p_1$ \nodepart{two} $t'_1$};
            \node[adam] at (5, -0.8)  (c2) {$p_2$ \nodepart{two} $t'_2$};
            \node[adam] at (1, -1.6)  (d2) {$p_3$ \nodepart{two} $t'_3$};
            \node[adam] at (3, -1.6)  (e2) {$p_4$ \nodepart{two} $t'_4$};
            \node[adam] at (5, -1.6)  (f2) {$p_5$ \nodepart{two} $t'_5$};
            \node[empty] at (1, -2.4)  (g2) {$\ldots$};
            \node[empty] at (3, -2.4)  (h2) {$\ldots$};
            \node[empty] at (5, -2.4)  (i2) {$\ldots$};
        
            \draw (a) edge[->] (b);
            \draw (a) edge[->] (c);
            \draw (b) edge[->] (d);
            \draw (b) edge[->] (e);
            \draw (c) edge[->] (f);
            \draw (d) edge[->] (g);
            \draw (e) edge[->] (h);
            \draw (f) edge[->] (i);

            \draw (a2) edge[->] (b2);
            \draw (a2) edge[->] (c2);
            \draw (b2) edge[->] (d2);
            \draw (b2) edge[->] (e2);
            \draw (c2) edge[->] (f2);
            \draw (d2) edge[->] (g2);
            \draw (e2) edge[->] (h2);
            \draw (f2) edge[->] (i2);
        \end{tikzpicture}
    \end{center}
    This means that the corresponding term $t_x$ for the node $x$ in $\F{T}$ has the same structure as the term $t_x'$ in $\F{T}'$,
    and additionally, it contains no annotations.
    We label the root of $\F{T}'$ with $\flat(t)$.
    Here, we have $t \doteq \flat(t)$ and $\posT(\flat(t)) = \emptyset$.
    As long as there is still an inner node $x \in X$ such that its successors are not contained in $X$, we do the following.
    Let $xE = \{y_1, \ldots, y_k\}$ be the set of its successors.
    Since $x$ is not a leaf, we have $t_x \itored{}{}{\DPair{\R}} \{\tfrac{p_{y_1}}{p_x}:t_{y_1}, \ldots, \tfrac{p_{y_k}}{p_x}:t_{y_k}\}$.
    This means that there is an ADP $\ell \to \{ p_1 : \annoD(r_1), \ldots, p_k : \annoD(r_k) \}^{\ttrue} \in \DPair{\R}$, a position $\pi$, and a substitution $\sigma$ such that $\flat({t_x}|_\pi) = \ell\sigma \in \ANF_{\R}$.
    Furthermore, we have $t_{y_j} = t_x[r_j \sigma]_{\pi}$ or $t_{y_j} = t_x[\annoD(r_j) \sigma]_{\pi}$ for all $1 \leq j \leq k$.
    
    The original rule for the ADP is $\ell \to \{ p_1 : r_1, \ldots, p_k : r_k \}$.
    Furthermore, we have $t_x \doteq_{(IH)} t_x'$.
    Hence, we can rewrite $t_x'$ with $\ell \to \{ p_1 : r_1, \ldots, p_k : r_k \}$, 
    using the position $\pi$
    and the substitution $\sigma$, since $t_x'|_{\pi} = \ell\sigma \in \ANF_{\R}$ (as $\posT(t_x') = \emptyset$).
    We get $t_x' \itor \{p_1: t_{y_1}', \ldots, p_k: t_{y_k}'\}$ with
    $t_{y_j} = t_x'[r_j \sigma]_{\pi}$.
    This means that we have $t_{y_j} \doteq_{(IH)} t_{y_j}'$ and $\posT(t_{y_j}') = \emptyset$ for all $1 \leq j \leq k$, which ends the induction proof for this direction.
\end{myproof}

Next, we prove the theorems regarding the processors 
that we adapted from \cite{kassinggiesl2023iAST} to our new ADP framework.
We will see that the proofs become way more readable compared to
\cite{kassinggiesl2023iAST}
(even though they are still more complicated than in the non-probabilistic setting).
First, we repeat two lemmas from \cite{kassinggiesl2023iAST} and 
prove the theorems on the processors afterwards.
We start with the \emph{P-partition lemma}.
This lemma was proven in \cite{kassinggiesl2023iAST} and still applies to our new ADP problems, since the structure of CTs are the same as in \cite{kassinggiesl2023iAST}.

\begin{lemma}[P-Partition Lemma, \cite{kassinggiesl2023iAST}]\label{lemma:p-partition}
  Let $(\PP, \R)$ be a DT problem and let $\F{T} = (V,E,L,P)$ be a $\PP$-CT that converges with probability $<1$.
    Assume that we can partition $P = P_1 \uplus P_2$ such that every sub-CT that only contains $P$-nodes from $P_1$ converges with probability $1$.
    Then there is a grounded sub-CT $\F{T}'$ that converges with probability $<1$ such that every infinite path has an infinite number of nodes from $P_2$.
\end{lemma}

\begin{myproof}
    Analogous to the proof for \Cref{lemma:cutting}.
\end{myproof}

Next, we recapitulate the \emph{starting lemma}.
W.l.o.g., we will often assume 
that we label the root of our CT with $(1:t)$
for an annotated term $t$ such that $\flat(t) = s \theta \in \ANF_{\PP}$ for a
substitution $\theta$ and an ADP $s \to \ldots \in \PP$, and $\posT(t) = \{\varepsilon\}$. 

\begin{lemma}[Starting Lemma, \cite{kassinggiesl2023iAST}]\label{lemma:starting}
    If an ADP problem $\PP$ is not iAST, then there exists a $\PP$-CT
    $\F{T}$ with $|\F{T}|_{\ctleaf} < 1$ that starts with $(1:t)$ with $\flat(t) = s \theta \in \ANF_{\PP}$ for a
    substitution $\theta$ and an ADP $s \to \ldots \in \PP$, and $\posT(t) = \{\varepsilon\}$. 
\end{lemma}

\begin{myproof}
    We prove the contraposition.
    Assume that every $\PP$-CT $\F{T}$
    converges with probability $1$ if it 
    starts with $(1:t)$ and $\flat(t) = s \theta \in \ANF_{\PP}$ for a
    substitution $\theta$ and an ADP $s \to \ldots \in \PP$, and $\posT(t) = \{\varepsilon\}$.
    We now prove that then also every $\PP$-CT $\F{T}$ 
    that starts with $(1:t)$ for some arbitrary term $t$ converges with probability $1$, 
    and thus $\PP$ is iAST\@.
    We prove the claim by induction on the number of annotations in the initial term $t$.

    If $t$ contains no annotation, then the CT
    starting with $(1:t)$ is trivially finite (it cannot contain an infinite path, since there are no nodes in $P$) and hence, it converges with probability $1$.
    Next, if $t$ contains exactly one annotation at position $\pi$, then we can ignore everything above the annotation, as we will never use a $P$-step above the annotated position.
    If $t|_{\pi}$ is in normal form, then the claim is again trivial. 
    If we have $t|_{\pi} = s \sigma$ for some ADP $s \to \ldots \in \PP$ and some substitution 
    $\sigma$ such that $t|_{\pi} = s \theta \in \ANF_{\PP}$ and $\posT(t|_{\pi}) = \{\varepsilon\}$,
    then we know by our assumption that such a CT converges with probability $1$.
    Otherwise, we can rewrite below $\pi$. 
    However, in this case, we have to eventually rewrite at position $\pi$ in every infinite path of the tree
    (as this is the only annotation in the beginning), hence again, such a CT converges with probability $1$ by assumption.

    Now we regard the induction step, and assume that for a term $t$ with $n > 1$ annotations, there is a CT $\F{T}$ that
    converges with probability $< 1$.
    Here, our induction hypothesis is that every $\PP$-CT $\F{T}$ that starts with $(1:t')$, where $t'$ contains $m$ annotations for some $1 \leq m < n$ converges with probability $1$.
    Let $\Phi_1 = \{\tau\}$ and $\Phi_2 = \{\chi \in \posT(t) \mid \chi \neq \tau\}$ for some $\tau \in \posT(t)$ and consider the two terms $\anno_{\Phi_1}(t)$ and $\anno_{\Phi_2}(t)$, which contain both strictly less than $n$ annotations.
    By our induction hypothesis, we know that every $\PP$-CT that starts with $(1:\anno_{\Phi_1}(t))$ or $(1:\anno_{\Phi_2}(t))$ converges with probability $1$.
    Let $\F{T}_1 = (V, E, L_1, P_1)$ be the tree that starts with $(1:\anno_{\Phi_1}(t))$ and uses the same rules as we did in $\F{T}$.
    We can partition $P$ into the sets $P_1$ and $P_2 = P \setminus P_1$.
    Note that every sub-CT of $\F{T}$ such that every infinite path has an infinite number of $P_1$-nodes is a $\PP$-CT again.
    In order to use the P-Partition Lemma (\Cref{lemma:p-partition}) for the tree $\F{T}$, we have to show that every sub-CT $\F{T}'_1$ of $\F{T}$ that only contains $P$-nodes from $P_1$ converges with probability $1$.
    Let $\F{T}'_1 = (V',E',L',P')$ be a sub-CT of $\F{T}$ that does not contain nodes from $P_2$.
    There exists a set $W$ satisfying the conditions of \cref{def:chain-tree-induced-sub} such that $\F{T}'_1 = \F{T}[W]$.
    Since $\F{T}$ and $\F{T}_1$ have the same tree structure, $\F{T}_1[W]$ is a sub-CT of $\F{T}_1$.
    Moreover, $\F{T}_1[W]$ is a $\PP$-CT, since the set $W$ does not contain any inner nodes from $P_2$.
    Finally, since $\F{T}_1[W]$ is a sub-CT of a $\PP$-CT that converges with probability $1$, we know that $\F{T}_1[W]$ must be
    converging with probability $1$ as well.

    Now, we have shown that the conditions for the P-Partition Lemma (\cref{lemma:p-partition}) are satisfied.
    We can now apply the P-Partition Lemma to get a grounded sub-CT
    $\F{T}'$ of $\F{T}$ with $|\F{T}'|_{\ctleaf} < 1$ such that on every infinite path, we have an infinite number of $P_2$ nodes.
    Let $\F{T}_2$ be the tree that starts with $\anno_{\Phi_2}(t)$
    and uses the same rules as we did in $\F{T}'$.

    Again, all local properties for a $\PP$-CT are satisfied.
    Additionally, this time we know that every infinite path has an infinite number of $P_2$-nodes in $\F{T}'$, hence we also know that the global property for $\F{T}_2$ is satisfied.
    This means that $\F{T}_2$ is a $\PP$-CT that starts with
    $\anno_{\Phi_2}(t)$ and with $|\F{T}_2|_{\ctleaf} < 1$.
    This is our desired contradiction, which proves the induction step.
\end{myproof}

Now we show soundness and completeness for all processors.

\ProbDepGraphProc*

\begin{myproof}
    Let $\overline{X} = X \cup \flat(\PP \setminus X)$ for $X \subseteq \PP$.
    \smallskip
   
    \noindent
    \underline{\emph{Completeness:}} Every $\overline{\PP_i}$-CT is also a $\PP$-CT with fewer annotations in the terms.
    So if some $\overline{\PP_i}$ is not iAST, then there exists a $\overline{\PP_i}$-CT $\F{T}$ that converges with probability $<1$. 
    By adding annotations to the terms of the tree, we result in a $\PP$-CT that converges with probability $<1$ as well.
    Hence, if $\overline{\PP_i}$ is not iAST, then $\PP$ is not iAST either.
     \medskip
   
    \noindent
    \underline{\emph{Soundness:}} Let $\F{G}$ be the $\PP$-dependency graph.
    Suppose that every $\overline{\PP_i}$-CT converges with probability $1$ for all $1 \leq i \leq n$.
    We prove that then also every $\PP$-CT converges with probability 1.
    Let $W = \{\PP_1, \ldots, \PP_n\} \cup \{\{v\} \subseteq \PP \mid v$ is not in an SCC  of $\F{G}\}$ be the set of all SCCs and all singleton sets of nodes that do not belong to any SCC\@.
    The core steps of this proof are the following:
    \begin{enumerate}
      \item[1.] We show that every ADP problem $\overline{X}$ with $X \in W$ is iAST\@.
      \item[2.] We show that composing SCCs maintains the iAST property.
      \item[3.] We show that for every $X \in W$, the ADP problem $\overline{\bigcup_{X >_{\F{G}}^* Y}Y}$ is iAST by induction on $>_{\F{G}}$.
      \item[4.] We conclude that $\PP$ must be iAST\@.
    \end{enumerate}
    Here, for two $X_1,X_2 \in W$ we say that $X_2$ is a direct successor of $X_1$ (denoted
    $X_1 >_{\F{G}} X_2$) if there exist nodes $v \in X_1$ and $w \in X_2$ such that there is
    an edge from $v$ to $w$ in $\F{G}$.
  
    \medskip
  
    \noindent
    \textbf{\underline{1. Every ADP problem $\overline{X}$ with $X \in W$ is iAST\@.}}
  
    \noindent
    We start by proving the following:
    \begin{equation}
      \label{W is iAST}
      \mbox{Every ADP problem $\overline{X}$ with $X \in W$ is iAST\@.}
    \end{equation}
    To prove~\eqref{W is iAST}, note that if $X$ is an SCC, then it follows from our assumption that $\overline{X}$ is iAST\@.
    If $X$ is a singleton set of a node that does not belong to any SCC,
    then assume for a contradiction that $\overline{X}$ is not iAST\@.
    By \cref{lemma:starting} there exists an $\overline{X}$-CT $\F{T} = (V,E,L,P)$ that converges with probability
    $< 1$ and starts with $(1:t)$ where $\flat(t) = s \theta \in \ANF_{\PP}$ for a
    substitution $\theta$ and some ADP $s \to \{p_1:r_1, \ldots, p_k:r_k\}^{m} \in \overline{X}$, and $\posT(t) = \{\varepsilon\}$.
    If $s \to \ldots \notin X$, then the resulting terms after the first rewrite step contain no annotations anymore and this cannot start a CT that converges with probability $<1$.
    Hence, we have $s \to \ldots \in X$.
    Assume for a contradiction that there exists a node $x \in P$ in $\F{T}$ that is not the root and introduces new annotations.
    W.l.o.G., let $x$ be reachable from the root without traversing any other node from $P$.
    This means that for the corresponding term $t_x$ for
    node $x$ there is a $t' \trianglelefteq_{\#} t_x$ such that $t' = s \sigma' \in \ANF_{\PP}$ for
    some substitution $\sigma'$ and the only ADP $s \to \ldots \in X$ (which is the only ADP that contains any annotations in the right-hand side).
    Let $(z_0, \ldots, z_m)$ with $z_m = x$ be the path from the root to $x$ in $\F{T}$.
    The first rewrite step at the root must be $s \theta \itored{}{}{\overline{X}} \{p_1:r_1 \theta, \ldots, p_k:r_k \theta\}$.
    After that, we only use steps at non-annotated positions in the path since all the nodes $z_{1}, \ldots, z_{m-1}$ are contained in $R$.
    Therefore, we must have an $1 \leq j \leq k$ and a $t'' \trianglelefteq_{\#} r_j$ such that 
    $t''^\# \theta \ito_{\nonprob(\PP)}^* s^\# \sigma'$, 
    which means that there must be a self-loop for the only
    ADP in $X$, which is a contradiction to our assumption that $X$ is a
    singleton consisting of an ADP that is not in any SCC of $\F{G}$.
  
    Now, we have proven that the $\overline{X}$-CT $\F{T}$ does not introduce new annotations.
    By definition of a $\PP$-CT, every infinite path must contain an infinite number of nodes in $P$, i.e., nodes where we rewrite at an annotation.
    Thus, every path in $\F{T}$ must be finite, which means that $\F{T}$ is finite itself,
    as the tree is finitely branching.
    But every finite CT converges with probability $1$, which is a contradiction to
    our assumption that $\F{T}$
    converges with probability $<1$. 
  
    \medskip
  
    \noindent
    \textbf{\underline{2. Composing SCCs maintains the iAST property.}}
  
    \noindent
    Next, we show that composing SCCs maintains the iAST property. More precisely, we prove
    the following:
    \begin{equation}
    \label{Composing iAST}
      \parbox{.9\textwidth}{Let $\hat{X} \subseteq W$ and $\hat{Y} \subseteq W$
        such that there are no $X_1,X_2 \!\in\! \hat{X}$ and $Y \!\in\! \hat{Y}$ which
        satisfy both $X_1 >_{\F{G}}^* Y >_{\F{G}}^* X_2$ and $Y \not\in \hat{X}$, and such that there are no $Y_1,Y_2 \!\in\! \hat{Y}$ and $X \!\in\! \hat{X}$ which
        satisfy both $Y_1 >_{\F{G}}^* X >_{\F{G}}^* Y_2$ and $X \not\in \hat{Y}$.
      If both $ \overline{\bigcup_{X \in \hat{X}} X} $ and $ \overline{\bigcup_{Y \in \hat{Y}} Y} $ are iAST, then $ \overline{\bigcup_{X \in \hat{X}} X \cup \bigcup_{Y \in \hat{Y}} Y} $ is iAST.}
    \end{equation}
    To show~\eqref{Composing iAST}, we assume that both $\overline{\bigcup_{X \in \hat{X}} X}$ and $\overline{\bigcup_{Y \in \hat{Y}} Y}$ are iAST\@.
    Let $\overline{Z} = \overline{\bigcup_{X \in \hat{X}} X \cup \bigcup_{Y \in \hat{Y}} Y}$.
    The property in~\eqref{Composing iAST} for $\hat{X}$ and $\hat{Y}$ says that a
    path between two nodes from $\bigcup_{X \in \hat{X}} X$ that only traverses nodes
    from $Z$ must also be a path that only traverses
    nodes from $\bigcup_{X \in \hat{X}} X$, so that $\bigcup_{Y \in \hat{Y}} Y$ cannot be used to ``create'' new paths between two nodes from $\bigcup_{X \in \hat{X}} X$, and vice versa.
    Assume for a contradiction that $\overline{Z}$ is not iAST\@.
    By \cref{lemma:starting} there exists a $\overline{Z}$-CT $\F{T} = (V,E,L,P)$ that converges with probability
    $< 1$ and starts with $(1:t)$ where $\flat(t) = s \theta \in \ANF_{\PP}$ for a
    substitution $\theta$ and an ADP $s \to \ldots \in \overline{Z}$, and $\posT(t) = \{\varepsilon\}$.
    
    If $s \to \ldots \notin \bigcup_{X \in \hat{X}} X \cup \bigcup_{Y \in \hat{Y}} Y$, then the resulting terms contain no annotations anymore and this cannot start a CT that converges with probability $<1$.
    W.l.o.g., we may assume that the ADP that is used for the rewrite step at
    the root is in $\bigcup_{X \in \hat{X}} X$.
    Otherwise, we simply swap $\bigcup_{X \in \hat{X}} X$ with $\bigcup_{Y \in \hat{Y}} Y$ in the following.
  
    We can partition the set $P$ of our $\overline{Z}$-CT $\F{T}$ into the sets
    \begin{itemize}
      \item[$\bullet$] $P_1 := \{x \in P \mid x$ together with the labeling and its successors represents a step with an ADP from $\bigcup_{X \in \hat{X}} X\}$
      \item[$\bullet$] $P_2 := P \setminus P_1$
    \end{itemize}
    Note that in the case of $x \in P_2$, we know that $x$ together with its successors and the labeling represents a step with an ADP from $\PP \setminus \bigcup_{X \in \hat{X}} X$.
    We know that every $\overline{\bigcup_{Y \in \hat{Y}} Y}$-CT converges with probability $1$, since $\overline{\bigcup_{Y \in \hat{Y}} Y}$ is iAST\@.
    Thus, also every $\overline{\bigcup_{Y \in \hat{Y}} Y \setminus \bigcup_{X \in \hat{X}} X}$-CT converges with probability $1$ (as it contains fewer annotations than $\overline{\bigcup_{Y \in \hat{Y}} Y}$).
    Furthermore, we have $|\F{T}|_{\ctleaf} < 1$ by our assumption.
    By the P-Partition Lemma (\cref{lemma:p-partition}) we can find a grounded sub $\overline{Z}$-CT $\F{T}' = (V',E',L',P')$ with $|\F{T}'|_{\ctleaf} < 1$ such that every infinite path has an infinite number of $P_1$-edges.
    Since $\F{T}'$ is a grounded sub-CT of $\F{T}$ it must also start with $(1:t)$.
  
    We now construct a $\overline{\bigcup_{X \in \hat{X}} X}$-CT $\F{T}'' = (V',E',L'',P'')$ 
    with $P_1 \cap P' \subseteq P''$
    that has the same underlying tree structure and adjusted labeling
    such that all nodes get the same probabilities as in $\F{T}'$.
    Since the tree structure and the probabilities are the same, we then get $|\F{T}'|_{\ctleaf} = |\F{T}''|_{\ctleaf}$.
    To be precise, the set of leaves in $\F{T}'$ is equal to the set of leaves in $\F{T}''$, and every leaf has the same probability.
    Since $|\F{T}'|_{\ctleaf} < 1$ we thus have $|\F{T}''|_{\ctleaf} < 1$, which is a contradiction to our
    assumption that $\overline{\bigcup_{X \in \hat{X}} X}$ is iAST\@.
  
    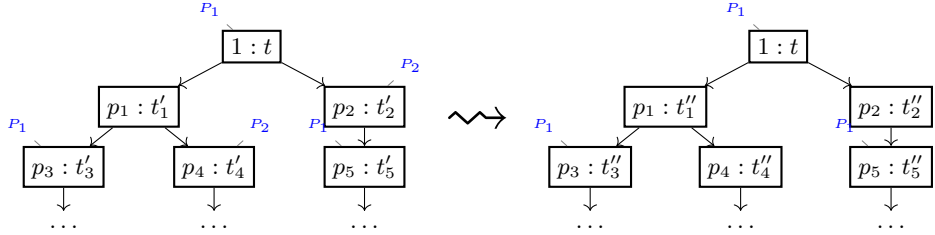
\begin{figure}[H]
      \centering
      \small
      \begin{tikzpicture}
        \tikzstyle{adam}=[rectangle,thick,draw=black!100,fill=white!100,minimum size=4mm]
        \tikzstyle{empty}=[rectangle,thick,minimum size=4mm]
        
        \node[adam,pin={[pin distance=0.1cm, pin edge={,-}] 135:\tiny \textcolor{blue}{$P_1$}}] at (-3.5, 0)  (a) {$1:t$};
        \node[adam] at (-5, -0.8)  (b) {$p_1:t_{1}'$};
        \node[adam,pin={[pin distance=0.1cm, pin edge={,-}] 45:\tiny \textcolor{blue}{$P_2$}}] at (-2, -0.8)  (c) {$p_2:t_{2}'$};
        \node[adam,pin={[pin distance=0.1cm, pin edge={,-}] 135:\tiny \textcolor{blue}{$P_1$}}] at (-6, -1.6)  (d) {$p_3:t_3'$};
        \node[adam,pin={[pin distance=0.1cm, pin edge={,-}] 45:\tiny \textcolor{blue}{$P_2$}}] at (-4, -1.6)  (e) {$p_4:t_4'$};
        \node[adam,pin={[pin distance=0.1cm, pin edge={,-}] 135:\tiny \textcolor{blue}{$P_1$}}] at (-2, -1.6)  (f) {$p_5:t_5'$};
        \node[empty] at (-6, -2.4)  (g) {$\ldots$};
        \node[empty] at (-4, -2.4)  (h) {$\ldots$};
        \node[empty] at (-2, -2.4)  (i) {$\ldots$};
  
        \node[empty] at (-0.5, -1)  (arrow) {\Huge $\leadsto$};
        
        \node[adam,pin={[pin distance=0.1cm, pin edge={,-}] 135:\tiny \textcolor{blue}{$P_1$}}] at (3.5, 0)  (a2) {$1:t$};
        \node[adam] at (2, -0.8)  (b2) {$p_1:t''_{1}$};
        \node[adam] at (5, -0.8)  (c2) {$p_2:t''_{2}$};
        \node[adam,pin={[pin distance=0.1cm, pin edge={,-}] 135:\tiny \textcolor{blue}{$P_1$}}] at (1, -1.6)  (d2) {$p_3:t''_3$};
        \node[adam] at (3, -1.6)  (e2) {$p_4:t''_4$};
        \node[adam,pin={[pin distance=0.1cm, pin edge={,-}] 135:\tiny \textcolor{blue}{$P_1$}}] at (5, -1.6)  (f2) {$p_5:t''_5$};
        \node[empty] at (1, -2.4)  (g2) {$\ldots$};
        \node[empty] at (3, -2.4)  (h2) {$\ldots$};
        \node[empty] at (5, -2.4)  (i2) {$\ldots$};
      
        \draw (a) edge[->] (b);
        \draw (a) edge[->] (c);
        \draw (b) edge[->] (d);
        \draw (b) edge[->] (e);
        \draw (c) edge[->] (f);
        \draw (d) edge[->] (g);
        \draw (e) edge[->] (h);
        \draw (f) edge[->] (i);
  
        \draw (a2) edge[->] (b2);
        \draw (a2) edge[->] (c2);
        \draw (b2) edge[->] (d2);
        \draw (b2) edge[->] (e2);
        \draw (c2) edge[->] (f2);
        \draw (d2) edge[->] (g2);
        \draw (e2) edge[->] (h2);
        \draw (f2) edge[->] (i2);
      \end{tikzpicture}
      \caption{Construction for this proof. Some nodes $x \in P_2$ in $\F{T}'$
        are removed from $P$, which yields $\F{T}''$.}\label{fig:dep-graph-proof-construction}
    \end{figure}
  
    The core idea of this construction is that annotations introduced by rewrite steps at a node $x \in P_2$ are not
    important for our computation.
    The reason is that if annotations are introduced using an ADP from 
    $\bigcup_{Y \in \hat{Y}} Y$ that is not contained in $\bigcup_{X \in \hat{X}} X$,
    then by the prerequisite of~\eqref{Composing iAST}, we know that such an ADP 
    has no path in the dependency graph to an ADP in $\bigcup_{X \in \hat{X}} X$.
    Hence, by definition of the dependency graph, we are never able to use these terms 
    for a rewrite step with an ADP from $\bigcup_{X \in \hat{X}} X$ to introduce new annotations.
    We can therefore apply the non-annotated ADP from $\bigcup_{Y \in \hat{Y}} Y$ to perform the rewrite step.
  
    We now construct the new labeling $L''$ for the $\overline{\bigcup_{X \in \hat{X}} X}$-CT $\F{T}''$ recursively.
    Let $Q \subseteq V$ be the set of nodes where we have already defined the labeling $L''$.
    During our construction, we ensure that the following property holds:
    \begin{equation}\label{dep-graph-construction-induction-hypothesis}
      \parbox{.9\textwidth}{For every $x \in Q$ we have $t'_x \doteq t''_x$ and 
      $\posT(t'_x) \setminus \Junk(t'_x, \hat{X}) \subseteq \posT(t''_{x})$.}
    \end{equation}
    Here, for any term $t'_x$, let $\Junk(t'_x, \hat{X})$ denote the positions of all annotated subterms $s \trianglelefteq_{\#} t'_x$ 
    that can never be used for a rewrite step with an ADP from $\hat{X}$, as indicated by the dependency graph.
    To be precise, we define $\pi \in \Junk(t'_x, \hat{X})$:$\Leftrightarrow$ there
    is no $A \in W$ with $A >_{\F{G}}^* X$
    for some $X \in \hat{X}$ such that there is
    an ADP $\ell \to \{p_1: r_1, \ldots, p_k: r_k\}^{m} \in A$, and a substitution $\sigma$ with $\annoEps(t_x'|_{\pi}) \ito_{\nonprob(\PP)}^* \ell^\# \sigma \in \ANF_{\PP}$.
  
    We start by setting $t''_v = t'_v$ for the root $v$ of $\F{T}'$.
    Here, our property~\eqref{dep-graph-construction-induction-hypothesis} is clearly satisfied.
    As long as there is still an inner node $x \in Q$ such that its successors are not contained in $Q$, we do the following.
    Let $xE = \{y_1, \ldots, y_k\}$ be the set of its successors.
    We need to define the corresponding terms for the nodes $y_1, \ldots, y_k$ in $\F{T}''$.
    Since $x$ is not a leaf and $\F{T}'$ is a $\overline{Z}$-CT, we have $t'_x \itored{}{}{\overline{Z}} \{\tfrac{p_{y_1}}{p_x}:t'_{y_1}, \ldots, \tfrac{p_{y_k}}{p_x}:t'_{y_k}\}$,
    and hence, we have to deal with the following two cases:
    \begin{enumerate}
        \item If we use an ADP from $\bigcup_{X \in \hat{X}} X$ in $\F{T}'$, then we perform
        the rewrite step with the same ADP, the same position $\pi$, and the same substitution in $\F{T}''$.
        Since we have $t'_x \doteq_{(IH)} t''_x$, we also get $t'_{y_j} \doteq t''_{y_j}$ for all $1 \leq j \leq k$.
        Furthermore, since we rewrite at position $\pi$ it cannot be in $\Junk(t'_x, \hat{X})$, and hence, 
        if $\pi \in \posT(t'_x)$, then also $\pi \in \posT(t''_{x})$ by \eqref{dep-graph-construction-induction-hypothesis}.
        Thus, whenever we create annotations in the rewrite step in $\F{T}'$ (a step with
    $(\mathbf{p})$ or $(\mathbf{pr})$), then we do the same in $\F{T}''$ (the step is also a $(\mathbf{p})$ or $(\mathbf{pr})$ step, respectively), and whenever we remove annotations in the rewrite step in $\F{T}''$ (a step with $(\mathbf{p})$ or $(\mathbf{irr})$), then we do the same in $\F{T}'$ (the step is also either a $(\mathbf{p})$ or $(\mathbf{irr})$ step).
        Therefore, we also get $\posT(t'_{y_j}) \setminus \Junk(t'_{y_j}, \hat{X}) \subseteq \posT(t''_{y_j})$ for all $1 \leq j \leq k$ and \eqref{dep-graph-construction-induction-hypothesis} is again satisfied.
        \item If we use an ADP from $\PP \setminus \bigcup_{X \in \hat{X}} X$ in $\F{T}'$, and we use the ADP $\ell \to \{p_1:r_1, \ldots, p_k:r_k\}^{m}$, 
        then we can use $\ell \to \{p_1:\flat(r_1), \ldots, p_k:\flat(r_k)\}^{m}$ instead, with the same position $\pi$, and the same substitution.
        Note that if $\pi \in \posT(t'_x)$, then all the annotations introduced by the ADP are in $\Junk(t'_{y_j}, \hat{X})$ for all $1 \leq j \leq k$, since the used ADP is not in $\bigcup_{X \in \hat{X}} X$ and by \eqref{Composing iAST} we cannot use another ADP to create a path in the dependency graph to a node in $\bigcup_{X \in \hat{X}} X$ again.
        Otherwise, we remove the annotations during the application of the rule anyway.
        Again, \eqref{dep-graph-construction-induction-hypothesis} is satisfied.
    \end{enumerate}
    We have now shown that~\eqref{Composing iAST} holds.
  
    \medskip
  
    \noindent
    \textbf{\underline{3. For every $X \in W$, the ADP problem $\overline{\bigcup_{X >_{\F{G}}^* Y}Y}$ is iAST\@.}}
  
    \noindent
    Using~\eqref{W is iAST} and~\eqref{Composing iAST}, by induction on $>_{\F{G}}$ we now prove that
    \begin{equation}
      \label{SCC induction} \mbox{for every $X \in W$, the ADP problem $\overline{\bigcup_{X >_{\F{G}}^* Y}Y}$ is iAST\@.}
    \end{equation}
    Note that $>_{\F{G}}$ is well founded, \pagebreak[2] since $\F{G}$ is finite.
    
    For the base case, we consider an $X \in W$ that is minimal w.r.t.\ $>_{\F{G}}$.
    Hence, we have $\bigcup_{X >_{\F{G}}^* Y} Y = X$.
    By~\eqref{W is iAST}, $\overline{X}$ is iAST\@.
  
    For the induction step, we consider an $X \in W$ and assume that
    $\overline{\bigcup_{Y >_{\F{G}}^* Z} Z}$ is iAST for every $Y \in W$ with $X >_{\F{G}}^+ Y$.
    Let $\mathtt{Succ}(X) = \{Y \in W \mid X >_{\F{G}} Y\} = \{Y_1, \ldots Y_m\}$ be the set of all direct successors of $X$.
    The induction hypothesis states that $\overline{\bigcup_{Y_u >_{\F{G}}^* Z} Z}$ is iAST for all $1 \leq u \leq m$.
    We first prove by induction that for all $1 \leq u \leq m$,
    $\overline{\bigcup_{1 \leq i \leq u} \bigcup_{Y_i >_{\F{G}}^* Z} Z}$ is iAST\@.
    
    In the inner induction base, we have $u = 1$ and hence $\overline{\bigcup_{1 \leq i \leq u} \bigcup_{Y_i >_{\F{G}}^* Z} Z} = \overline{\bigcup_{Y_1 >_{\F{G}}^* Z} Z}$.
    By our outer induction hypothesis we know that $\overline{\bigcup_{Y_1 >_{\F{G}}^* Z} Z}$ is iAST\@.
    
    In the inner induction step, assume that the claim holds for some $1 \leq u < m$.
    Then $\overline{\bigcup_{Y_{u+1} >_{\F{G}}^* Z} Z}$ is iAST by our outer induction hypothesis and\linebreak $\overline{\bigcup_{1 \leq i \leq u} \bigcup_{Y_{i} >_{\F{G}}^* Z} Z}$ is iAST by our inner induction hypothesis.
    By~\eqref{Composing iAST}, we know that then $\overline{\bigcup_{1 \leq i \leq u+1} \bigcup_{Y_{i} >_{\F{G}}^* Z} Z}$ is iAST as well.
    The conditions for~\eqref{Composing iAST} are clearly satisfied, as we use the
    reflexive, transitive closure $>_{\F{G}}^*$ of the direct successor relation 
    in both $\bigcup_{1 \leq i \leq u} \bigcup_{Y_{i} >_{\F{G}}^* Z} Z$ and $\bigcup_{Y_{u+1} >_{\F{G}}^* Z} Z$.
    
    Now we have shown that $\overline{\bigcup_{1 \leq i \leq m} \bigcup_{Y_i >_{\F{G}}^* Z} Z}$ is iAST\@.
    We know that $\overline{X}$ is iAST by our assumption and that $\overline{\bigcup_{1 \leq i \leq m} \bigcup_{Y_i >_{\F{G}}^* Z} Z}$ is iAST\@.
    Hence, by~\eqref{Composing iAST} we obtain that $\overline{\bigcup_{X >_{\F{G}}^* Y} Y}$ iAST\@.
    Again, the conditions of~\eqref{Composing iAST} are satisfied, since $X$ is strictly
    greater w.r.t.\ $>_{\F{G}}^+$ than all $Z$ with $Y_i >_{\F{G}}^* Z$.

    \medskip
  
    \noindent
    \textbf{\underline{4. $\PP$ is iAST\@.}}
  
    \noindent
    In~\eqref{SCC induction} we have shown that $\overline{\bigcup_{X >_{\F{G}}^* Y} Y}$ for every $X \in W$ is iAST\@.
    Let $X_1, \ldots, X_m\linebreak \in W$ be the maximal elements of $W$ w.r.t.\ $>_{\F{G}}$.
    By induction, one can prove that $\overline{\bigcup_{1 \leq i \leq u} \bigcup_{X_i >_{\F{G}}^* Y}
    Y}$ is iAST for all $1 \leq u \leq m$ by~\eqref{Composing iAST}, analogous to the previous induction.
    Again, the conditions of~\eqref{Composing iAST} are satisfied as we use
    the reflexive, transitive closure of $>_{\F{G}}$.
    In the end, we know that $\overline{\bigcup_{1 \leq i \leq m} \bigcup_{X_i >_{\F{G}}^* Y} Y} = \PP$ is iAST and this ends the proof.
\end{myproof}

\UsableTermsProc*

\begin{myproof}
    \smallskip
   
    \noindent
    \underline{\emph{Completeness:}} Every $\mathcal{T}_\mathtt{UT}(\PP)$-CT is also a $\PP$-CT with fewer annotations in the terms.
    So if $\mathcal{T}_\mathtt{UT}(\PP)$ is not iAST, then there exists a $\mathcal{T}_\mathtt{UT}(\PP)$-CT $\F{T}$ that converges with probability $<1$. 
    By adding annotations to the terms of the tree, we result in a $\PP$-CT that converges with probability $<1$ as well.
    Hence, if $\mathcal{T}_\mathtt{UT}(\PP)$ is not iAST, then $\PP$ is not iAST either.
    \medskip
   
    \noindent
    \underline{\emph{Soundness:}} Let $\PP$ be not iAST.
    Then by \Cref{lemma:starting} there exists a $\PP$-CT $\F{T} = (V,E,L,P)$ that converges with probability
    $< 1$ whose root is labeled with $(1: t)$ and $\flat(t) = s \theta \in \ANF_{\PP}$ for a
    substitution $\theta$ and an ADP $s \to \ldots \in \PP$, and $\posT(t) = \{\varepsilon\}$. 
    We will now create a $\mathcal{T}_\mathtt{UT}(\PP)$-CT $\F{T}' = (V,E,L',P)$, with the same underlying tree structure, and an adjusted labeling such that $p_x^{\F{T}} = p_x^{\F{T}'}$ for all $x \in V$.
    Since the tree structure and the probabilities are the same, we then get $|\F{T}'|_{\ctleaf} = |\F{T}|_{\ctleaf} < 1$, and hence $\mathcal{T}_\mathtt{UT}(\PP)$ is not iAST either.
    
    We now construct the new labeling $L'$ for the $\mathcal{T}_\mathtt{UT}(\PP)$-CT $\F{T}'$ recursively.
    Let $X \subseteq V$ be the set of nodes where we have already defined the labeling $L'$.
    During our construction, we ensure that the following property holds for every node $x \in X$:
    \begin{equation}\label{usable-terms-soundness-induction-hypothesis}
        \parbox{.9\textwidth}{For every $x \in X$ we have $t_x \doteq t'_x$ and $\posT(t_x) \setminus \Junk(t_x) \subseteq \posT(t'_x)$.}
    \end{equation}
    Here, for any annotated term $t_x$, let $\Junk(t_x)$ denote the set of all positions of annotations in $t_x$ that will never be used for a rewrite step in $\F{T}$.
    To be precise, we define $\Junk(t_x)$ recursively:
    For the term $t$ at the root, we define $\Junk(t) = \emptyset$.
    For a node $y_j$ for some $1 \leq j \leq h$ with predecessor $x$ such that
    $t_x \itored{}{}{\PP} \{\tfrac{p_{y_1}}{p_x}:t_{y_1}, \ldots, \tfrac{p_{y_h}}{p_x}:t_{y_h}\}$
    at a position $\pi$, we define $\Junk(t_{y_j})
    = \{ \rho \mid \rho \in \Junk(t_{x}), \pi \nless \rho\}$ if $\pi \notin \posT(t_x)$,
    and otherwise, we define $\Junk(t_{y_j})
    = \{ \rho \mid \rho \in \Junk(t_{x}), \pi \nless \rho\}
    \cup \Junk((t_x)_+^j,\F{T})$.
    Here, we have $\rho \in \Junk((t_x)_+^j,\F{T})$, if $\rho = \pi.\tau$ and $\tau \in \posT(\hat{r}_j)$, for the used ADP $\hat{\ell} \to \{ \hat{p}_1:\hat{r}_1, \ldots, \hat{p}_h:\hat{r}_h\}^{m'}$, and there is no (not necessarily direct) successor node in $\F{T}$ that rewrites at position $\rho$ without rewriting above position $\rho$ before.
  
    We start with the same term $t$ at the root.
    Here, our property~\eqref{usable-terms-soundness-induction-hypothesis} is clearly satisfied.
    As long as there is still an inner node $x \in X$ such that its successors are not contained in $X$, we do the following.
    Let $xE = \{y_1, \ldots, y_k\}$ be the set of its successors.
    We need to define the terms for the nodes $y_1, \ldots, y_k$ in $\F{T}'$.
    Since $x$ is not a leaf and $\F{T}$ is a $\PP$-CT, we have $t_x \itored{}{}{\PP} \{\tfrac{p_{y_1}}{p_x}:t_{y_1}, \ldots, \tfrac{p_{y_k}}{p_x}:t_{y_k}\}$.
    If we performed a step with $\itored{}{}{\PP}$ using the ADP $\ell \to \{ p_1: r_1, \ldots, p_k:r_k\}^{m}$, 
    the position $\pi$, and the substitution $\sigma$ in $\F{T}$,
    then we can use the ADP $\ell \to \{ p_1: \flat_{\ell,\PP}(r_1), \ldots,
    p_k:\flat_{\ell,\PP}(r_k)\}^{m}$ with the same position $\pi$ and the same substitution $\sigma$.
    Now, we directly get $t_{y_j} \doteq t'_{y_j}$ and $\posT(t_{y_j}) \setminus \Junk(t_{y_j}) \subseteq \posT(t'_{y_j})$ for all $1 \leq j \leq k$ since the original rule contains the same terms with more annotations, 
    but all missing annotations are in $\Junk(t_x)$ by definition of $\flat_{\ell,\PP}(r_j)$ for each $1 \leq j \leq k$.
  \end{myproof}

\ProbUsRulesProc*

\begin{myproof}
    Let $\overline{\PP} = \urules(\PP) \cup \{\ell \to \mu^{\tfalse} \mid \ell \to \mu^{m} \in \PP \setminus \urules(\PP)\}$.
    \smallskip
   
    \noindent
    \underline{\emph{Completeness:}} Every $\overline{\PP}$-CT is also a $\PP$-CT with fewer annotations in the terms.
    So if $\overline{\PP}$ is not iAST, then there exists a $\overline{\PP}$-CT $\F{T}$ that converges with probability $<1$. 
    By adding annotations into the terms of the tree, we result in a $\PP$-CT that converges with probability $<1$ as well.
    Hence, if $\overline{\PP}$ is not iAST, then $\PP$ is not iAST either.

    \smallskip
   
    \noindent
    \underline{\emph{Soundness:}} Assume that $\PP$ is not iAST\@.
    Then by \Cref{lemma:starting} there exists a $\PP$-CT $\F{T} = (V,E,L,P)$ that converges with probability
    $< 1$ whose root is labeled with $(1: t)$ and $\flat(t) = s \theta \in \ANF_{\PP}$ for a
    substitution $\theta$ and an ADP $s \to \ldots \in \PP$, and $\posT(t) = \{\varepsilon\}$. 
  
    By the definition of usable rules, as in the non-probabilistic case, 
    rules $\ell \to \mu \in \PP$ that are not usable 
    (i.e., $\ell \to \mu \not\in \overline{\PP}$) will never be used below an annotated symbol in such a $\PP$-CT.
    Hence, we can also view $\F{T}$ as a $\overline{\PP}$-CT that
    converges with probability $<1$ and thus $\overline{\PP}$ is not iAST\@.
  \end{myproof}

\ProbRPP*

\begin{myproof}
    Let $\overline{\PP} = \PP_{\geq} \cup \flat(\PP_{>})$.
    \smallskip
   
    \noindent
    \underline{\emph{Completeness:}} Every $\overline{\PP}$-CT is also a $\PP$-CT with fewer annotations in the terms.
    So if $\overline{\PP}$ is not iAST, then there exists a $\overline{\PP}$-CT $\F{T}$ that converges with probability $<1$. 
    By adding annotations to the terms of the tree, we result in a $\PP$-CT that converges with probability $<1$ as well.
    Hence, if $\overline{\PP}$ is not iAST, then $\PP$ is not iAST either.

    \smallskip
   
    \noindent
    \underline{\emph{Soundness:}} This proof uses the proof idea for AST from~\cite{mciver2017new}.
    The core steps of the proof are the following:
	\begin{enumerate}
		\item[(I)] We extend the conditions (1), (2), and (3) to rewrite steps instead of just rules (and thus, to edges of a CT).
		\item[(II)] We create a CT $\F{T}^{\leq N}$ for any $N \in \IN$.
		\item[(III)] We prove that $|\F{T}^{\leq N}|_{\ctleaf} \geq p_{min}^{N}$ for any $N \in \IN$.
		\item[(IV)] We prove that $|\F{T}^{\leq N}|_{\ctleaf}=1$ for any $N \in \IN$.
		\item[(V)] Finally, we prove that $|\F{T}|_{\ctleaf}=1$.
	\end{enumerate}
	Parts (II) to (V) remain completely the same as in \cite{kassinggiesl2023iAST}.
    We only show that we can adjust part (I) to our new rewrite relation and new annotated
    dependency pairs.

\smallskip

    \noindent
    \textbf{\underline{(I) We extend the conditions to rewrite steps instead of just rules}}

    \noindent
    We set $V(s) = \sum_{t \trianglelefteq_{\#} s} \Pol(t^\#)$, and show that the conditions (1), (2), and (3) of the lemma extend to rewrite steps instead of just rules:
	\begin{enumerate}
		\item[(a)] If $s \ito \{ p_1:t_1, \ldots, p_k:t_k \}$ using a rewrite rule $\ell \to \{ p_1:r_1, \ldots, p_k:r_k \}$\linebreak with $\Pol(\ell) \geq \Pol(r_j)$ for some $1 \leq j \leq k$, then we have $Pol(s) \geq Pol(t_j)$.
		\item[(b)] If $a \itored{}{}{\PP} \{ p_1:b_1, \ldots, p_k:b_k \}$ using the rule $\ell \to \{ p_1:r_1, \ldots, p_k:r_k \}^{m} \in \PP_>$ at a position $\pi \in \posT(s)$, then $V(a) > V(b_j)$ for some $1 \leq j \leq k$.
		\item[(c)] If $s \ito \{ p_1:t_1, \ldots, p_k:t_k \}$ using a rewrite rule $\ell \to \{ p_1:r_1, \ldots, p_k:r_k \}$\linebreak with $Pol(\ell) \geq \sum_{1 \leq j \leq k} p_j \cdot \Pol(r_j)$, then $Pol(s) \geq \sum_{1 \leq j \leq k} p_j \cdot \Pol(t_j)$.
		\item[(d)] If $a \itored{}{}{\PP} \{ p_1:b_1, \ldots, p_k:b_k \}$ using the rule $\ell \to \{ p_1:r_1, \ldots, p_k:r_k \}^{m} \in \PP$, then $V(a) \geq \sum_{1 \leq j \leq k} p_j \cdot V(b_j)$.
	\end{enumerate}

	\begin{itemize}
    	\item[(a)] 
        In this case, there exist a rule $\ell \to \{ p_1:r_1, \ldots, p_k:r_k \}$ with $\Pol(\ell) \geq \Pol(r_j)$ for some $1 \leq j \leq k$, a substitution $\sigma$, and a position $\pi$ of $s$ such that $s|_\pi =\ell\sigma \in \ANF_{\PP}$, and $t_j = s[r_j \sigma]_\pi$ for all $1 \leq j \leq k$.
        
        We perform structural induction on $\pi$.
        So in the induction base, let $\pi = \varepsilon$.
        Hence, we have $s = \ell\sigma \ito \{ p_1:r_1 \sigma, \ldots, p_k:r_k \sigma\}$.
        By assumption, we have $\Pol(\ell) \geq \Pol(r_j)$ for some $1 \leq j \leq k$.
        As these inequations hold for all instantiations of the occurring variables, for $t_j = r_j\sigma$ we have
        \[ \Pol(s) = \Pol(\ell\sigma) \geq \Pol(r_j\sigma) = \Pol(t_j). \]
        
        In the induction step, we have $\pi = i.\pi'$, $s = f(s_1,\ldots,s_i,\ldots,s_n)$, 
        $f \in \Sigma$, $s_i \ito \{ p_1:t_{i,1}, \ldots, p_k:t_{i,k} \}$, and $t_j =
        f(s_1,\ldots,t_{i,j},\ldots,s_n)$ with $t_{i,j} = s_i[r_j\sigma]_{\pi'}$ for all $1 \leq j \leq k$.
        Then by the induction hypothesis we have $Pol(s_i) \geq Pol(t_{i,j})$.
        For $t_j = f(s_1,\ldots,t_{i,j},\ldots,s_n)$ we obtain
        \[
        \begin{array}{lcl}
            \Pol(s) & = & \Pol(f(s_1,\ldots,s_i,\ldots,s_n)) \\
                & = & f_{\Pol}(\Pol(s_1),\ldots,\Pol(s_i),\ldots,\Pol(s_n)) \\
                & \geq & f_{\Pol}(\Pol(s_1),\ldots,\Pol(t_{i,j}),\ldots,\Pol(s_n)) \\
                &  & \hspace*{1cm} \text{ \textcolor{blue}{(by weak monotonicity of $f_{\Pol}$ and $Pol(s_i) \geq Pol(t_{i,j})$)}} \\
            & = & \Pol(f(s_1,\ldots,t_{i,j},\ldots,s_n)) \\
                & = & \Pol(t_j).
        \end{array}
        \]

	    \item[(b)] In this case, there exist an ADP $\ell \to \{ p_1:r_1, \ldots, p_k:r_k \}^{m} \in \PP_>$, 
        a substitution $\sigma$, and position $\pi \in \posT(a)$ with $\flat(a|_{\pi}) = \ell \sigma \in \ANF_{\PP}$ and $b_j \doteq a[r_j \sigma]_{\pi}$.
        First, assume that $m = \ttrue$.
        Let $I_1 = \{\tau \in \posT(a) \mid \tau < \pi\}$ be the set of positions of all annotations strictly above $\pi$, $I_2 = \{\tau \in \posT(a) \mid \tau > \pi\}$ be the set of positions of all annotations strictly below $\pi$, and let $I_3 = \{\tau \in \posT(a) \mid \tau \bot \pi\}$ be the set of positions of all annotations orthogonal to $\pi$.
        Furthermore, for each $i \in I_1$ let $\tau_{i}$ be the positions such that $i.\tau_{i} = \pi$.
        By Requirement (3), there exists a $1 \leq j \leq k$ with $\Pol(\ell^\#) > \sum_{t \trianglelefteq_{\#} r_j} \Pol(t^\#)$ and, additionally, $\Pol(\ell) \geq \Pol(\flat(r_j))$ since $m = \ttrue$.
        As these inequations hold for all instantiations of the occurring variables, we have

{\scriptsize\[
            \begin{array}{lcl}
            V(a) & = & \sum_{s \trianglelefteq_{\#} a} \Pol(s^\#) \\
                & = & \Pol(\annoEps(s|_{\pi})) + \sum_{i \in I_1} \Pol(\annoEps(a|_{i})) + \sum_{i' \in I_2} \Pol(\annoEps(a|_{i'})) + \sum_{i' \in I_3} \Pol(\annoEps(a|_{i'}))\\
                & \geq & \Pol(\annoEps(s|_{\pi})) + \sum_{i \in I_1} \Pol(\annoEps(a|_{i})) + \sum_{i' \in I_3} \Pol(\annoEps(a|_{i'})) \\
                & = & \Pol(\annoEps(\ell) \sigma) + \sum_{i \in I_1} \Pol(\annoEps(a|_{i})) + \sum_{i' \in I_3} \Pol(\annoEps(a|_{i'})) \\
                    && \hspace*{.5cm} \text{\textcolor{blue}{(as $\annoEps(s|_{\pi}) = \annoEps(\ell) \sigma$)}} \\
                & > & \sum_{s \trianglelefteq_{\#} r_j} \Pol(\annoEps(s)\sigma) + \sum_{i \in I_1} \Pol(\annoEps(a|_{i})) + \sum_{i' \in I_3} \Pol(\annoEps(a|_{i'})) \\
                    && \hspace*{.5cm} \text{\textcolor{blue}{(as $\Pol(\annoEps(\ell)) > \sum_{s \trianglelefteq_{\#} r_j} \Pol(\annoEps(s))$, hence $\Pol(\annoEps(\ell)\sigma) > \sum_{s \trianglelefteq_{\#} r_j} \Pol(\annoEps(s)\sigma)$)}}\\
                & \geq & \sum_{s \trianglelefteq_{\#} b_j|_{\pi}} \Pol(\annoEps(s)) + \sum_{i \in I_1} \Pol(\annoEps(a|_{i}[r_j \sigma]_{\tau_{i}})) + \sum_{i' \in I_3} \Pol(\annoEps(a|_{i'})) \\
                                && \hspace*{.5cm} \text{\textcolor{blue}{(by $\Pol(\ell) \geq \Pol(r_j)$ and (a))}}\\
                & = & \sum_{s \trianglelefteq_{\#} b_j} \Pol(s^\#) \\
            & = & V(b_j)\!    
            \end{array}
        \]}

\noindent
        In case of $m = \tfalse$ we additionally remove $\sum_{i \in I_1} \Pol(\annoEps(a|_{i}[r_j \sigma]_{\tau_{i}}))$, so that the inequation remains correct.

	    \item[(c)] In this case, there exists a rule $\ell \to \{ p_1:r_1, \ldots, p_k:r_k \}$ with $\Pol(\ell) \geq \sum_{1 \leq j \leq k} p_j \cdot \Pol(r_j)$, a substitution $\sigma$, and a position $\pi$ of $s$ such that $s|_\pi =\ell\sigma \in \ANF_{\PP}$, and $t_h = s[r_h \sigma]_\pi$ for all $1 \leq h \leq k$.
        
        We perform structural induction on $\pi$.
        So in the induction base $\pi = \varepsilon$ we have $s = \ell\sigma \ito \{ p_1:r_1\sigma, \ldots, p_k:r_k\sigma \}$.
        As $\Pol(\ell) \geq \sum_{1 \leq j \leq k} p_j \cdot \Pol(r_j)$ holds for all instantiations of the occurring variables, for $t_j = r_j\sigma$ we obtain
        \[
            \Pol(s) \;=\; \Pol(\ell\sigma) \;\geq\;\sum_{1 \leq j \leq k} p_j \cdot \Pol(r_j\sigma) \;=\; \sum_{1 \leq j \leq k} p_j \cdot \Pol(t_j). 
        \]
        
        In the induction step, we have $\pi = i.\pi'$, $s = f(s_1,\ldots,s_i,\ldots,s_n)$,
        $s_i \ito \{ p_1:t_{i,1}, \ldots, p_k:t_{i,k} \}$, and $t_j =
        f(s_1,\ldots,t_{i,j},\ldots,s_n)$
        with $t_{i,j} = s_i[r_j\sigma]_{\pi'}$ for all $1 \leq j \leq k$.
        Then by the induction hypothesis we have $\Pol(s_i) \geq \sum_{1 \leq j \leq k} p_j \cdot \Pol(t_{i,j})$.
        Thus, \pagebreak we have
        \[
            \begin{array}{lcl}
            \Pol(s) & = & \Pol(f(s_1,\ldots,s_i,\ldots,s_n)) \\
                & = & f_{\Pol}(\Pol(s_1),\ldots,\Pol(s_i),\ldots,\Pol(s_n)) \\
                & \geq & f_{\Pol}(\Pol(s_1),\ldots,\sum_{1 \leq j \leq k} p_j \cdot \Pol(t_{i,j}),\ldots,\Pol(s_n)) \\
                &   & \; \text{\textcolor{blue}{(by weak monotonicity of $f_{\Pol}$ and $\Pol(s_i) \geq \sum_{1 \leq j \leq k} p_j \cdot \Pol(t_{i,j})$)}} \\
                & = & \sum_{1 \leq j \leq k} p_j \cdot f_{\Pol}(\Pol(s_1),\ldots,\Pol(t_{i,j}),\ldots,\Pol(s_n)) \\
                &   & \; \text{\textcolor{blue}{(as $f_{\Pol}$ is multilinear)}} \\
                & = & \sum_{1 \leq j \leq k} p_j \cdot \Pol(f(s_1,\ldots,t_{i,j},\ldots,s_n))\\
                & = & \sum_{1 \leq j \leq k} p_j \cdot
                \Pol(t_j).
            \end{array}
        \]

	    \item[(d)] In this case, there exist an ADP $\ell \to \{ p_1:r_1, \ldots, p_k:r_k \}^{m} \in \PP$, 
        a substitution $\sigma$, and position $\pi$ with $\flat(a|_{\pi}) = \ell \sigma \in \ANF_{\PP}$ and $b_j \doteq a[r_j \sigma]_{\pi}$.
        First, assume that $m = \ttrue$ and $\pi \in \posT(a)$.
        Let $I_1 = \{\tau \in \posT(a) \mid \tau < \pi\}$ be the set of positions of all annotations strictly above $\pi$, $I_2 = \{\tau \in \posT(a) \mid \tau > \pi\}$ be the set of positions of all annotations strictly below $\pi$, and let $I_3 = \{\tau \in \posT(a) \mid \tau \bot \pi\}$ be the set of positions of all annotations orthogonal to $\pi$.
        Furthermore, for each $i \in I_1$ let $\tau_{i}$ be the position such that $i.\tau_{i} = \pi$.
        By Requirement (2), we have $\Pol(\annoEps(\ell)) \geq \sum_{1 \leq j \leq k} p_j \cdot \sum_{t \trianglelefteq_{\#} r_j} \Pol(\annoEps(t))$ and by (1) we have $\Pol(\ell) \geq \sum_{1 \leq j \leq k} p_j \cdot \Pol(\flat(r_j))$.
        As these inequations hold for all instantiations of the occurring variables, we have

{\scriptsize 
        \begin{longtable}{lcl}
            $V(a)$ & $=$ & $\sum_{t \trianglelefteq_{\#} a} \Pol(t^\#)$\\
            & $=$ & $\Pol(\annoEps(a|_{\pi})) + \sum_{i \in I_1} \Pol(\annoEps(a|_{i})) + \sum_{i' \in I_2} \Pol(\annoEps(a|_{i'})) + \sum_{i' \in I_3} \Pol(\annoEps(a|_{i'}))$ \\
            & $\geq$ & $\Pol(\annoEps(a|_{\pi})) + \sum_{i \in I_1} \Pol(\annoEps(a|_{i})) + \sum_{i' \in I_3} \Pol(\annoEps(a|_{i'}))$ \\
            & $=$ & $\Pol(\annoEps(\ell)\sigma) + \sum_{i \in I_1} \Pol(\annoEps(a|_{i})) + \sum_{i' \in I_3} \Pol(\annoEps(a|_{i'}))$ \\
                && \hspace*{2cm} \text{\textcolor{blue}{(as $a|_{\pi} = \annoEps(\ell) \sigma$)}} \\
            & $\geq$ & $\sum_{1 \leq j \leq k} p_j \cdot \sum_{t \trianglelefteq_{\#} r_j \sigma} \Pol(\annoEps(t)) + \sum_{i \in I_1} \Pol(\annoEps(a|_{i})) + \sum_{i' \in I_3} \Pol(\annoEps(a|_{i'}))$ \\
                &  &  \hspace*{2cm} \text{\textcolor{blue}{(by $\Pol(\annoEps(\ell)) \geq \sum_{1 \leq j \leq k} p_j \cdot \sum_{t \trianglelefteq_{\#} r_j} \Pol(\annoEps(t))$,}}  \\
                &  &  \hspace*{3cm} \text{\textcolor{blue}{hence $\Pol(\annoEps(\ell)\sigma) \geq \sum_{1 \leq j \leq k} p_j \cdot \sum_{t \trianglelefteq_{\#} r_j \sigma} \Pol(\annoEps(t))$)}}  \\
            & $\geq$ & $\sum_{1 \leq j \leq k} p_j \cdot \sum_{t \trianglelefteq_{\#} r_j \sigma} \Pol(\annoEps(t)) + \sum_{i \in I_1} \sum_{1 \leq j \leq k} p_j \cdot \Pol(\annoEps(a|_{i}[r_j \sigma]_{\tau_{i}}))$\\
            &  & $+ \sum_{i' \in I_3} \Pol(\annoEps(a|_{i'}))$\\
                &  &  \hspace*{2cm} \text{\textcolor{blue}{(by $\Pol(\ell) \geq \sum_{1 \leq j \leq k} p_j \cdot \Pol(r_j)$ and (c))}} \\
            & $=$ & $\sum_{1 \leq j \leq k} p_j \cdot \sum_{t \trianglelefteq_{\#} r_j \sigma} \Pol(\annoEps(t)) + \sum_{1 \leq j \leq k} \sum_{i \in I_1} p_j \cdot \Pol(\annoEps(a|_{i}[r_j \sigma]_{\tau_{i}}))$\\
            &  & $+ \sum_{i' \in I_3} \Pol(\annoEps(a|_{i'}))$\\
            & $=$ & $\sum_{1 \leq j \leq k} p_j \cdot \sum_{t \trianglelefteq_{\#} r_j \sigma} \Pol(\annoEps(t)) + \sum_{1 \leq j \leq k} p_j \cdot \sum_{i \in I_1} \Pol(\annoEps(a|_{i}[r_j \sigma]_{\tau_{i}}))$\\
            &  & $+ \sum_{i' \in I_3} \Pol(\annoEps(a|_{i'}))$\\
            & $=$ & $\sum_{1 \leq j \leq k} p_j \cdot \sum_{t \trianglelefteq_{\#} r_j \sigma} \Pol(\annoEps(t)) + \sum_{1 \leq j \leq k} p_j \cdot \sum_{i \in I_1} \Pol(\annoEps(a|_{i}[r_j \sigma]_{\tau_{i}}))$\\
            &  & $+ \sum_{1 \leq j \leq k} p_j \cdot \sum_{i' \in I_3} \Pol(\annoEps(a|_{i'}))$\\
            & $=$ & $\sum_{1 \leq j \leq k} p_j \cdot (\sum_{t \trianglelefteq_{\#} r_j \sigma} \Pol(\annoEps(t)) + \sum_{i \in I_1} \Pol(\annoEps(a|_{i}[r_j \sigma]_{\tau_{i}})) + \sum_{i' \in I_3} \Pol(\annoEps(a|_{i'})))$ \\
            & $=$ & $\sum_{1 \leq j \leq k} p_j \cdot \sum_{t \trianglelefteq_{\#} b_j} \Pol(t^\#)$ \\
            & $=$ & $V(b_j) \!$
        \end{longtable}
        }

\noindent
        In case of $\pi \notin \posT(a)$, we remove $\sum_{t \trianglelefteq_{\#}
	r_j \sigma} \Pol(t^\#)$ in the end, and in case of $m = \tfalse$ we remove $\sum_{i \in I_1} \Pol(\annoEps(a|_{i}[r_j \sigma]_{\tau_{i}}))$.
	\end{itemize}
    The rest is completely analogous to the proof in \cite{kassinggiesl2023iAST}.
\end{myproof}

Finally, we prove soundness of the new rewriting processor.

\setcounter{auxctr}{\value{theorem}}
\setcounter{theorem}{\value{soundnessRewriteCtr}}

\begin{theorem}[Soundness of the Rewriting Processor]
    $\Proc_{\mathtt{r}}$ as in \Cref{def:RewritingProcessor} is \defemph{sound} 
    if one of the following cases holds:
    \begin{enumerate}
        \item $\urules_{\PP}(r_j|_{\tau})$ is NO, and the rule used for rewriting 
        $r_j|_{\tau}$ is L and \pagebreak NE.
        \item $\urules_{\PP}(r_j|_{\tau})$ is NO, and all its rules
        have the form $\ell' \to \{1:r'\}$ for some $\ell'$, $r'$.
        \item $\urules_{\PP}(r_j|_{\tau})$ is NO, $r_j|_\tau$
        is a ground term, and $r_j \itored{}{}{\tau} \{q_1:e_{1}, \ldots, q_h:e_{h}\}$ is an innermost step.
    \end{enumerate}
\end{theorem}

\setcounter{theorem}{\value{auxctr}}

\begin{myproof}
Let $\overline{\PP'} = \PP' \cup N \cup \{\ell \ruleArr{}{}{} \{ p_1:\flat(r_{1}), \ldots, p_k: \flat(r_{k})\}^{m}\}$ and $\overline{\PP} = \overline{\PP'} \cup \{\ell \ruleArr{}{}{} \{ p_1:r_{1}, \ldots, p_k:r_{k}\}^{m}\}$.
    We call $\ell \ruleArr{}{}{} \{ p_1:r_{1}, \ldots, p_k: r_k\}^{m}$ 
    the \emph{old} ADP, we call
     $\ell \to \{p_1:r_1, \ldots, p_k:r_k\} \setminus \{p_j:r_j\} \cup \{p_j \cdot
    q_1:e_1, \ldots, p_j \cdot q_h:e_h\}^{m}$ the \emph{new} ADP, and
    $\ell \ruleArr{}{}{} \{ p_1:\flat(r_{1}), \ldots, p_k: \flat(r_{k})\}^{m}$ is called
    the \emph{non-annotated old} ADP.

    \noindent
    \underline{\textbf{First Case}}

    \noindent  
    We start with the case where $\urules_{\PP}(r_j|_{\tau})$ is NO and the used rule 
    $\hat{\ell} \to \{ \hat{p}_1:\hat{r}_1, \ldots, \hat{p}_h:\hat{r}_h\}$ is L and NE.
    First, note that the rule used for rewriting $r_j|_{\tau}$ in the rewrite processor
    is contained in $\urules_{\PP}(r_j|_{\tau})$.\footnote{The rules that are applicable at position $\tau$ 
    do not have to be applicable in an innermost RST, see \cite[Sect.\ 6]{AAECC01} and
    \cite[Ex.\ 5.14]{thiemanndiss2007}.}
    Since $\urules_{\PP}(r_j|_{\tau})$ is NO, we can therefore be sure that there is only 
    a single rule applicable at position $\tau$.

    \smallskip
   
    \noindent
    Let $\PP$ be not iAST.
    Then there exists a $\PP$-CT $\F{T} = (V,E,L,P)$ that converges with probability $c < 1$.
    We will now create a $\overline{\PP'}$-CT $\F{T}' = (V',E',L',P')$ such that $|\F{T}'|_{\ctleaf} \leq |\F{T}|_{\ctleaf} < 1$, and hence $\overline{\PP'}$ is not iAST either.

    The core steps of the proof are the following:
    \begin{enumerate}
        \item[1.] We iteratively remove usages of the old ADP using a construction $\Phi(\circ)$. 
        The limit of this iteration, namely $\F{T}^{(\infty)}$ is a $\overline{\PP'}$-CT that converges with probability at most $c < 1$, hence, $\overline{\PP'}$ is not iAST.
        \begin{enumerate}
            \item[1.1.] For a $\PP$-CT $\F{T}_x$ that uses the old ADP at the root $x$ at a position \linebreak $\pi \in \posT(t_x)$, we create a new $\overline{\PP}$-CT $\Phi(\F{T}_x)$ that uses the new ADP at the root.
            \item[1.2.] For a $\PP$-CT $\F{T}_x$ that uses the old ADP at the root $x$ at a position \linebreak $\pi \not\in \posT(t_x)$, we create a new $\overline{\PP'}$-CT $\Phi(\F{T}_x)$ that uses the non-annotated old ADP at the root.
        \end{enumerate}
    \end{enumerate}
  This proof structure will also be used in the other
    cases for soundness.

    \smallskip
          
    \noindent 
    \textbf{\underline{1. We iteratively remove usages of the old ADP}}

    \noindent 
    W.l.o.g., in $\F{T}$ there exists at least one rewrite step performed at some node $x$ with the old ADP (otherwise, $\F{T}$ would already be a $\overline{\PP'}$-CT).
    Furthermore, we can assume that this is the first such rewrite step in the path from the root to the node $x$ and that $x$ is a node of minimum depth with this property.
    We will now replace this rewrite step with a rewrite step using the new ADP such that we result in a CT $\F{T}^{(1)}$ with the following connections between $\F{T}$ and $\F{T}^{(1)}$.
    Here, for an infinite path $p = v_1, v_2, \ldots$ in a CT $\F{T} = (V,E,L,P)$, by $\mathcal{C}(p) =
    w_1, w_2, \ldots$ we denote the sequence of $\{P,R\}$-labels for these nodes, i.e., we
    have $w_i = P$ if $v_i \in P$ or $w_i = R$ otherwise for all $i$.
    \begin{itemize}
        \item[(a)] $|\F{T}^{(1)}|_{\ctleaf} \leq |\F{T}|_{\ctleaf} = c$, and
        \item[(b)] for every infinite path $p$ in $\F{T}^{(1)}$ we can find an infinite
          path $p'$ in $\F{T}$ such that $\mathcal{C}(p)$ and $\mathcal{C}(p')$ only differ in a
          finite number of $R$-nodes, i.e., we can remove and add a finite number
          of $R$-nodes \pagebreak from $\mathcal{C}(p')$ to get $\mathcal{C}(p)$.
    \end{itemize}
    This construction only works because $\urules_{\PP}(r_j|_{\tau})$ is NO and there
    exists no annotation below $\tau$, and the
    resulting CT has only at most
    the same probability of termination as the original one due to
    the fact that $\hat{\ell} \to \{ \hat{p}_1:\hat{r}_1, \ldots, \hat{p}_h:\hat{r}_h\}^{m'}$ is also L and NE.
    Let $\F{T}_x$ be the induced sub-CT that starts at node $x$, i.e. $\F{T}_x = \F{T}[xE^*]$.
    The construction defines a new tree $\Phi(\F{T}_x)$ such that (a) and (b)
    w.r.t.\ $\F{T}_x$ and $\Phi(\F{T}_x)$ holds, and where we use the new ADP at the root
    node $x$ instead of the old one (i.e., we pushed the first use of the old ADP deeper into the tree).
    Then, by replacing the subtree $\F{T}_x$ with the new tree $\Phi(\F{T}_x)$ in $\F{T}$, we get a $\PP$-CT $\F{T}^{(1)}$, with (a) and (b) w.r.t.\ $\F{T}$ and $\F{T}^{(1)}$, and where we use the new ADP at node $x$ instead of the old one.
    We can then do this replacement iteratively for every use of the old ADP, i.e., we again replace the first use of the old ADP in $\F{T}^{(1)}$ to get $\F{T}^{(2)}$ with (a) and (b) w.r.t.\ $\F{T}^{(1)}$ and $\F{T}^{(2)}$, and so on.
    In the end, the limit of all these CTs $\lim_{i \to \infty} \F{T}^{(i)}$ is a $\overline{\PP'}$-CT, that we denote by $\F{T}^{(\infty)}$ and that converges with probability at most $c < 1$, 
    and hence, $\overline{\PP'}$ is not iAST.
    
    To see that $\F{T}^{(\infty)}$ is indeed a valid $\overline{\PP'}$-CT, note that in
    every iteration of the construction we turn a use of the old ADP at minimum depth
    into a use of the new one.
    Hence, for every depth $H$ of the tree, we eventually turned every use of the old ADP up to
    depth $H$ into a use of the new one so that the construction will not change the tree above depth $H$ anymore, i.e.,
    there exists an $m_H$ such that $\F{T}^{(\infty)}$ and $\F{T}^{(m_H)}$ are the same trees up to depth $H$.
    This means that the sequence $\lim_{i \to \infty} \F{T}^{(i)}$ really converges into a tree that satisfies the first five conditions of a $\overline{\PP'}$-CT.
    We only have to show that the last condition of a CT, namely that every infinite path in $\lim_{i \to \infty} \F{T}^{(i)}$ contains infinitely many nodes from $P$, holds as well.
    First, by induction on $n$ one can prove that all trees $\F{T}^{(i)}$ for all $1 \leq i \leq
    n$ satisfy Condition (6), because due to (b) we can find for
    each infinite path $p \in \F{T}^{(i)}$ an infinite path $p'$ in $\F{T}$ such that
    $\mathcal{C}(p)$ and $\mathcal{C}(p')$ only differ in a finite number of $R$-labels.
    Hence, if $p$ contains no nodes from $P$, then $p'$ contains no nodes from $P$, which is a contradiction to $\F{T}$ satisfying Condition (6).
    Moreover, we only replace subtrees after a node in $P$, the node itself remains in $P$, and after we replaced the subtree at a node $v$, we will never replace a predecessor of $v$ anymore.
    This means that the subsequence between the $i$-th and $(i+1)$-th occurrence of $P$ in $p$ and the subsequence between the $i$-th and $(i+1)$-th occurrence of $P$ in $p'$ only differ in a finite number of $R$ nodes again, for every $i$.
    Now, let $p = v_1, v_2, \ldots$ be an infinite path that starts at the root in $\F{T}^{(\infty)}$ and only contains finitely many nodes from $P$, i.e., there exists an $1 \leq i$ such that $v_i, v_{i+1}, \ldots$ contains no node from $P$.
    Again, let $m_H \in \IN$ such that $\F{T}^{(\infty)}$ and $\F{T}^{(m_H)}$ are the same trees up to depth $H$.
    The path $v_1, \ldots, v_H$ must be a path in $\F{T}^{(m_H)}$ as well.
    For two different $H_1, H_2$ with $i < H_1 < H_2$ we know that since the path $v_i, \ldots, v_{H_2}$ contains no node from $P$, the path must also exist in $\F{T}^{(m_{H_1})}$.
    We can now construct an infinite path in $\F{T}^{(m_i)}$ that contains no nodes from $P$, which is a contradiction.

    Next, we want to prove that we really have $|\F{T}^{(\infty)}|_{\ctleaf} \leq c$.
    Again, by induction on $n$ one can prove that $|\F{T}^{(i)}|_{\ctleaf} \leq c$ for all $1 \leq i \leq
    n$.
    Assume for a contradiction that $\F{T}^{(\infty)}$ converges with probability greater than $c$, i.e. $|\F{T}^{(\infty)}|_{\ctleaf} > c$.
    Then there exists an $H \in \IN$ for the depth such that $\sum_{x \in \ctleaf^{\F{T}^{(\infty)}}, d(x) \leq H} p_x > c$.
    Here, $d(x)$ denotes the depth of node $x$.
    Again, let $m_H \in \IN$ such that $\F{T}^{(\infty)}$ and $\F{T}^{(m_H)}$ are the same trees up to depth $H$.
    But this would \pagebreak mean that $|\F{T}^{(m_H)}|_{\ctleaf} \geq \sum_{x \in \ctleaf^{\F{T}^{(m_H)}}, d(x) \leq H} p_x = \sum_{x \in \ctleaf^{\F{T}^{(\infty)}}, d(x) \leq H} p_x > c$, which is a contradiction to $|\F{T}^{(m_H)}|_{\ctleaf} \leq c$.

    It remains to show the mentioned construction $\Phi(\circ)$.
  
    \smallskip
          
    \noindent 
    \textbf{\underline{1.1. Construction of $\Phi(\circ)$ if $\pi \in \posT(t_x)$}}

    \noindent 
    Let $\F{T}_x$ be a $\PP$-CT that uses the old ADP at the root node $x$, i.e, $t_x \itored{}{}{\PP} \{p_{y_1}:t_{y_1}, \ldots, p_{y_k}:t_{y_k}\}$ using the ADP $\ell \ruleArr{}{}{} \{ p_1:r_{1}, \ldots, p_k: r_k\}^{m}$, the position $\pi$ with $\pi \in \posT(t_x)$, and a substitution $\sigma$ such that $\flat(t_x|_{\pi}) = \ell \sigma \in \ANF_{\PP}$.
    Then $t_{y_j} = t_x[r_j\sigma]_{\pi}$ if $m = \ttrue$, or $t_{y_j} = \disannoPos{\pi}(t_x[r_j\sigma]_{\pi})$, otherwise.
  
    \smallskip
          
    \noindent 
    \textbf{\underline{1.1.1 General construction of $\Phi(\F{T}_x)$}}

    \noindent 
    Instead of applying only the old ADP at the root $x$
    \begin{center}
        \scriptsize
        \begin{tikzpicture}
            \tikzstyle{adam}=[thick,draw=black!100,fill=white!100,minimum size=4mm, shape=rectangle split, rectangle split parts=2,rectangle split
            horizontal]
            \tikzstyle{empty}=[rectangle,thick,minimum size=4mm]
            
            \node[adam] at (1.5, 2)  (1) {$p_x$ \nodepart{two} $t_x$};
            \node[adam] at (-1.5, 1)  (11) {$p_{y_1}$ \nodepart{two} $t_{y_1}$};
            \node[empty] at (0, 1)  (12) {$\ldots$};
            \node[adam] at (1.5, 1)  (13) {$p_{y_j}$ \nodepart{two} $t_{y_j}$};
            \node[empty] at (3, 1)  (14) {$\ldots$};
            \node[adam] at (4.5, 1)  (15) {$p_{y_k}$ \nodepart{two} $t_{y_k}$};
            
            \draw (1) edge[->] (11);
            \draw (1) edge[->] (12);
            \draw (1) edge[->] (13);
            \draw (1) edge[->] (14);
            \draw (1) edge[->] (15);
        \end{tikzpicture}
    \end{center}
    where we use an arbitrary rewrite step at node $y_j$ afterwards, we want to directly
    apply the rewrite rule at position $\pi.\tau$ of the term $t_{y_j}$ in our CT, which we performed on $r_j$ at position $\tau$ to transform the old into the new ADP, to get
    \begin{center}
        \scriptsize
        \begin{tikzpicture}
            \tikzstyle{adam}=[thick,draw=black!100,fill=white!100,minimum size=4mm, shape=rectangle split, rectangle split parts=2,rectangle split
            horizontal]
            \tikzstyle{empty}=[rectangle,thick,minimum size=4mm]
            
            \node[adam] at (1.5, 2)  (1) {$p_x$ \nodepart{two} $t_x$};
            \node[adam] at (-1.5, 1)  (11) {$p_{y_1}$ \nodepart{two} $t_{y_1}$};
            \node[empty] at (0, 1)  (12) {$\ldots$};
            \node[adam] at (1.5, 1)  (13) {$p_{y_j}$ \nodepart{two} $t_{y_j}$};
            \node[empty] at (3, 1)  (14) {$\ldots$};
            \node[adam] at (4.5, 1)  (15) {$p_{y_k}$ \nodepart{two} $t_{y_k}$};
            \node[adam] at (0, 0)  (131) {$p_{y_{j_1}}$ \nodepart{two} $t_{y_{j_1}}$};
            \node[empty] at (1.5, 0)  (132) {$\ldots$};
            \node[adam] at (3, 0)  (133) {$p_{y_{j_h}}$ \nodepart{two} $t_{y_{j_h}}$};
            
            \draw (1) edge[->] (11);
            \draw (1) edge[->] (12);
            \draw (1) edge[->] (13);
            \draw (1) edge[->] (14);
            \draw (1) edge[->] (15);
            \draw (13) edge[->] (131);
            \draw (13) edge[->] (132);
            \draw (13) edge[->] (133);
        \end{tikzpicture}
    \end{center}
    Then, we can contract the edge $(x,y_j)$ to get
    \begin{center}
        \scriptsize
        \begin{tikzpicture}
            \tikzstyle{adam}=[thick,draw=black!100,fill=white!100,minimum size=4mm, shape=rectangle split, rectangle split parts=2,rectangle split
            horizontal]
            \tikzstyle{empty}=[rectangle,thick,minimum size=4mm]
            
            \node[adam] at (3, 3)  (1) {$p_x$ \nodepart{two} $t_x$};
            \node[adam] at (-1.5, 1)  (11) {$p_{y_1}$ \nodepart{two} $t_{y_1}$};
            \node[empty] at (0, 1)  (12) {$\ldots$};
            \node[adam] at (1.5, 1)  (13) {$p_{y_{j_1}}$ \nodepart{two} $t_{y_{j_1}}$};
            \node[empty] at (3, 1)  (14) {$\ldots$};
            \node[adam] at (4.5, 1)  (15) {$p_{y_{j_h}}$ \nodepart{two} $t_{y_{j_h}}$};
            \node[empty] at (6, 1)  (16) {$\ldots$};
            \node[adam] at (7.5, 1)  (17) {$p_{y_{k}}$ \nodepart{two} $t_{y_{k}}$};
            
            \draw (1) edge[->] (11);
            \draw (1) edge[->] (12);
            \draw (1) edge[->] (13);
            \draw (1) edge[->] (14);
            \draw (1) edge[->] (15);
            \draw (1) edge[->] (16);
            \draw (1) edge[->] (17);
        \end{tikzpicture}
    \end{center}
    and this is equivalent to applying the new ADP.
    Note that the rewrite step at position $\pi.\tau$ may not be an innermost rewrite step in the CT.

    \begin{wrapfigure}[9]{r}{0.37\textwidth}
        \vspace*{-1.5cm}
        \begin{center}
            \begin{tikzpicture}[scale=0.5]
                \begin{pgfonlayer}{nodelayer}
                    \node [style=target,pin={[pin distance=0.05cm, pin edge={,-}] 140:\tiny \textcolor{blue}{$x$}}] (0) at (0, 6) {};
                    \node [style=none] (1) at (0, 3) {};
                    \node [style=target] (2) at (-3, 3) {};
                    \node [style=target,pin={[pin distance=0.05cm, pin edge={,-}] 140:\tiny \textcolor{blue}{$y_j$}}] (3) at (0, 3) {};
                    \node [style=none] (6) at (1.5, 0) {};
                    \node [style=none] (7) at (-1.5, 0) {};
                    \node [style=none] (9) at (-2, -1) {};
                    \node [style=none] (10) at (2, -1) {};
                    \node [style=moveBlock] (12) at (-0.75, 0.75) {};
                    \node [style=moveBlock] (13) at (0, 0) {};
                    \node [style=moveBlock] (14) at (0.75, 0.25) {};
                    \node [style=moveBlock] (15) at (-0.75, 0.75) {};
                    \node [style=none] (16) at (-0.5, 1.5) {};
                    \node [style=none] (17) at (0, 1.25) {};
                    \node [style=none] (18) at (0.5, 1.5) {};
                    \node [style=none] (19) at (-0.75, 0) {};
                    \node [style=none] (20) at (0, -0.75) {};
                    \node [style=none] (21) at (0.75, -0.5) {};
                    \node [style=target] (22) at (3, 3) {};
                \end{pgfonlayer}
                \begin{pgfonlayer}{edgelayer}
                    \draw (0) to (1.center);
                    \draw (0) to (2);
                    \draw (3) to (6.center);
                    \draw (3) to (7.center);
                    \draw [style=dotWithoutHead] (7.center) to (9.center);
                    \draw [style=dotWithoutHead] (6.center) to (10.center);
                    \draw [style=dotWithoutHead, in=15, out=-105, looseness=0.50] (3) to (16.center);
                    \draw [style=dotWithoutHead, in=120, out=-90, looseness=0.75] (3) to (17.center);
                    \draw [style=dotWithoutHead, in=135, out=-75] (3) to (18.center);
                    \draw [style=dotHead, in=90, out=-30, looseness=0.75] (18.center) to (14);
                    \draw [style=dotHead, in=90, out=-150, looseness=0.75] (16.center) to (15);
                    \draw [style=dotHead, in=90, out=-45] (17.center) to (13);
                    \draw [style=dashHead, bend right=75, looseness=2.00] (14) to (3);
                    \draw [style=dashHead, bend left=75, looseness=1.75] (15) to (3);
                    \draw [style=dashHead, bend right=105, looseness=2.75] (13) to (3);
                    \draw [style=dotWithoutHead] (15) to (19.center);
                    \draw [style=dotWithoutHead] (13) to (20.center);
                    \draw [style=dotWithoutHead] (14) to (21.center);
                    \draw (0) to (22);
                \end{pgfonlayer}
            \end{tikzpicture}
        \end{center}
    \end{wrapfigure}
    The subtrees that start at the nodes $y_1, \ldots, y_{j-1}, y_{j+1}, \ldots, y_{k}$ remain completely the same.
    We only have to construct a new subtree for node $y_j$, i.e., the node that really
    changed when applying the rewriting processor.
    To be precise, let $\F{T}_{y_j} = \F{T}_{x}[y_jE^*]$ be the subtree starting at node $y_j$.
    The construction first creates a new subtree $\Psi(\F{T}_{y_j})$ such that (a) and (b) hold w.r.t.\ $\F{T}_{y_j}$ and $\Psi(\F{T}_{y_j})$, and that directly performs the first rewrite step at position $\pi.\tau$ at the root of the tree, by pushing it from the original position in the tree $\F{T}_{y_j}$ to the root.
    This can be seen in the diagram above.
    Again, this push only results in the exact same termination probability due to our
    restriction that $\hat{\ell} \to \{ \hat{p}_1:\hat{r}_1, \ldots, \hat{p}_h:\hat{r}_h\}$ is L and NE.
    Then, by replacing $\F{T}_{y_j}$ by $\Psi(\F{T}_{y_j})$ in $\F{T}_x$ we result in a tree $\F{T}_x'$ such that (a) and (b) hold w.r.t.\ $\F{T}_x$ and $\F{T}_x'$, and such that we perform the desired rewrite step at node $y_j$.
    Finally, we contract the edge $(x,y_j)$ in $\F{T}_x'$, in order to get $\Phi(\F{T}_x)$.
    Again, (a) and (b) hold w.r.t.\ $\F{T}_x$ and $\Phi(\F{T}_x)$, and we use the new ADP at the root $x$ in $\Phi(\F{T}_x)$.
    It only remains to explain the construction of $\Psi(\F{T}_{y_j})$.
    \smallskip
          
    \noindent 
    \textbf{\underline{1.1.2. Construction of $\Psi(\F{T}_{y_j})$}}

    \noindent 
    We will move the first rewrite step that takes place at position $\pi.\tau$ from the original tree $\F{T}_{y_j}$ (example on the left below) to the top of the new tree $\Psi(\F{T}_{y_j})$ (example on the right below) and show that (a) and (b) both hold after this construction.
    Below, the circled nodes represent the nodes where we perform a rewrite step at
    position $\pi.\tau$.

\vspace*{-.7cm}

    \begin{center}
        \centering
        \scriptsize
		\begin{tikzpicture}
			\tikzstyle{adam}=[rectangle,thick,draw=black!100,fill=white!100,minimum size=3mm]
			\tikzstyle{empty}=[shape=circle,thick,minimum size=8mm]
			\tikzstyle{circle}=[shape=circle,draw=black!100,fill=white!100,thick,minimum size=3mm]
			
			\node[empty] at (-3.5, 0.5)  (name) {$\F{T}_{y_j}$};
			\node[adam] at (-3.5, 0)  (la) {$v_0$};

			\node[circle] at (-4.5, -1)  (lb1) {$v_1$};
			\node[adam] at (-3.5, -1)  (lb2) {$v_2$};
			\node[adam] at (-2.5, -1)  (lb3) {$v_3$};

			\node[adam] at (-4.5, -2)  (lc1) {$v_4$};

			\node[circle] at (-3, -2)  (ld1) {$v_5$};
			\node[circle] at (-2, -2)  (ld3) {$v_6$};
		
			\node[adam] at (-3, -3)  (lf1) {$v_7$};
			\node[adam] at (-2, -3)  (lf3) {$v_8$};

			\node[adam] at (-3, -4)  (lg1) {$v_9$};

			\draw (la) edge[->] (lb1);
			\draw (la) edge[->] (lb2);
			\draw (la) edge[->] (lb3);
			\draw (lb1) edge[->] (lc1);
			\draw (lb3) edge[->] (ld1);
			\draw (lb3) edge[->] (ld3);
			\draw (ld1) edge[->] (lf1);
			\draw (ld3) edge[->] (lf3);
			\draw (lf1) edge[->] (lg1);

			\node[empty] at (-5.1,-1.3)  (Z) {$Z$};

            \draw [dashed] (-5,-1.5) -- (-4,-1.5) -- (-4,-2.5) -- (-1,-2.5);

			\node[empty] at (-0.5, -1.5)  (lead) {\huge $\leadsto$};
			
			\node[empty] at (2.5, 0.5)  (name2) {$\Psi(\F{T}_{y_j})$};
			\node[circle] at (2.5, 0)  (a) {$\hat{v}$};

			\node[adam] at (2.5, -1)  (b1) {$1.v_0$};

			\node[adam] at (1.5, -2)  (c1) {$1.v_1$};
			\node[adam] at (2.5, -2)  (c2) {$1.v_2$};
			\node[adam] at (3.5, -2)  (c3) {$1.v_3$};

			\node[adam] at (3, -3)  (d1) {$1.v_5$};
			\node[adam] at (4, -3)  (d2) {$1.v_6$};

			\node[adam] at (3, -4)  (f1) {$v_9$};

			\draw (a) edge[->] (b1);
			\draw (b1) edge[->] (c1);
			\draw (b1) edge[->] (c2);
			\draw (b1) edge[->] (c3);
			\draw (c3) edge[->] (d1);
			\draw (c3) edge[->] (d2);
			\draw (d1) edge[->] (f1);
		\end{tikzpicture}
    \end{center}
    We will define the $\PP$-CT $\Psi(\F{T}_{y_j})$ that satisfies the properties (a) and (b) w.r.t.\ $\F{T}_{y_j}$ and $\Psi(\F{T}_{y_j})$, and that directly performs the rewrite step $t_{y_j} \itored{}{}{\PP, \pi.\tau} \{\hat{p}_{1}:t_{y_{j_1}}, \ldots, \hat{p}_{h}:t_{y_{j_h}}\}$, with the rule $\hat{\ell} \to \{ \hat{p}_1:\hat{r}_1, \ldots, \hat{p}_h:\hat{r}_h\}^{\hat{m}} \in \PP$, a substitution $\hat{\sigma}$, and the position $\pi.\tau$, at the new root $\hat{v}$.
    Here, we have $\flat(t_{y_j}|_{\pi.\tau}) = \flat(r_j\sigma|_{\tau}) = \flat(r_j|_{\tau} \sigma) = \hat{\ell} \hat{\sigma}$.
    Let $Z$ be the set of all nodes $v$ of $\F{T}_{y_j}$ where we did not perform a 
    rewrite step at position $\pi.\tau$ in the path from 
    the root $x$ to the node $v$, or $v$ is the first node in the path that 
    performs a rewrite step at position $\pi.\tau$.
    In the example we have $Z = \{v_0, \ldots, v_6\} \setminus \{v_4\}$.
    For each of these nodes $z \in Z$ and each $1 \leq e \leq h$,
    we create a new node $e.z \in V'$ with edges as in $\F{T}_{y_j}$ for the nodes in $Z$, e.g., for the node $1.v_3$ we create an edge to $1.v_5$ and $1.v_6$.
    Furthermore, we add the edges from the new root $\hat{v}$ to the nodes $e.{y_j}$ for all $1 \leq e \leq h$.
    Remember that $y_j$ was the root in the tree $\F{T}_{y_j}$ and has to be contained in $Z$.
    For example, for the node $\hat{v}$ we create an edge to $1.v_0$.
    For all these new nodes in $Z$, we show the following:        
    \begin{itemize}
        \item[] (T1) $t_z[\hat{r}_e \gamma]_{\pi.\tau} \doteq t'_{e.z}$ for the substitution $\gamma$ such that $\flat(t_z|_{\pi.\tau}) = \hat{\ell} \gamma$
        \item[] (T2) $\posT(t_z[\flat(\hat{r}_e) \gamma]_{\pi.\tau}) \subseteq \posT(t'_{e.z})$.
        \item[] (T3) $p_{e.z}^{\Psi(\F{T}_{y_j})} = p_z^{\F{T}_{y_j}} \cdot \hat{p}_e$ 
    \end{itemize}
    Note that we only regard the subtree of the $j$-th child of the root. Because of the 
    prerequisite in the definition of the rewrite processor (there is no
    annotation below or at position $\tau$), there are no
    annotations on or below $\pi.\tau$ for all \pagebreak nodes in $Z$. 
    
    Now, for a leaf $e.z \in V'$ either $z \in V$ is also a leaf (e.g., node $v_2$) or we rewrite the position $\pi.\tau$ at node $z$ in $\F{T}_{y_j}$ (e.g., node $v_1$).
    If we rewrite $t_{z} \itored{}{}{\PP, \pi.\tau} \{\hat{p}_{1}:t_{w_1}, \ldots,
    \hat{p}_{h}:t_{w_h}\}$, then we have $t_{w_e} = t_x[\flat(\hat{r}_e)\gamma]_{\pi.\tau}$ if $m = \ttrue$ or $t_{w_e} = \disannoPos{\pi.\tau}(t_x[\flat(\hat{r}_e)\gamma]_{\pi.\tau})$, otherwise.
    In both cases, we get $t_{w_e} \doteq t_x[\flat(\hat{r}_e)\gamma]_{\pi.\tau} \doteq_{(T1)} t'_{e.z}$, $\posT(t_{w_e}) \subseteq \posT(t_z[\flat(\hat{r}_e) \gamma]_{\pi.\tau}) \subseteq_{(T2)} \posT(t'_{e.z})$ and $p_{e.z}^{\Psi(\F{T}_{y_j})} = p_z^{\F{T}_{y_j}} \cdot \hat{p}_e =_{(T3)} p_{w_e}^{\F{T}_{y_j}}$, 
    and we can again copy the rest of this subtree of $\F{T}_{y_j}$ in
    our newly generated tree $\Psi(\F{T}_{y_j})$.
    In our example, $v_1$ has the only successor $v_4$, hence we can copy the subtree starting at node $v_4$, which is only the node itself, to the node $1.v_1$ in $\Psi(\F{T}_{y_j})$.
    For $v_5$ we have the only successor $v_7$, hence we can copy the subtree starting at node $v_7$, which is the node itself together with its successor $v_9$, to the node $1.v_5$ in $\Psi(\F{T}_{y_j})$.
    So essentially, we just have to define the part of the tree before we reach the rewrite step in $\F{T}_{y_j}$, 
    and then, we have to show that (a) and (b) for $\F{T}_{y_j}$ and $\Psi(\F{T}_{y_j})$ are satisfied.
    We show the latter first, and then explain the proof that this label gives us indeed a valid $\PP$-CT.
    
    We start by showing (a) for $\F{T}_{y_j}$ and $\Psi(\F{T}_{y_j})$.
    Let $u$ be a leaf in $\Psi(\F{T}_{y_j})$.
    If $u = e.v$ for some node $v \in Z$ that is a leaf in $\F{T}_{y_j}$ (e.g., node $1.v_2$), then also $e.v$ must be a leaf in $\Psi(\F{T}_{y_j})$ for every $1 \leq e \leq h$.
    Here, we get $\sum_{1 \leq e \leq h} p_{e.v}^{\Psi(\F{T}_{y_j})} \stackrel{\text{(T3)}}{=} \sum_{1 \leq e \leq h} p_v^{\F{T}_{y_j}} \cdot \hat{p}_e = p_v^{\F{T}_{y_j}} \cdot \sum_{1 \leq e \leq h} \hat{p}_e = p_v^{\F{T}_{y_j}} \cdot 1 = p_v^{\F{T}_{y_j}}$.
    If $u = e.v$ for some node $v \in Z$ that is not a leaf in $\F{T}_{y_j}$ (e.g., node $1.v_1$), then we know by construction 
    that all successors of $v$ in $\F{T}_{y_j}$ are not contained in $Z$ and are leaves.
    Here, we get $p_{e.v}^{\Psi(\F{T}_{y_j})} \stackrel{\text{(T3)}}{=} p_v^{\F{T}_{y_j}} \cdot \hat{p}_e = p_w^{\F{T}_{y_j}}$ for the (unique) $e$-th successor $w$ of $v$.
    Finally, if $u$ does not have the form $u = e.v$, then $u$ is also a leaf in $\F{T}_{y_j}$ with $p_{u}^{\Psi(\F{T}_{y_j})} = p_{u}^{\F{T}_{y_j}}$.
    Note that these cases cover no leaf of $\F{T}_{y_j}$ twice.
    This implies that we have
    {\small
    \allowdisplaybreaks
    \begin{align*}
        & |\Psi(\F{T}_{y_j})|_{\ctleaf}\\
        = & \sum_{v \in \ctleaf^{\Psi(\F{T}_{y_j})}} p_{e.v}^{\Psi(\F{T}_{y_j})}\\
        = & \sum_{\substack{e.v \in \ctleaf^{\Psi(\F{T}_{y_j})}\\ v \in Z}} p_{e.v}^{\Psi(\F{T}_{y_j})} + \sum_{\substack{e.v \in \ctleaf^{\Psi(\F{T}_{y_j})}\\ v \in wE, v \not\in Z, w \in Z}} p_{e.v}^{\Psi(\F{T}_{y_j})} + \sum_{\substack{v \in \ctleaf^{\Psi(\F{T}_{y_j})}\\v \in \ctleaf^{\F{T}_{y_j}}, v \in wE, w \not\in Z}} p_v^{\Psi(\F{T}_{y_j})} \\
        \leq & \sum_{\substack{v \in \ctleaf^{\F{T}_{y_j}}\\ v \in Z}} \left( \sum_{1 \leq e \leq h} p_{e.v}^{\Psi(\F{T}_{y_j})} \right) + \sum_{\substack{v \in \ctleaf^{\F{T}_{y_j}}, 1 \leq e \leq h\\ v \in wE, v \not\in Z, w \in Z}} p_{e.v}^{\Psi(\F{T}_{y_j})} + \sum_{\substack{v \in \ctleaf^{\F{T}_{y_j}}\\ v \in wE, w \not\in Z}} p_v^{\Psi(\F{T}_{y_j})} \\
        \leq & \sum_{\substack{v \in \ctleaf^{\F{T}_{y_j}}\\ v \in Z}} \!\! p_v^{\F{T}_{y_j}} + \sum_{\substack{v \in \ctleaf^{\F{T}_{y_j}}\\ v \in wE, v \not\in Z, w \in Z}} p_v^{\F{T}_{y_j}} + \sum_{\substack{v \in \ctleaf^{\F{T}_{y_j}}\\ v \in wE, w \not\in Z}} p_v^{\F{T}_{y_j}} \\
        = & \sum_{v \in \ctleaf^{\F{T}_{y_j}}} p_v^{\F{T}_{y_j}} \\
        = & |\F{T}_{y_j}|_{\ctleaf} 
    \end{align*}
    }
    Next,
    we show (b) for $\F{T}_{y_j}$ and $\Psi(\F{T}_{y_j})$.
    Let $p = u_0, u_1, \ldots$ be an infinite path in $\Psi(\F{T}_{y_j})$  that starts at the root $\hat{v}$.
    If for all $1 \leq i$ we have $u_i = e.v_i$ for some node $v_i \in Z$ and $1 \leq
    e \leq h$, \pagebreak
    then $p = v_1, \ldots$ is our desired path in $\F{T}_{y_j}$.
    Otherwise, there is a maximal $1 \leq o$ such that for all $1 \leq i \leq o$ we have $u_i = e.v_i$ for some node $v_i \in Z$ and $1 \leq e \leq h$.
    Then our desired path is $v_1, \ldots, v_o, w, u_{o+1}, \ldots$.
    Here, $w$ is the $e$-th successor of $v_o$ in $\F{T}_{y_j}$.
    Note that in the first case, we remove one $R$-label at the start of our path, while in the other case, we just move an $R$-label from the first position to a later one in the path.
    This shows that (b) is satisfied.

    Finally, we prove that $\Psi(\F{T}_{y_j})$ is a valid $\PP$-CT.
    We only need to prove the construction that satisfies all of our conditions for the nodes in $Z$.
    As the rest of the tree is copied, we can be sure that all Conditions (1)-(5) of a $\PP$-CT are satisfied.
    Additionally, due to (b) w.r.t.\ $\F{T}_{y_j}$ and $\Psi(\F{T}_{y_j})$, we get (6) as well, because $\F{T}_{y_j}$ satisfies (6).

    \smallskip
    
    At the root, after applying the old ADP, we directly perform the rewrite
    step
    $t_{y_j} \itored{}{}{\PP} \{\hat{p}_{1}:t'_{1.y_{j}}, \ldots, \hat{p}_{h}:t'_{h.y_{j}}\}$, with the rule $\hat{\ell} \to \{ \hat{p}_1:\hat{r}_1, \ldots, \hat{p}_h:\hat{r}_h\}^{\hat{m}} \in \PP$, a substitution $\hat{\sigma}$, and the position $\tau$.
    Then, $t_{e.y_{j}} = t_{y_j}[\flat(\hat{r}_e) \hat{\sigma}]_{\pi.\tau}$ if $m = \ttrue$ or $t_{w_e} = \disannoPos{\pi.\tau}(t_{y_j}[\flat(\hat{r}_e) \hat{\sigma}]_{\pi.\tau})$, otherwise.
    Here, the conditions (T1)-(T3) are clearly satisfied.
    We now construct the rest of this subtree by mirroring the rewrite steps in
    the original tree (which is always possible due to the fact that $\urules_{\PP}(r_j|_{\tau})$ is NO), and once we encounter the rewrite step that we moved to the top, i.e., once we use a rewrite step at position $\pi.\tau$, we skip this rewrite step and directly go on with the $e$-th successor if we are in a path that went to the $e$-th successor in the initial rewrite step, as described above.
    In the following, we distinguish between two different cases for a rewrite step at a node $u$:
    \begin{enumerate}
        \item[(A)] We use a step with $\itored{}{}{\PP}$ in $\F{T}_{y_j}$ at a position orthogonal to $\pi.\tau$.
        \item[(B)] We use a step with $\itored{}{}{\PP}$ in $\F{T}_{y_j}$ at a position below $\pi.\tau$.
        Note that this is the more interesting case, where we need to use the properties L and NE.
    \end{enumerate}
    Note that we cannot rewrite above $\pi.\tau$ before rewriting at position $\pi.\tau$ due to the innermost restriction.
    
    \noindent
    \textbf{(A) If we have} $t_u \itored{}{}{\PP} \{\tfrac{p_{g_1}}{p_u}:t_{g_1}, \ldots,
    \tfrac{p_{g_{\overline{h}}}}{p_u}:t_{g_{\overline{h}}}\}$, then there is a rule
    $\bar{\ell} \to \{ \bar{p}_1:\bar{r}_1, \ldots, \bar{p}_{{\overline{h}}}:\bar{r}_{{\overline{h}}}\}^{\overline{m}} \in \PP$, 
    a substitution $\delta$, and a position $\zeta \in \IN^+$ with
    $\flat(t_u|_{\zeta}) = \bar{\ell} \delta \in \ANF_{\PP}$. 
    Furthermore, let $\zeta \bot \pi.\tau$. 
    Then, we have $t'_{e.u}|_{\zeta} \doteq_{(T1)} t_u[\hat{r}_e \gamma]_{\pi.\tau}|_{\zeta} = t_{u}|_{\zeta}$ for the substitution $\gamma$ such that $t_u|_{\pi.\tau} = \hat{\ell} \gamma$, and we can rewrite $t'_{e.u}$ using the same rule, same substitution, and same position.
    Then (T3) is again satisfied.
    Furthermore, we have $t'_{e.{g_j}} = t'_{e.u}[\bar{r}_e \delta]_{\zeta}$ if $\zeta \in \posT(t'_{e.{u}})$ and $m = \ttrue$, $t'_{e.{g_j}} = \disannoPos{\zeta}(t'_{e.u}[\bar{r}_e \delta]_{\zeta})$ if $\zeta \in \posT(t'_{e.{u}})$ and $m = \tfalse$, $t'_{e.{g_j}} = t'_{e.u}[\flat(\bar{r}_e) \delta]_{\zeta}$ if $\zeta \notin \posT(t'_{e.{u}})$ and $m = \ttrue$, or $t'_{e.{g_j}} = \disannoPos{\zeta}(t'_{e.u}[\flat(\bar{r}_e) \delta]_{\zeta})$ if $\zeta \notin \posT(t'_{e.{u}})$ and $m = \tfalse$.
    In all four cases we get (T1) (using the same substitution $\gamma$) and (T2) as well.
    
    \noindent
    \textbf{(B) If we have} $t_u \itored{}{}{\PP} \{\tfrac{p_{g_1}}{p_u}:t_{g_1}, \ldots,
    \tfrac{p_{g_{\overline{h}}}}{p_u}:t_{g_{\overline{h}}}\}$, then there is a rule
    $\bar{\ell} \to \{ \bar{p}_1:\bar{r}_1, \ldots, \bar{p}_{{\overline{h}}}:\bar{r}_{{\overline{h}}}\}^{\overline{m}} \in \PP$, 
    a substitution $\delta$, and a position $\zeta \in \IN^+$ with
    $\flat(t_u|_{\zeta}) = \bar{\ell} \delta \in \ANF_{\PP}$. 
    Furthermore, let $\zeta > \pi.\tau$. 
    Since $\hat{\ell} \to \{ \hat{p}_1:\hat{r}_1, \ldots, \hat{p}_h:\hat{r}_h\}$ is L and NE, we know that $\hat{\ell}$ contains exactly the same variables as $\hat{r}_e$ and all of them exactly once.
    Furthermore, since $\urules_{\PP}(r_j|_{\tau})$ is
    non-overlapping, we know that the rewriting must be completely inside the
    substitution $\gamma$ for the substitution $\gamma$ such that $t_u|_{\pi.\tau} = \hat{\ell} \gamma$, 
    i.e., there is a position $\alpha_c$ of a variable $c$ in $\hat{\ell}$ 
    and another position $\beta_c$  \pagebreak with $\pi.\tau.\alpha_c.\beta_c = \zeta$.
  Let $\varphi_e(c)$ be the (unique) variable position of $c$ in $\hat{r}_e$.
    Then, we have $t'_{e.u}|_{\pi.\tau.\varphi_e(c).\beta_c} \doteq_{(T1)} t_u[\hat{r}_e \gamma]_{\pi.\tau}|_{\pi.\tau.\varphi_e(c).\beta_c} = \hat{r}_e \gamma|_{\varphi_e(c).\beta_c} = \gamma(c)|_{\beta_c} = \hat{\ell} \hat{\sigma}|_{\alpha_c.\beta_c} = t_{u}|_{\pi.\tau.\alpha_c.\beta_c}$, and we can rewrite $t'_{e.u}$ using the same rule, same substitution, and position $\pi.\tau.\varphi_e(c).\beta_c$.
    Again, in all four cases (T1)-(T3) are satisfied.

    \smallskip

    \noindent 
    \textbf{\underline{1.2. $\pi \notin \posT(t_x)$}}

    \noindent 
    If we have $\pi \notin \posT(t_x)$, then we can simply use the non-annotated old ADP instead of the old one.
    Since we would remove all annotations in the right-hand side of the rule anyway, due to $\pi \notin \posT(t_x)$, this leads to the same labels in the resulting $\overline{\PP}$-CT.

    \medskip

    \noindent
    \underline{\textbf{Second Case}}

    \noindent
    Next, we prove the theorem in the case where $\urules_{\PP}(r_j|_{\tau})$ is NO and all rules in $\urules_{\PP}(r_j|_{\tau})$ have the form $\ell' \to \{1:r'\}$ for some terms $\ell'$ and $r'$.
    \smallskip
   
    \noindent
    The proof uses the same idea as in the first case but the construction of $\Phi(\circ)$ 
    is easier since $\urules_{\PP}(r_j|_{\tau})$ is non-probabilistic.
    First assume that $\urules_{\PP}(r_j|_{\tau})$ is not weakly innermost terminating.
    This means that after using the ADP 
    $\ell \ruleArr{}{}{} \{ p_1:r_{1}, \ldots, p_k: r_k\}^{m}$ as we did in the soundness proof for the first case, 
    every subterm at a position above $\pi.\tau$ will never be in $\NF_{\PP}$ 
    in the subtree starting at the $j$-th successor $y_j$.
    Furthermore, since all rules in $\urules_{\PP}(r_j|_{\tau})$ have the form $\ell' \to \{1:r'\}$, 
    we can simply remove all nodes that perform a rewrite step below $\pi.\tau$.
    To be precise, if there is a node $v$ that performs a rewrite step below position $\pi.\tau$,
    then we have $t_v \itored{}{}{\PP} \{1 : t_w\}$ for the only successor $w$ of $v$.
    Here, we have $t_w \doteq t_v[r' \sigma]_{\zeta}$ for the used substitution $\sigma$ and position $\zeta$ below $\pi.\tau$.
    The construction $\Psi(\F{T}_{y_j})$ contracts all edges $(x,y)$ where we use a rewrite step at a position below $\pi.\tau$.
    This only removes $R$-nodes, as there is no annotation below or at position $\pi.\tau$.
    Furthermore, we adjust the labeling such that the subterm at position $\pi.\tau$ remains the same for the whole CT.
    Finally, we exchange the rewrite step at the root $x$ from using the old ADP to using the new ADP.
    Since all subterms at a position above $\pi.\tau$ 
    will never be reduced to normal forms in the original CT,
    it does not matter which subterms really occur.
    It is easy to see that $\Phi(\F{T}_x)$ is a valid $\PP$-CT and that (a) and (b) hold w.r.t.\ $\F{T}_x$ and $\Phi(\F{T}_x)$, and this ends the proof if $\urules_{\PP}(r_j|_{\tau})$ is not weakly innermost terminating.
    
    If $\urules_{\PP}(r_j|_{\tau})$ is weakly innermost
    terminating, then it follows directly that it is also confluent and terminating
    \cite[Thm.\ 3.2.11]{GramlichDiss}.
    We can use the construction $\Psi(\circ)$ from the first case to iteratively push the next innermost rewrite step that is performed below position $\pi.\tau$ to a higher position in the tree, until we reach the node that performs the rewrite step at position $\pi.\tau$.
    Note that we do not need the conditions L or NE for the used rule here, because only
    Case (A) of the construction can happen.
    There is no rewrite step below possible (Case (B)), since we move an innermost rewrite step further up and there is no rewrite step above possible, since before doing this construction it was a valid innermost $\PP$-CT.
    Since $\urules_{\PP}(r_j|_{\tau})$ is terminating, this construction ends 
    after a finite number of $\Psi(\circ)$-applications in a tree $\F{T}_{y_j}^{(\infty)}$.
    Now $\F{T}_{y_j}^{(\infty)}$ first rewrites below or at position $\pi.\tau$ until it is a normal form.
    Since all rules in $\urules_{\PP}(r_j|_{\tau})$ have the form $\ell' \to \{1:r'\}$,
    these rewrite steps are a single path in $\F{T}_{y_j}^{(\infty)}$, and it does not
    matter how long the path is. 
    Furthermore, it does not \pagebreak matter which rewrite strategy we use, since $\urules_{\PP}(r_j|_{\tau})$ is confluent.
    We will always reach the same normal form at the end of this path.
    Hence, we can replace the steps with the old ADP and the
    corresponding innermost rewrite steps in the CT by a step with the new
    ADP (where the rewriting does not necessarily correspond to the innermost strategy), i.e.,
    we move this non-innermost step directly to the point where the ADP is applied.
    We will reach the same normal form and can copy the rest of the tree again.

    \medskip

    \noindent
    \underline{\textbf{Third Case}}

    \noindent
    Finally, we prove the theorem in the case where
    $\urules_{\PP}(r_j|_{\tau})$ is NO, $r_j|_{\tau}$ is a ground term, 
    and we have $D_j \itored{}{}{\PP, \pi.\tau} \{q_1:E_{1}, \ldots, q_h:E_{h}\}$, 
    i.e., it is an innermost step.
    \smallskip

    \noindent   
    Once again, we use the same idea as in the proof for the second case but the construction of $\Phi(\circ)$ is, again, easier.
    Note that if $\urules_{\PP}(r_j|_{\tau})$ is non-overlapping, $r_j$ contains no variable below position $\tau$, and we perform an innermost rewrite step, then this is always an innermost rewrite step in every possible CT and this is the only possible rewrite step at this position.
    Hence, we can move this innermost step directly after the use of the ADP using the construction $\Psi(\circ)$. 
    Again, here only Case (A) can happen.
    The reason is that we have to perform this rewrite step eventually 
    if we want to rewrite above position $\tau$, and all other rewrite steps 
    that we can perform in such a situation would be at orthogonal positions.
    So we get the same normal forms in the leaves with the same probability.
\end{myproof}

Next, we show why the rewriting processor needs the new requirement L
that was not imposed in the non-probabilistic setting.
More precisely, we give counterexamples for soundness if the used rule is not left-linear, i.e.,
a variable occurs more than once in the left-hand side, and if the used rule is
 not right-linear, i.e., a variable occurs more than once in a term on the right-hand side.
The other new requirement, namely NE, is currently used in the soundness proof, but we were 
unable to find a counterexample to soundness if the used rule is not NE.
In fact, we conjecture that one can omit this requirement, but then one needs a much more
complicated construction and estimation of the resulting termination probability in the soundness proof.
The reason is that with only L we can guarantee that performing this rewrite step
at a (possibly) non-innermost redex can only increase the probability of innermost termination for
the rewritten subterm but not decrease it.
Increasing the probability of termination for a proper subterm without any annotations means that we have a higher
probability to apply a rewrite step at the position of an annotated symbol.
Remember that we have to rewrite redexes with annotated root symbol on each path of the CT infinitely often (the nodes that are labeled $P$), 
hence a higher probability of termination of the proper subterm leads to a lower probability of
the leaves in the actual CT as the probability to rewrite redexes with annotated root is higher.
However, proving this requires a much more involved approximation of the probability for
termination than our current proof, where we additionally require NE.

\begin{example}[Left-Linearity for Soundness]\label{example:rew-proc-ll-for-soundness}
    To see why left-linearity is required\linebreak for soundness in the probabilistic
    setting, consider the ADP problem\footnote{Here, we have already applied the usable rules processor to
    turn the flag of the $\tg$-ADP to $\tfalse$, and we have moved the $\tf$-ADP with $\tF$ on its
    right-hand side to another ADP problem via the dependency graph processor.}  with\pagebreak 
    
    {\small
    \vspace*{-0.7cm}
    \begin{minipage}[t]{5cm}
        \vspace*{0.3cm}
        \begin{align*}
            \tg(\tf(x,y)) &\to \{1:\td(\tG(\tf(\ta,\ta)),\tG(\tf(\ta,\ta)),\tG(\tf(\ta,\ta))) \}^{\tfalse}\!
        \end{align*}
    \end{minipage}
    \begin{minipage}[t]{4.5cm}
        \begin{align*}
            \tf(x,x) &\to \{1:\te(\tf(\ta,\ta))\}^{\ttrue}\\
            \ta &\to \{\nicefrac{1}{2}: \tb_1, \nicefrac{1}{2}: \tb_2\}^{\ttrue}\!
        \end{align*}
    \end{minipage}}
    
    \vspace*{.2cm}
    \noindent
    This example could also be made non-erasing by adding $x$ as an additional argument to $\te$ and by instantiating $x,y$ by all possible values from $\{\tb_1, \tb_2 \}$ in the $\tg$-rule.
    This ADP problem is not iAST, as it allows for the following CT whose leaves
    have a probability $<1$.
    \begin{center}
        \scriptsize
        \begin{tikzpicture}
            \tikzstyle{adam}=[thick,draw=black!100,fill=white!100,minimum size=4mm,shape=rectangle split, rectangle split parts=2,rectangle split horizontal]
            \tikzstyle{adam2}=[thick,draw=red!100,fill=white!100,minimum size=4mm,shape=rectangle split, rectangle split parts=2,rectangle split horizontal]
            \tikzstyle{empty}=[rectangle,thick,minimum size=4mm]
            
            \node[adam] at (0, -0.7)  (b) {$1$ \nodepart{two} $\tG(\tf(\ta,\ta))$};
            \node[adam] at (-3, -1.4)  (c) {$\nicefrac{1}{2}$ \nodepart{two} $\tG(\tf(\tb_1,\ta))$};
            \node[adam] at (3, -1.4)  (d) {$\nicefrac{1}{2}$ \nodepart{two} $\tG(\tf(\tb_2,\ta))$};
            \node[adam] at (-4.5, -2.1)  (e1) {$\nicefrac{1}{4}$ \nodepart{two} $\tG(\tf(\tb_1,\tb_1))$};
            \node[adam] at (-1.5, -2.1)  (e2) {$\nicefrac{1}{4}$ \nodepart{two} $\tG(\tf(\tb_1,\tb_2))$};
            \node[adam] at (1.5, -2.1)  (e3) {$\nicefrac{1}{4}$ \nodepart{two} $\tG(\tf(\tb_2,\tb_1))$};
            \node[adam] at (4.5, -2.1)  (e4) {$\nicefrac{1}{4}$ \nodepart{two} $\tG(\tf(\tb_2,\tb_2))$};
            \node[adam2] at (-4.5, -2.8)  (e11) {$\nicefrac{1}{4}$ \nodepart{two} $\tG(\te(\tf(\ta,\ta)))$};
            \node[adam] at (-1.5, -2.8)  (e22) {$\nicefrac{1}{4}$ \nodepart{two} $\tG^3(\tf(\ta,\ta))$};
            \node[adam] at (1.5, -2.8)  (e33) {$\nicefrac{1}{4}$ \nodepart{two} $\tG^3(\tf(\ta,\ta))$};
            \node[adam2] at (4.5, -2.8)  (e44) {$\nicefrac{1}{4}$ \nodepart{two} $\tG(\te(\tf(\ta,\ta)))$};
            \node[empty] at (-1.5, -3.5)  (e222) {$\ldots$};
            \node[empty] at (1.5, -3.5)  (e333) {$\ldots$};
            
            \draw (b) edge[->] (c);
            \draw (b) edge[->] (d);
            \draw (c) edge[->] (e1);
            \draw (c) edge[->] (e2);
            \draw (d) edge[->] (e3);
            \draw (d) edge[->] (e4);
            \draw (e1) edge[->] (e11);
            \draw (e2) edge[->] (e22);
            \draw (e3) edge[->] (e33);
            \draw (e4) edge[->] (e44);
            \draw (e22) edge[->] (e222);
            \draw (e33) edge[->] (e333);
        \end{tikzpicture}
    \end{center}
    Here,$\tG^3(\tf(\ta,\ta))$ is an abbreviation for
    the term $\td(\tG(\tf(\ta,\ta)),\tG(\tf(\ta,\ta)),\tG(\tf(\ta,\ta)))$. 
    The paths starting in $\tG(\te(\tf(\ta,\ta)))$ 
    can never use a rewrite step with the $\tg$-ADP anymore and therefore, they converge with probability $1$ in our CT.
    So we can rewrite a single $\tG$-term to a leaf with a probability of $\nicefrac{1}{2}$
    and to three copies of itself with a probability of $\nicefrac{1}{2}$. Hence, this CT
    corresponds to a
    random walk that terminates with probability $<1$. 
    But without the restriction to left-linearity, we could apply the rewriting
    processor and replace the $\tg$-ADP by
    {\small$\tg(\tf(x,y)) \to \{1:\td(\tG(\te(\tf(\ta,\ta))),\tG(\te(\tf(\ta,\ta))),\tG(\te(\tf(\ta,\ta))))\}^{\tfalse}$}.
    Now in every path of the CT this ADP can be used at most once
    and hence, the resulting ADP problem
    is iAST, which shows unsoundness of the rewriting processor without left-linearity.
\end{example}

\begin{example}[Right-Linearity for Soundness]\label{example:rew-proc-rl-for-soundness}
    For right-linearity consider the ADP problem\footnote{
        Again, this ADP problem results from an actual PTRS, where we already applied
         the usable rules processor 
        and the usable terms processor before.} $\PP$ with 

    {\small
    \vspace*{-0.7cm}
\hspace*{-.5cm}\begin{minipage}[t]{5cm}
        \vspace*{0.3cm}
        \begin{align*}
            \tf(\te(\tb_1,\tb_1)) &\to \{1:\th(\tF(\td(\tg)),\tF(\td(\tg)), \tF(\td(\tg)),\tF(\td(\tg)))\}^{\tfalse}\!
        \end{align*}
    \end{minipage}
    \begin{minipage}[t]{4.5cm}
        \begin{align*}
            \td(x) &\to \{1: \te(x,x)\}^{\ttrue}\\
            \tg &\to \{\nicefrac{1}{2}:\tb_1, \nicefrac{1}{2}:\tb_2\}^{\ttrue}\!
        \end{align*}
    \end{minipage}}
    
    \vspace*{.2cm}
    \noindent
    Note that we have the following innermost $\PP$-rewrite sequence tree for $\td(\tg)$:
    \begin{center}
        \scriptsize
        \begin{tikzpicture}
            \tikzstyle{adam}=[thick,draw=black!100,fill=white!100,minimum size=4mm,shape=rectangle split, rectangle split parts=2,rectangle split horizontal]
            \tikzstyle{adam2}=[thick,draw=red!100,fill=white!100,minimum size=4mm,shape=rectangle split, rectangle split parts=2,rectangle split horizontal]
            \tikzstyle{empty}=[rectangle,thick,minimum size=4mm]
            
            \node[adam] at (0, 0)  (a) {$1$ \nodepart{two} $\td(\tg)$};
            \node[adam] at (-1.5, -0.7)  (c) {$\nicefrac{1}{2}$ \nodepart{two} $\td(\tb_1)$};
            \node[adam] at (1.5, -0.7)  (d) {$\nicefrac{1}{2}$ \nodepart{two} $\td(\tb_2)$};
            \node[adam] at (-1.5, -1.4)  (c1) {$\nicefrac{1}{2}$ \nodepart{two}
              $\te(\tb_1, \tb_1)$};
            \node[adam] at (1.5, -1.4)  (d1) {$\nicefrac{1}{2}$ \nodepart{two} $\te(\tb_2,\tb_2)$};

            \draw (a) edge[->] (c);
            \draw (a) edge[->] (d);
            \draw (c) edge[->] (c1);
            \draw (d) edge[->] (d1);
                 \end{tikzpicture}
    \end{center}
    Hence, the term $\tF(\td(\tg))$ can rewrite to $\tF(\te(\tb_1,\tb_1))$ and then to four
    occurrences of $\tF(\td(\tg))$ again with a chance of $\nicefrac{1}{2}$, or it reaches a
    leaf with a chance of $\nicefrac{1}{2}$ as it can never rewrite an annotated redex again. 
    This is once again a random walk that terminates with probability $<1$, so that our ADP
    problem is not iAST. 
    But without \pagebreak the restriction to right-linearity of the used rule, it would be possible to
    apply the rewriting processor and replace the $\tf$-ADP by 
    $\tf(\te(\tb_1,\tb_1)) \to \{1:\th(\tF(\te(\tg,\tg)),\tF(\te(\tg,\tg)),\tF(\te(\tg,\tg)),\tF(\te(\tg,\tg)))\}^{\tfalse}$.
    The term $\te(\tg,\tg)$ can now be rewritten to the term $\te(\tb_1,\tb_1)$ with a probability of
    $\nicefrac{1}{4}$, whereas one obtains a term of the form
    $\te(\tb_i,\tb_j)$ with $i \neq 1$ or $j \neq 1$ 
    with a probability of $\nicefrac{3}{4}$. 
    Hence, now $\tF(\te(\tg,\tg))$ can rewrite to a term with four
    subterms $\tF(\te(\tg,\tg))$ only with a probability of $\nicefrac{1}{4}$, or it reaches a
    leaf with a chance of $\nicefrac{3}{4}$. 
    This is now a random walk that terminates with probability $1$ and the same happens for
    all possible CTs. 
    Hence, the resulting ADP problem is iAST, which shows unsoundness of the rewriting
    processor without right-linearity.
\end{example}

As mentioned in \Cref{RewritingComplete}, the rewriting processor is
complete in the non-probabilistic setting. The reason is that there one used a different
definition to determine when a DP problem $(\mathcal{P},\R)$ is ``not terminating'' (in
\cite{giesl2006mechanizing}, this was called
``infinite''). There, a problem is not only considered to be non-terminating if there is
an infinite $(\mathcal{P},\R)$-chain, but also if $\R$ is not terminating.
In the future, we will examine whether such a modified definition is also useful in the
probabilistic setting. However, as in 
\cite{giesl2006mechanizing}, this modified definition would mean that there can be ADP problems
that are both ``iAST'' and ``non iAST''.

    \section{Further ADP Transformation Processors}\label{moreTrans}

Now we present three further transformational ADP processors which we adapted to the probabilistic
setting, viz., the \emph{instantiation}, \emph{forward instantiation}, and \emph{rule overlap instantiation}
processors.
The latter is a weaker version of the \emph{narrowing} processor, which is
unsound in the probabilistic setting in general, as we will see in \cref{counterexample narrowing}.

\subsection{Instantiation}\label{Instantiation}

In the non-probabilistic setting \cite{giesl2006mechanizing}, the idea of the instantiation processor is to consider all possible \emph{predecessors} $s \to t$ of a dependency pair $u
\to v$ in a chain and to compute the skeleton $\capterm(t)$ of $t$ that remains unchanged when we reduce $t\sigma$
to $u\sigma$, i.e., when going from one DP in a chain to the next. Then $\capterm(t)$ and $u$ must
unify with some mgu $\delta$, and $t\sigma \itorstar u\sigma$ implies that $\sigma$ is an instance
of $\delta$. Hence, the instantiation processor replaces the DP $u \to v$ by $u\delta \to
v\delta$.

As in \cite{giesl2006mechanizing}, $\capterm_\PP(t)$ results from replacing all those subterms
of $t$ by different fresh variables
whose root is a defined
symbol of $\PP$.
Here, multiple occurrences of the same subterm are
also replaced by pairwise different variables. So if $\tg \in \SignatureD$ and $\tc \in
\SignatureC$, then $\capterm_\PP(\tc(\tg(x),\tg(x))) = \tc(x_1, x_2)$.
There exist several improvements to replace this definition of $\capterm_\PP$ by more
precise approximations of the ``skeleton'', see, e.g., 
\cite{giesl2006mechanizing,hirokawa2005automating,thiemanndiss2007}. Moreover, one could
also improve $\capterm_\PP$ by only regarding the root symbols
of left-hand sides of ADPs with the flag $m = \true$
as ``defined''.

To adapt the instantiation processor to ADPs, 
we consider all terms in the distributions on the
right-hand sides of all predecessors.
Note that in the ordinary DP framework, \pagebreak one only instantiates the DPs, but the rules are
left unchanged. Since our ADPs represent both DPs and rules, when instantiating an ADP, we
add a copy of the original ADP without any annotations (i.e., this corresponds to the
original non-instantiated rule which can now only be used for $R$-steps).
In the following, $\vr(\PP)$ is a variable-renamed copy of $\PP$ where all variables
are replaced by fresh ones.

\setcounter{soundnessInstantiationCtr}{\value{theorem}}

\begin{restatable}[Instantiation Processor]{theorem}{Inst} \label{theorem:inst-proc}
    Let $\PP$ be an ADP problem with $\PP = \PP' \uplus \{\ell \ruleArr{}{}{} \{ p_1:r_{1}, \ldots, p_k: r_k\}^{m}\}$. 
    Then $\Proc_{\mathtt{i}}$ is sound and complete, where
    $\Proc_{\mathtt{i}}(\PP)\!=\!\{\PP' \cup N \cup \{\ell \ruleArr{}{}{} \{ p_1:\flat(r_{1}), \ldots, p_k: \flat(r_{k})\}^{m}\}\}$ with

    \vspace*{-.4cm}
    
    \[
        \begin{array}{rl}
            N = & \{ \ell \delta \ruleArr{}{}{} \{ p_1:r_{1} \delta, \ldots, p_k: r_k \delta\}^{m} \\ 
                &\hspace*{1cm}\Bigg| 
                \begin{array}{c}
                    \ell' \ruleArr{}{}{} \{ p_1':r_{1}', \ldots, p_h': r_h'\}^{m'} \in \vr(\PP), 1 \leq j \leq h,\\
                    t \trianglelefteq_{\#} r_{j}', \delta =
                    mgu(\capterm_\PP(t^\#), \ell^\#),
                    \{\ell' \delta, \ell \delta\} \subseteq \ANF_{\PP} \! 
                \end{array}\Biggr\} \!
        \end{array}
        \]
\end{restatable}

\begin{example}\label{example:inst}
    Consider the PTRS $\R$ with the rules
    $\tf(x,y,z) \to \{1 : \tg(x,y,z)\}$ and 
    $\tg(\ta,\tb,u) \to \{\nicefrac{1}{2} : \tf(u,u,u), \nicefrac{1}{2} : \tg(\ta,\tb,u)\}$.
    Its ADPs are $\tf(x,y,z) \to \{1:\tG(x,y,z)\}^{\ttrue}$ and
    $\tg(\ta,\tb,u) \to \{\nicefrac{1}{2}:\tF(u,u,u), \nicefrac{1}{2}:\tG(\ta,\tb,u)\}^{\ttrue}$.
    Using only the processors of
\cref{The ADP Framework}, we cannot prove that $\R$ is iAST.
However, we can apply the instantiation processor on the $\tf$-ADP.

There is a term $t = \tF(u,u,u)$ in the right-hand side of the $\tg$-ADP. As it does not
contain defined symbols, we have $\capterm_\R(\annoEps(t)) = \capterm_\R(t) = t$. For the
left-hand side $\ell = \tf(x,y,z)$, $\delta = mgu(t, \annoEps(\ell)) = mgu(t, \tF(x,y,z))$
instantiates $x$, $y$, and $z$ by $u$. So the instantiation processor
replace the $\tf$-ADP by
$\tf(u,u,u) \to \{1:\tG(u,u,u)\}^{\ttrue}$ (and moreover, we add $\tf(x,y,z) \to \{1:\tg(x,y,z)\}^{\ttrue}$). 
We can now remove the annotation in the transformed $\tf$-ADP by the 
dependency graph processor, and afterwards remove the annotation in the $\tg$-ADP by applying the reduction pair processor with 
the polynomial interpretation that maps every function symbol to the constant $0$
except for $\tG$ that is mapped to $1$.
As we removed all annotations, $\DPair{\R}$, and hence $\R$ must be iAST.
\end{example}

Before we can prove soundness and completeness of the new instantiation processor,
we start with a corollary that expresses the essential property of the $\capterm_\PP$-function.

\begin{corollary}[Property of $\capterm_\PP$] \label{lemma:cap-props}
    Let $t,u \in \TSet{\SignatureADC}{\VSet}$.
    If $t \sigma \ito_{\normalfont{\nonprob}(\PP)}^* u$ for some substitution $\sigma$, then $u = \capterm_\PP(t)
    \delta$ for some substitution $\delta$ which only differs from $\sigma$ on the fresh
    variables that are introduced by $\capterm_\PP$. 
\end{corollary}

\begin{myproof}
    Let $t \sigma \ito_{\nonprob(\PP), \rho_1} u_1 \ito_{\nonprob(\PP), \rho_2} \ldots \ito_{\nonprob(\PP), \rho_n} u_n = u$.
    Let $\{\pi_1, \ldots , \pi_m\}$ be the set of positions where $\capterm_\PP$ replaces the subterms of $t$ by corresponding fresh variables $x_1,\ldots,x_m$. 
    By the definition of $\capterm_\PP$ for each $\rho_i$ there is a higher position $\pi_j \leq \rho_i$. 
    Hence, $u$ can at most differ from $t \sigma$ on positions below a $\pi_j$. 
    We define $\delta$ to be like $\sigma$ but on the fresh
    variables $x_1,\ldots, x_m$ we define $\delta(x_j) = u|_{\pi_j}$. 
    Then by construction $\capterm_\PP(t) \delta = u$ and $\delta$ and $\sigma$ differ only on the fresh variables.
\end{myproof}

We now prove \Cref{theorem:inst-proc}, i.e., we prove
soundness and completeness of the instantiation processor.

\medskip

\begin{myproof}
We will use the following two observations.
  As in the proof of \Cref{theorem:ptrs-rewriting-proc}, let
  $\overline{\PP'} = \PP' \cup N \cup \{\ell \ruleArr{}{}{} \{ p_1:\flat(r_{1}), \ldots,
  p_k: \flat(r_{k})\}^{m}\}$ \pagebreak and $\overline{\PP} = \overline{\PP'} \cup \{\ell \ruleArr{}{}{} \{ p_1:r_{1}, \ldots, p_k:r_{k}\}^{m}\}$.
    First, note that $\ANF_{\PP} = \ANF_{\overline{\PP}} = \ANF_{\overline{\PP'}}$, since the left-hand sides in $\overline{\PP}$ and $\overline{\PP'}$ are either already from rules in $\PP$ or instantiated left-hand sides from $\PP$.
    So it suffices to consider only $\ANF_{\PP}$.
    Second, assume that there exists a $\overline{\PP}$-CT $\F{T}$ that converges with probability
    $< 1$ whose root is labeled with $(1 : t)$ and $\flat(t) = s \theta \in \ANF_{\PP}$ for a
    substitution $\theta$ and an ADP $s \to \ldots \in \overline{\PP}$, and $\posT(t) = \{\varepsilon\}$. 
    If in addition, the ADP
    $\ell \ruleArr{}{}{} \{ p_1:r_{1}, \ldots, p_k: r_k\}^{m}$
    may only be used at the root, then we know that not only $\overline{\PP}$ is not iAST but also that 
    $\overline{\PP'}$ is not iAST.
    The reason is that if only the root $x$ uses the ADP 
    $\ell \ruleArr{}{}{} \{ p_1:r_{1}, \ldots, p_k: r_k\}^{m}$
    then all the subtrees starting at one of its direct successors $xE = \{y_1, \ldots, y_k\}$ 
    are $\overline{\PP'}$-CTs.
    Furthermore, since $\F{T}$ converges with probability $<1$, 
    there must be at least one subtree $\F{T}[y_iE^*]$ that starts at the node $y_i$ 
    for some $1 \leq i \leq k$ and also converges with probability $<1$.
   
    \noindent
    \underline{\emph{Soundness:}} Let $\PP$ be not iAST.
    Then by \Cref{lemma:starting} there exists a $\PP$-CT $\F{T} = (V,E,L,P)$ that converges with probability
    $< 1$ whose root is labeled with $(1: t)$ and $\flat(t) = s \theta \in \ANF_{\PP}$ for a
    substitution $\theta$ and an ADP $s \to \ldots \in \PP$, and $\posT(t) = \{\varepsilon\}$. 
    We will now create a $\overline{\PP}$-CT $\F{T}' = (V,E,L',P)$, with the same underlying tree structure, and an adjusted labeling such that $p_x^{\F{T}} = p_x^{\F{T}'}$ for all $x \in V$.
    Furthermore, we will at most use the ADP $\ell \ruleArr{}{}{} \{ p_1:r_{1}, \ldots, p_k: r_k\}^{m}$ at the root.
    Since the tree structure and the probabilities are the same, we then get $|\F{T}'|_{\ctleaf} = |\F{T}|_{\ctleaf} < 1$, and hence, by our previous discussion, $\overline{\PP'}$ is not iAST either.
    
    The core idea is that every rewrite step with $\ell \ruleArr{}{}{} \{ p_1:r_{1}, \ldots, p_k: r_k\}^{m}$ at a node $v$ that is not the root can also be done with a rule from $N$, or we can use $\ell \ruleArr{}{}{} \{ p_1:\flat(r_{1}), \ldots, p_k: \flat(r_{k})\}^{m}$ if the annotations do not matter, e.g., we rewrite at a position that is not annotated.
    We construct the new labeling $L'$ for the $\overline{\PP}$-CT $\F{T}'$ inductively
    such that for all nodes $x \in V \setminus \ctleaf$
    with children nodes $xE = \{y_1,\ldots,y_h\}$
    we have $t'_x \itored{}{}{\overline{\PP}} \{\tfrac{p_{y_1}}{p_x}:t'_{y_1}, \ldots, \tfrac{p_{y_h}}{p_x}:t'_{y_h}\}$ and for all non-root nodes in $P$ we even have $t'_x \itored{}{}{\overline{\PP'}} \{\tfrac{p_{y_1}}{p_x}:t'_{y_1}, \ldots, \tfrac{p_{y_h}}{p_x}:t'_{y_h}\}$.
    Let $X \subseteq V$ be the set of nodes $x$ where we have already defined the labeling $L'(x)$.
    During our construction, we ensure that the following property holds:
    \begin{equation} \label{soundness-inst-induction-hypothesis}
        \parbox{.9\textwidth}{For every node $x \in X$ we have $t_x \doteq t_x'$ and $\posT(t_x) \subseteq \posT(t_x')$.}
    \end{equation}
    This means that the corresponding term $t_x$ for the node $x$ in $\F{T}$ has the same structure as the term $t_x'$ in $\F{T}'$,
    and additionally, every annotation in $t_x$ also exists in $t_x'$.
    The second condition ensures that if we rewrite using Case $(\mathbf{p})$ or $(\mathbf{pr})$ of \Cref{def:ADPs-and-Rewriting} in $\F{T}$, we do the same in $\F{T}'$, i.e., the corresponding node $x$ remains in $P$ in $\F{T}'$.
    We label the root of $\F{T}'$ by $(1:t)$.
    Here, \eqref{soundness-inst-induction-hypothesis} obviously holds.
    As long as there is still an inner node $x \in X$ such that its successors are not contained in $X$, we do the following.
    Let $xE = \{y_1, \ldots, y_h\}$ be the set of its successors.
    We need to define the corresponding terms $t_{y_1}', \ldots, t_{y_m}'$ for the nodes $y_1, \ldots, y_h$.
    Since $x$ is not a leaf and $\F{T}$ is a $\PP$-CT, we have $t_x \itored{}{}{\PP} \{\tfrac{p_{y_1}}{p_x}:t_{y_1}, \ldots, \tfrac{p_{y_h}}{p_x}:t_{y_h}\}$.
    We have the following three different cases:

    \smallskip
    
    \noindent
    \textbf{(A) We have} $t_x \itored{}{}{\PP} \{\tfrac{p_{y_1}}{p_x}:t_{y_1}, \ldots, \tfrac{p_{y_h}}{p_x}:t_{y_h}\}$ with an ADP $\ell' \ruleArr{}{}{} \{ p_1:r_{1}', \ldots, p_h: r_h'\}^{m'}$ that is either different to $\ell \ruleArr{}{}{} \{ p_1:r_{1}, \ldots, p_k: r_k\}^{m}$ or the node $x$ is the root, using the position $\pi$, and a substitution $\sigma$ such that $\flat(t_x|_\pi)=\ell\sigma \in \ANF_\PP$.
    Then, $t_x \doteq_{(IH)} t_x'$, and hence also $\flat(t_x'|_\pi)=\ell\sigma \in \ANF_\PP$.
    Thus, \pagebreak we can rewrite the term $t'_{x}$ using the same ADP, the same position and the same substitution.
    This means that we have $t'_x \itored{}{}{\overline{\PP}} \{\tfrac{p_{y_1}}{p_x}:t'_{y_1}, \ldots, \tfrac{p_{y_h}}{p_x}:t'_{y_h}\}$.
    Let $1 \leq j \leq h$.
    If $\pi \in \posT(t_x)$, then also $\pi \in \posT(t'_{x})$ by \eqref{soundness-inst-induction-hypothesis}.
    Whenever we create annotations in the rewrite step in $\F{T}$ (a step with $(\mathbf{p})$ or
    $(\mathbf{pr})$), then we do the same in $\F{T}'$ (the step is also a $(\mathbf{p})$- or $(\mathbf{pr})$-step,
    respectively), and whenever we remove annotations in the rewrite step in $\F{T}'$ (a
    step with $(\mathbf{r})$ or $(\mathbf{irr})$), then we do the same in $\F{T}$ (the step is also either a
    $(\mathbf{r})$- or $(\mathbf{irr})$-step).
    Therefore, we also get $\posT(t_{y_j}) \subseteq \posT(t'_{y_j})$ for all $1 \leq j \leq k$ and \eqref{soundness-inst-induction-hypothesis} is again satisfied.

\smallskip
    
    \noindent
    \textbf{(B) We have} $t_x \itored{}{}{\PP} \{\tfrac{p_{y_1}}{p_x}:t_{y_1}, \ldots, \tfrac{p_{y_k}}{p_x}:t_{y_k}\}$ using the ADP $\ell \ruleArr{}{}{} \{ p_1:r_{1}, \ldots, p_k: r_k\}^{m}$, the position $\pi$, and a substitution $\sigma$ such that $\flat(t_x|_\pi)=\ell\sigma \in \ANF_\PP$, and $\pi \not\in \posT(t_x)$.
    Since $t_x \doteq_{(IH)} t_x'$, we can rewrite the term $t'_{x}$ using the ADP 
    $\ell \ruleArr{}{}{} \{ p_1:\flat(r_{1}), \ldots, p_k: \flat(r_{k})\}^{m}$, 
    the same position, and the same substitution.
    This means that we have $t'_x \itored{}{}{\overline{\PP}} \{\tfrac{p_{y_1}}{p_x}:t'_{y_1}, \ldots, \tfrac{p_{y_k}}{p_x}:t'_{y_k}\}$ and \eqref{soundness-inst-induction-hypothesis} is again satisfied.
    In order to prove this, one can do a similar analysis as above.
    Note that only cases $(\mathbf{irr})$ and $(\mathbf{r})$ can be applied in $\F{T}$ (since $\pi \not\in \posT(t_x)$), and we would remove the annotations of the terms $r_j$ anyway in those cases.

\smallskip
    
    \noindent
    \textbf{(C) Finally,  we have} $t_x \itored{}{}{\PP} \{\tfrac{p_{y_1}}{p_x}:t_{y_1}, \ldots, \tfrac{p_{y_k}}{p_x}:t_{y_k}\}$ using the ADP $\ell \ruleArr{}{}{} \{ p_1:r_{1}, \ldots, p_k: r_k\}^{m}$, the position $\pi$, and a substitution $\sigma$ such that $\flat(t_x|_\pi)=\ell\sigma \in \ANF_\PP$, $\pi \in \posT(t_x)$, and $x$ is not the root.
    Then $t_{y_j} = \disannoPos{\pi}(t_x[r_j\sigma]_{\pi})$ if $m = \tfalse$ and $t_{y_j} = t_x[r_j\sigma]_{\pi}$, otherwise.

    We now look at the (not necessarily direct) predecessor $v$ of $x$ that is in $P$,
    where an ADP is applied on a position above or equal to $\pi$, and where in the path from $v$ to
    $x$, no ADP is applied on a position on or above $\pi$.
    There is always such a node $v$. The reason is that $x$ is not the first node in $P$
    and by the Starting Lemma (\Cref{lemma:starting}) we can assume that a step at the root of the term takes place at the root of
    the CT, i.e., $t_x$ results from right-hand sides of $\PP$. 
    Furthermore, we only use rules with the flag $m = \ttrue$ as otherwise, 
    the position $\pi$ would not be annotated in $t_x$.
    We show that the path from $v$ to $x$ can also be taken when using one of the new instantiations
    of $\ell \ruleArr{}{}{} \{ p_1:r_{1}, \ldots, p_k: r_k\}^{m}$ instead.

    Let this predecessor $v \in P$
    use the ADP $\ell' \to \{p_1':r_1', \ldots, p_h':r_h'\}^{m}$, the position
    $\pi'$, and the substitution $\sigma'$.
    Furthermore, let $\tau$ be the position such that $\pi'.\tau = \pi$.
    Because we never rewrite at a position above $\pi$ before reaching node $x$, 
    we have $\annoEps(r_j'|_{\tau}) \sigma' = \annoEps(t_v|_{\pi}) \ito_{\nonprob(\PP)}^* \annoEps(t_x|_{\pi}) = \annoEps(\ell) \sigma$.

    \vspace{-0.2cm}
    \begin{center}
        \scriptsize
        \begin{tikzpicture}
            \tikzstyle{adam}=[thick,draw=black!100,fill=white!100,minimum size=4mm, shape=rectangle split, rectangle split parts=2,rectangle split horizontal]
            \tikzstyle{empty}=[rectangle,thick,minimum size=4mm]
            
            \node[adam,pin={[pin distance=0.1cm, pin edge={,-}] 140:\tiny \textcolor{blue}{$P$}}] at (0, 0)  (a) {$p_v$ \nodepart{two} $t_v$};
            \node[empty] at (2.0, 1)  (b) {$\ldots$};
            \node[empty] at (2.0, 0)  (c) {$\ldots$};
            \node[empty] at (4.0, 1)  (d) {$\ldots$};
            \node[adam,pin={[pin distance=0.1cm, pin edge={,-}] 140:\tiny \textcolor{blue}{$P$}}] at (4.0, 0)  (e) {$p_x$ \nodepart{two} $t_x$};
            \node[empty] at (4.0, 1)  (f) {$\ldots$};
            \node[empty] at (6.0, 0)  (g) {$\ldots$};
            \node[empty] at (6.0, 0.5)  (h) {$\ldots$};
            \node[empty] at (6.0, 1.0)  (i) {$\ldots$};
            
            \draw (a) edge[->, in = 180, out = 0] (b);
            \draw (a) edge[->, in = 180, out = 0] (c);
            \draw (c) edge[->, in = 180, out = 0] (d);
            \draw (c) edge[->, in = 180, out = 0] (e);
            \draw (e) edge[->, in = 180, out = 0] (g);
            \draw (e) edge[->, in = 180, out = 0] (h);
            \draw (e) edge[->, in = 180, out = 0] (i);
            \end{tikzpicture}
    \end{center}
    \vspace{-0.2cm}

    From $\annoEps(r_j'|_{\tau}) \sigma' \ito_{\nonprob(\PP)}^* \annoEps(\ell) \sigma$,
 \cref{lemma:cap-props} implies
    $\annoEps(\ell) \sigma = \capterm_\PP(\annoEps(r_j'|_{\tau})) \delta$
    for some substitution $\delta$ that differs from $\sigma'$ at most on the variables that are introduced by $\capterm_\PP$.
    W.l.o.g. we can assume that $\sigma'$ is equal to $\delta$ on all these fresh variables, and since the ADPs are variable-renamed, we can also assume that $\sigma$ is equal to $\delta$ on all the fresh variables and all the variables from $\ell'$.
    Hence, $\annoEps(\ell) \delta = \capterm_\PP(\annoEps(r_j'|_{\tau})) \delta$ shows there is an mgu $\gamma$ of $\annoEps(\ell)$ and $\capterm_\PP(\annoEps(r_j'|_{\tau}))$ with $\sigma = \gamma \zeta$ for some substitution $\zeta$. 
    Moreover, the property $\{\ell' \sigma', \ell \sigma\} \subseteq
    \ANF_{\PP}$ must remain true when replacing $\sigma$ and $\sigma'$ by the more general substitution $\gamma$, i.e., $\{\ell' \gamma, \ell \gamma\} \subseteq \ANF_{\PP}$. 
    Hence, we can apply \pagebreak the new ADP $\ell \gamma \ruleArr{}{}{} \{ p_1:r_{1}\gamma, \ldots, p_k: r_k\gamma\}^{m} \in N$ with the position $\pi$ and the substitution $\zeta$.
    This means that we have $t'_x \itored{}{}{\PP} \{\tfrac{p_{y_1}}{p_x}:t'_{y_1},
    \ldots, \tfrac{p_{y_k}}{p_x}:t'_{y_j}\}$ with $t_{y_j}' = t_x'[r_j\gamma \zeta]_{\pi}$ if $m = \ttrue$, or $t_{y_j}' = \disannoPos{\pi}(t_x'[r_j\gamma \zeta]_{\pi})$, otherwise.
    Since, $\sigma = \gamma \zeta$ we directly get $t_{y_j} \doteq t_{y_j}'$ and $\posT(t_{y_j}) \subseteq \posT(t_{y_j}')$ so that \eqref{soundness-inst-induction-hypothesis} is satisfied again, and this ends the proof.

    \medskip
   
    \noindent
    \underline{\emph{Completeness:}} Let $\overline{\PP'}$ be not iAST.
    Then there exists a $\overline{\PP'}$-CT $\F{T} = (V,E,L,P)$ that converges with probability $< 1$.
    We will now create a $\PP$-CT $\F{T}' = (V,E,L',P)$, with the same underlying tree structure, and an adjusted labeling such that $p_x^{\F{T}} = p_x^{\F{T}'}$ for all $x \in V$.
    Since the tree structure and the probabilities are the same, we then get $|\F{T}'|_{\ctleaf} = |\F{T}|_{\ctleaf} < 1$, and thus $\PP$ is not iAST either.
    
    The core idea of this construction is that every rewrite step with an ADP from $N$ or the ADP $\ell \ruleArr{}{}{} \{ p_1:\flat(r_{1}), \ldots, p_k: \flat(r_k)\}^{m}$ is also possible with the more general ADP $\ell \ruleArr{}{}{} \{ p_1:r_{1}, \ldots, p_k: r_k\}^{m}$ that may also contain more annotations.
    We construct the new labeling $L'$ for the $\PP$-CT $\F{T}'$ inductively such
    that for all inner nodes $x \in V \setminus \ctleaf$ with children nodes $xE =
    \{y_1,\ldots,y_m\}$ we have $t'_x \itored{}{}{\PP} \{\tfrac{p_{y_1}}{p_x}:t'_{y_1}, \ldots, \tfrac{p_{y_m}}{p_x}:t'_{y_h}\}$.
    Let $X \subseteq V$ be the set of nodes $x$ where we have already defined the labeling $L'(x)$.
    During our construction, we ensure that the following property holds (analogous to the soundness construction):
    \begin{equation} \label{completeness-inst-induction-hypothesis}
        \parbox{.9\textwidth}{For every node $x \in X$ we have $t_x \doteq t_x'$ and $\posT(t_x) \subseteq \posT(t_x')$.}
    \end{equation}
    We label the root of $\F{T}'$ exactly as the root of $\F{T}$.
    Here, \eqref{completeness-inst-induction-hypothesis} obviously holds.
    As long as there is still an inner node $x \in X$ such that its successors are not contained in $X$, we do the following.
    Let $xE = \{y_1, \ldots, y_h\}$ be the set of its successors.
    We need to define the corresponding terms $t_{y_1}', \ldots, t_{y_h}'$ for the nodes $y_1, \ldots, y_h$.
    Since $x$ is not a leaf and $\F{T}$ is a $\overline{\PP'}$-CT, we have $t_x \itored{}{}{\overline{\PP'}} \{\tfrac{p_{y_1}}{p_x}:t_{y_1}, \ldots, \tfrac{p_{y_h}}{p_x}:t_{y_h}\}$.
    We have the following three different cases:
    \begin{enumerate}
        \item[(A)] If it is a step with $\itored{}{}{\PP}$ using an ADP from $\PP'$ in $\F{T}$, 
        then we perform a rewrite step with the same ADP, the same position, and the same substitution in $\F{T}'$.
        This is analogous to Case (A) of the soundness proof.
        \item[(B)] If it is a step with $\itored{}{}{\PP}$ using the ADP $\ell \ruleArr{}{}{} \{ p_1:\flat(r_{1}), \ldots, p_k: \flat(r_k)\}^{m}$ in $\F{T}$, 
        then we use the ADP $\ell \ruleArr{}{}{} \{ p_1:r_{1}, \ldots, p_k: r_k\}^{m}$ that contain more annotations in $\F{T}'$.
        Since we use the same rule but with more annotations, we end up with $t_{y_j} \doteq t_{y_j}'$ and $\posT(t_{y_j}) \subseteq \posT(t_{y_j}')$ again.
        \item[(C)] If it is a step with $\itored{}{}{\PP}$ using an ADP from $N$ in $\F{T}$, 
        then we use the more general ADP $\ell \ruleArr{}{}{} \{ p_1:r_{1}, \ldots, p_k: r_k\}^{m}$ in $\F{T}'$.
        For this rewrite step, we use the substitution $\delta$ such that $\gamma \delta = \sigma$.
    \end{enumerate}
\end{myproof}

\subsection{Forward Instantiation}\label{ForwardInstantiation}

Next we adapt the \emph{forward instantiation} processor and prove its soundness and completeness.
In the non-probabilistic setting, the
idea of the forward instantiation processor \cite{giesl2006mechanizing}
is to consider all possible \emph{successors} $u \to v$ of a DP $s \to t$ in a chain,
again, in order to find the skeleton that remains unchanged when rewriting $t\sigma$ to $u\sigma$.
To find this skeleton, we reverse the rules of the TRS and then \pagebreak proceed as for the
instantiation processor.
Moreover, we can restrict ourselves to the (reversed)
usable rules of $t$. Note that 
these reversed rules might violate the variable conditions of TRSs, i.e., 
the right-hand side
 of a rule may contain variables that do not occur in the left-hand side or the
 left-hand side may be a variable.

\begin{restatable}[Forward Instantiation Processor]{theorem}{FInst} \label{theorem:forward-inst-proc}
    Let $\PP$ be an ADP problem with $\PP = \PP' \uplus \{\ell \ruleArr{}{}{} \{ p_1:r_{1}, \ldots, p_k: r_k\}^{m}\}$. 
    Then $\Proc_{\mathtt{f}}$ is sound and complete,
    where 
    $\Proc_{\mathtt{f}}\!=\!\{\PP' \cup N \cup \{\ell \ruleArr{}{}{} \{ p_1:\flat(r_{1}), \ldots, p_k: \flat(r_{k})\}^{m}\}\}$.
    Here,
    {\small
    \[
        \begin{array}{rl}
            N = & \Biggl\{ \ell \delta \ruleArr{}{}{} \{ p_1:r_{1} \delta, \ldots, p_k: r_k \delta\}^{m} \\ 
                &\hspace*{0.5cm}\Bigg| 
                \begin{array}{c}
                    \ell' \ruleArr{}{}{} \{ p_1':r_{1}', \ldots, p_h': r_h'\}^{m'} \in \vr(\PP), 1 \leq j \leq k,t \trianglelefteq_{\#} r_{j},\\
                    \QQ = \bigl(\nonprob(\urules_{\PP}(t^\#))\bigr)^{-1},\\
                    \delta = mgu(t^\#, \capterm_{\QQ}(\ell'^\#)), 
                    \{\ell \delta, \ell' \delta \} \subseteq \ANF_{\PP} \! 
                \end{array}\Biggr\} \!
        \end{array}
    \]}
\end{restatable}

\begin{example}\label{example:forward-inst}
    Consider the PTRS $\PP$ with the rules 

    {\small
    \vspace*{-0.7cm}
    \begin{minipage}[t]{5cm}
        \begin{align*}
            \tf(x) &\to \{\nicefrac{1}{2}:\tg(x), \nicefrac{1}{2}:\th(x)\}\\
            \tg(\ta) &\to \{1:\tf(\tq(\ta))\}\\
            \th(\tb) &\to \{1:\tf(\tq(\tb))\}\!
        \end{align*}
    \end{minipage}
    \begin{minipage}[t]{4.5cm}
        \vspace*{0.3cm}
        \begin{align*}
            \tq(\ta) &\to \{1:\ta\}\\
            \tq(\tb) &\to \{1:\tb\}\!
        \end{align*}
    \end{minipage}}
    
    \vspace*{.2cm}
    \noindent
    After the usable rules and the usable terms processor, we obtain the
    ADPs

    {\small
    \vspace*{-0.7cm}
    \begin{minipage}[t]{5cm}
        \begin{align*}
            \tf(x) &\to \{\nicefrac{1}{2}:\tG(x), \nicefrac{1}{2}:\tH(x)\}^{\tfalse}\\
            \tg(\ta) &\to \{1:\tF(\tq(\ta))\}^{\tfalse}\\
            \th(\tb) &\to \{1:\tF(\tq(\tb))\}^{\tfalse}\!
        \end{align*}
    \end{minipage}
    \begin{minipage}[t]{4.5cm}
        \vspace*{0.3cm}
        \begin{align*}
            \tq(\ta) &\to \{1:\ta\}^{\ttrue}\\
            \tq(\tb) &\to \{1:\tb\}^{\ttrue}\!
        \end{align*}
    \end{minipage}}
    
    \vspace*{.2cm}
    \noindent
    In this case, the instantiation processor is useless because both, 
    the mgu of \linebreak $\capterm_{\PP}(\tF(\tq(\ta)))
    = \tF(y)$ and $\tF(x)$, and the mgu of $\capterm_{\PP}(\tF(\tq(\tb))) = \tF(z)$ 
    and $\tF(x)$, do not modify $\tF(x)$. 
    However, the forward instantiation processor can replace the original $\tf$-ADP with 
    $\tf(\ta) \to \{\nicefrac{1}{2}:\tG(\ta), \nicefrac{1}{2}:\tH(\ta)\}^{\tfalse}$ and 
    $\tf(\tb) \to \{\nicefrac{1}{2}:\tG(\tb), \nicefrac{1}{2}:\tH(\tb)\}^{\tfalse}$.
    We can now remove the annotations of the normal forms $\tG(\tb)$ and $\tH(\ta)$ 
    with the usable terms processor and apply the reduction pair processor with 
    the polynomial interpretation that maps every function symbol to the constant $1$, 
    in order to remove all annotations of both new $\tf$-ADPs.
    Finally, we can remove all remaining annotations and prove iAST using the dependency graph processor.
    Note that without the forward instantiation processor, we would have to find a polynomial interpretation
    that is at least linear.
\end{example}

\cref{example:forward-inst} shows that the processors that instantiate a given ADP are not only needed in some cases for a 
successful innermost almost-sure termination proof (as in \cref{example:inst}), but sometimes they can also ease the search 
for polynomials in the reduction pair processor, the most time-consuming part of the ADP framework.

We now prove \cref{theorem:forward-inst-proc}, i.e.,
soundness and completeness of \Cref{theorem:forward-inst-proc}.
The construction in the proof is very similar to the one for the proof of \Cref{theorem:inst-proc}.

\medskip

\begin{myproof}
    As in the proof of \Cref{theorem:inst-proc}, let $\overline{\PP'} = \PP' \cup N \cup \{\ell \ruleArr{}{}{} \{ p_1:\flat(r_{1}), \ldots, p_k: \flat(r_{k})\}^{m}\}$ and $\overline{\PP} = \overline{\PP'} \cup \{\ell \ruleArr{}{}{} \{ p_1:r_{1}, \ldots, p_k:r_{k}\}^{m}\}$.
    \smallskip
    
    \noindent
    \underline{\emph{Soundness:}} The core proof idea, the construction itself, and even the induction hypothesis are
    completely the same as for the soundness proof of \cref{theorem:inst-proc}, only the
    third case \pagebreak (C) changes.

    \noindent
    \textbf{(C) If we have} $t_x \itored{}{}{\PP} \{\tfrac{p_{y_1}}{p_x}:t_{y_1}, \ldots, \tfrac{p_{y_k}}{p_x}:t_{y_k}\}$ using the ADP $\ell \ruleArr{}{}{} \{ p_1:r_{1}, \ldots, p_k: r_k\}^{m}$, the position $\pi \in \posT(s)$, and a substitution $\sigma$ such that $\flat(t_x|_\pi)=\ell\sigma \in \ANF_\PP$, and $x$ is not the root,
    then $t_{y_j} = \disannoPos{\pi}(t_x[r_j\sigma]_{\pi})$ if $m = \tfalse$ and $t_{y_j} = t_x[r_j\sigma]_{\pi}$, otherwise.
    
    First consider the case, where there is no successor $v$ of $x$ where an ADP is applied at a position $\tau \geq \pi$ with $\tau \in \posT(t_v)$,
    or an ADP is applied on a position above $\pi$ before reaching such a node $v$.
    Then, we can use $\ell \ruleArr{}{}{} \{ p_1:\flat(r_{1}), \ldots, p_k: \flat(r_{k})\}^{m}$ instead, because the annotations will never be used.

    Otherwise,  there exists a successor $v$ of $x$ where an ADP is applied at a position $\tau \geq \pi$ with $\tau \in \posT(t_v)$,
    and no ADP is applied on a position above $\pi$.
    We show that the path to this successor can also be taken when using one of the new instantiations
    of $\ell \ruleArr{}{}{} \{ p_1:r_{1}, \ldots, p_k: r_k\}^{m}$
    instead.\footnote{It suffices to consider only \emph{one} of those successors $v$ of
    $x$ to find an instantiation of the ADP that could be used instead of the old ADP in
    order to reach \emph{all} such successors. The reason is that this instantiation is equal or
    more general than the actual concrete instantiation of the ADP that was used to perform the ADP-step
    in the actual CT. Hence, every other such successor can also be used with the more general instantiation of the ADP.}
    At node $v$ we use an ADP $\ell' \ruleArr{}{}{} \{ p_1':r_{1}', \ldots, p_h': r_h'\}^{m'}$, a position $\tau$ with $\tau \geq \pi$ and $\tau \in \posT(t_v)$, a substitution $\sigma'$, and on the path from $x$ to $v$ there is no ADP used with a position above $\tau$, and no ADP used with a position below $\tau$ that has a flag $m = \tfalse$.
    This means we have $\annoEps(r_j|_{\tau}) \sigma \ito_{\nonprob(\PP)}^* \annoEps(\ell') \sigma'$.
    \begin{center}
        \scriptsize
        \begin{tikzpicture}
            \tikzstyle{adam}=[thick,draw=black!100,fill=white!100,minimum size=4mm, shape=rectangle split, rectangle split parts=2,rectangle split horizontal]
            \tikzstyle{empty}=[rectangle,thick,minimum size=4mm]
            
            \node[adam,pin={[pin distance=0.1cm, pin edge={,-}] 140:\tiny \textcolor{blue}{$P$}}] at (0, 0)  (a) {$p_x$ \nodepart{two} $t_x$};
            \node[empty] at (2.0, 1)  (b) {$\ldots$};
            \node[empty] at (2.0, 0)  (c) {$\ldots$};
            \node[empty] at (4.0, 1)  (d) {$\ldots$};
            \node[adam,pin={[pin distance=0.1cm, pin edge={,-}] 140:\tiny \textcolor{blue}{$P$}}] at (4.0, 0)  (e) {$p_v$ \nodepart{two} $t_v$};
            \node[empty] at (4.0, 1)  (f) {$\ldots$};
            \node[empty] at (6.0, 0)  (g) {$\ldots$};
            \node[empty] at (6.0, 0.5)  (h) {$\ldots$};
            \node[empty] at (6.0, 1.0)  (i) {$\ldots$};
            
            \draw (a) edge[->, in = 180, out = 0] (b);
            \draw (a) edge[->, in = 180, out = 0] (c);
            \draw (c) edge[->, in = 180, out = 0] (d);
            \draw (c) edge[->, in = 180, out = 0] (e);
            \draw (e) edge[->, in = 180, out = 0] (g);
            \draw (e) edge[->, in = 180, out = 0] (h);
            \draw (e) edge[->, in = 180, out = 0] (i);
            \end{tikzpicture}
    \end{center}
    Let $\QQ' = \nonprob(\urules_{\PP}(\annoEps(r_j|_{\tau})))$ and let $\QQ = \QQ'^{-1}$. 
    Note that $\annoEps(r_j|_{\tau}) \sigma \ito_{\QQ'}^* \annoEps(\ell') \sigma'$ and hence, $\annoEps(\ell') \sigma' \ito_{\QQ}^* \annoEps(r_j|_{\tau}) \sigma$. 
    By \cref{lemma:cap-props} one obtains \linebreak $\annoEps(r_j|_{\tau}) \sigma = \capterm_{\QQ}(\annoEps(\ell')) \delta$ for some substitution $\delta$ that differs from $\sigma'$ at most on the variables that are introduced by $\capterm_{\QQ}$. 
    W.l.o.g.\ we can assume that $\sigma'$ is equal to $\delta$ on all these fresh variables, and since the ADPs are variable-renamed, we can also assume that $\sigma$ is equal to $\delta$ on all the fresh variables and all the variables from $\annoEps(\ell')$.
    Hence, $\annoEps(r_j|_{\tau}) \sigma = \capterm_{\QQ}(\annoEps(\ell')) \sigma$ shows there is an mgu $\gamma$ of $\annoEps(r_j|_{\tau})$ and $\capterm_{\QQ}(\annoEps(\ell'))$ with $\sigma = \gamma \zeta$ for some substitution $\zeta$.
    Moreover, the property $\{\ell \sigma, \ell' \sigma'\} \subseteq
    \ANF_{\PP}$ must remain true when replacing $\sigma$ and $\sigma'$ by the more general substitution $\gamma$, i.e., $\{\ell \gamma, \ell' \gamma\} \subseteq \ANF_{\PP}$.
    Hence, we can apply the new ADP $\ell \gamma \ruleArr{}{}{} \{ p_1:r_{1}\gamma, \ldots, p_k: r_k\gamma\}^{m} \in N$ with the position $\pi$ and the substitution $\zeta$.
    This means that we have $t'_x \itored{}{}{\PP} \{\tfrac{p_{y_1}}{p_x}:t'_{y_1},
    \ldots, \tfrac{p_{y_k}}{p_x}:t'_{y_j}\}$ with $t_{y_j}' = \disannoPos{\pi}(t_x'[r_j\gamma \zeta]_{\pi})$ if $m = \ttrue$, or $t_{y_j}' = t_x'[r_j\gamma \zeta]_{\pi}$, otherwise.
    Since, $\sigma = \gamma \zeta$ we directly get $t_{y_j} \doteq t_{y_j}'$ and $\posT(t_{y_j}) \subseteq \posT(t_{y_j}')$ so that \eqref{soundness-inst-induction-hypothesis} is satisfied again, and this ends the proof for this case.
    The rest of the proof is completely analogous to the one of \cref{theorem:inst-proc}.

    \medskip

    \noindent
    \underline{\emph{Completeness: }} Completely the \pagebreak same as for \cref{theorem:inst-proc}.
\end{myproof}

\subsection{Narrowing}\label{Narrowing}

The narrowing processor \cite{arts2000termination,giesl2006mechanizing} can only be used in a weaker version in the probabilistic setting.
Let $\PP = \PP' \uplus \{\ell \ruleArr{}{}{} \{ p_1:r_{1}, \ldots, p_k: r_k\}^{m}\}$ be an ADP problem.
For each $1 \leq j \leq k$ and each $t \trianglelefteq_{\#} r_j$, we define its
\emph{narrowing substitutions} and its \emph{narrowing results} (as in \cite{noschinski2013analyzing}, where narrowing was adapted to dependency tuples for complexity analysis of ordinary term rewriting).
If we have to perform rewrite steps on (an instance of) $t$
in order to reach the next ADP usage at an annotated position, then the idea of the narrowing processor is to perform
the first step of this reduction already on the ADP 
$\ell \ruleArr{}{}{} \{ p_1:r_{1}, \ldots, p_k: r_k\}^{m}$.
So whenever
there is a $t \trianglelefteq_{\#} r_j$ and a position $\tau$
  in $t$ such that
$t|_\tau$ unifies with the left-hand side $\ell'$ of some rule $\ell' \ruleArr{}{}{} \{
p_1:r_{1}', \ldots, p_k: r_k'\}^{m'} \in \vr(\PP)$ using an mgu
$\delta$ such that $\ell \delta, \ell' \delta \in \ANF_{\PP}$,
then $\delta$ is a \emph{narrowing substitution} of $t$.
While the corresponding \emph{narrowing result}
could also be defined for probabilistic rules, to simplify the presentation let us assume
for the moment that the rule just has the form 
$\ell' \to \{1:r'\}$. Then the corresponding \emph{narrowing result} is
$s = t[r']_{\tau} \delta$ if we rewrite at an annotated position of $t$, and $s =
t[\flat(r')]_{\tau} \delta$ otherwise.

If $\delta_1, \ldots ,\delta_d$ are all narrowing substitutions of
$t$ with the corresponding narrowing results
$s_1,\ldots,s_d$,
then one would like to define a narrowing processor that replaces
$\ell \ruleArr{}{}{} \{ p_1:r_{1}, \ldots, p_k: r_k\}^{m}$ by
$\ell \delta_e \to \{p_1:r_1 \delta_e, \ldots, p_j:s_e, \ldots,
p_k:r_k \delta_e\}$, for all $1 \leq e \leq d$.

In addition, there could be another subterm $t' \trianglelefteq_{\#} r_j$ (with $t' \neq
t$)
  which was
  involved in a CT
(i.e., $t'^\# \sigma \ito^*_{\nonprob(\PP)} \ell'\sigma'$ for some substitutions
  $\sigma, \sigma'$), but this CT is no longer possible when instantiating $t'$
to $t' \delta_1, \ldots, t' \delta_d$. 
We say that $t'$ is \emph{captured} by $\delta_1, \ldots, \delta_d$ if for each narrowing
substitution $\rho$ of $t'$, there is a $\delta_e$
with $1 \leq e \leq d$ such that $\delta_e$ is more general than
$\rho$ (i.e., $\rho = \delta_e \rho'$ for some substitution $\rho'$). 
So the narrowing processor has to add another ADP $\ell \ruleArr{}{}{} \{ p_1:\anno_{\capt_1(\delta_1,\ldots,\delta_d)}(r_{1}), \ldots, p_k: \anno_{\capt_k(\delta_1,\ldots,\delta_d)}(r_k)\}^{m}$, where $\capt_i(\delta_1,\ldots,\delta_d)$ contains all positions of subterms $t' \trianglelefteq_{\#} r_i$ which are not captured by the narrowing substitutions $\delta_1, \ldots,
\delta_d$ of $t$. (Therefore, in contrast to instantiation and forward
      instantiation, here we do not have to add another copy of the original rule without
      annotations.)

However, the main idea of the narrowing processor, i.e., performing the first rewrite step directly on the ADPs,
is unsound for probabilistic ADP problems, as shown by the following example.

\begin{example}\label{counterexample narrowing}
    Consider the PTRS $\R$ with the rules

    {\small
    \vspace*{-0.5cm}
    \begin{minipage}[t]{5cm}
        \begin{align*}
            \tf(\tb_1,\td_1) &\to \{1: \tf(\ta,\te)\}\\
            \tf(\tb_2,\td_2) &\to \{1: \tf(\ta,\te)\}\\
            \ta &\to \{\nicefrac{1}{2}:\tb_1, \nicefrac{1}{2}:\tb_2\}\!
        \end{align*}
    \end{minipage}
    \begin{minipage}[t]{4.5cm}
        \vspace*{0.3cm}
        \begin{align*}
            \te &\to \{1:\td_1\}\\
            \te &\to \{1:\td_2\}\!
        \end{align*}
    \end{minipage}}
    
    \vspace*{.2cm}
    \noindent
    This PTRS is not iAST. The usable rules and the usable terms processor 
    transform the initial ADP problem into $\PP$ with

    {\small
    \vspace*{-0.5cm}
    \begin{minipage}[t]{5cm}
        \begin{align*}
            \tf(\tb_1,\td_1) &\to \{1: \tF(\ta,\te)\}^{\tfalse}\\
            \tf(\tb_2,\td_2) &\to \{1: \tF(\ta,\te)\}^{\tfalse}\\
            \ta &\to \{\nicefrac{1}{2}:\tb_1, \nicefrac{1}{2}:\tb_2\}^{\ttrue}\!
        \end{align*}
    \end{minipage}
    \begin{minipage}[t]{4.5cm}
        \vspace*{0.3cm}
        \begin{align*}
            \te &\to \{1:\td_1\}^{\ttrue}\\
            \te &\to \{1:\td_2\}^{\ttrue}\!
        \end{align*}
    \end{minipage}}

    \vspace*{.2cm}
    \noindent
    Indeed, \pagebreak this ADP problem allows for the CT below on the left without any leaves.
    Note that all occurring terms are ground terms, hence all narrowing substitutions
    are just the identity function.

    But if we apply the narrowing processor to the ADPs in order to rewrite 
    $\te$, then we obtain the four new ADPs 

    {\small
    \vspace*{-0.7cm}
    \begin{minipage}[t]{5cm}
        \begin{align*}
            \tf(\tb_1,\td_1)\to \{1:\tF(\ta,\td_1)\}^{\tfalse}\\
            \tf(\tb_1,\td_1)\to \{1:\tF(\ta,\td_2)\}^{\tfalse}\!
        \end{align*}
    \end{minipage}
    \begin{minipage}[t]{5cm}
        \begin{align*}
            \tf(\tb_2,\td_2)\to \{1:\tF(\ta,\td_1)\}^{\tfalse}\\
            \tf(\tb_2,\td_2)\to \{1:\tF(\ta,\td_2)\}^{\tfalse}\!
        \end{align*}
    \end{minipage}}
    
    \vspace*{.2cm}
    \noindent
    This new ADP problem is iAST, as we will reach a normal form with a probability of
    $\nicefrac{1}{2}$ after each application of an ADP.
    For example, if we use the first ADP, then we get the CT below on the right.
    There we reach the normal form $\tF(\tb_2,\td_1)$ with probability $\nicefrac{1}{2}$.
    \begin{center}
        \scriptsize
        \begin{tikzpicture}
            \tikzstyle{adam}=[thick,draw=black!100,fill=white!100,minimum size=4mm,shape=rectangle split, rectangle split parts=2,rectangle split horizontal]
            \tikzstyle{adam2}=[thick,draw=red!100,fill=white!100,minimum size=4mm,shape=rectangle split, rectangle split parts=2,rectangle split horizontal]
            \tikzstyle{empty}=[rectangle,thick,minimum size=4mm]
            
            \node[adam] at (-3, 0)  (a) {$1$ \nodepart{two} $\tF(\tb_1,\td_1)$};
            \node[adam] at (-3, -0.7)  (b) {$1$ \nodepart{two} $\tF(\ta,\te)$};
            \node[adam] at (-4.5, -1.4)  (c) {$\nicefrac{1}{2}$ \nodepart{two} $\tF(\tb_1,\te)$};
            \node[adam] at (-1.5, -1.4)  (d) {$\nicefrac{1}{2}$ \nodepart{two} $\tF(\tb_2,\te)$};
            \node[adam] at (-4.5, -2.1)  (e1) {$\nicefrac{1}{2}$ \nodepart{two} $\tF(\tb_1,\td_1)$};
            \node[adam] at (-1.5, -2.1)  (e2) {$\nicefrac{1}{2}$ \nodepart{two} $\tF(\tb_2,\td_2)$};
            \node[empty] at (-4.5, -2.8)  (e222) {$\ldots$};
            \node[empty] at (-1.5, -2.8)  (e333) {$\ldots$};
            
            \draw (a) edge[->] (b);
            \draw (b) edge[->] (c);
            \draw (b) edge[->] (d);
            \draw (c) edge[->] (e1);
            \draw (d) edge[->] (e2);
            \draw (e1) edge[->] (e222);
            \draw (e2) edge[->] (e333);
            
            \node[adam] at (3, 0)  (a) {$1$ \nodepart{two} $\tF(\tb_1,\td_1)$};
            \node[adam] at (3, -0.7)  (b) {$1$ \nodepart{two} $\tF(\ta,\td_1)$};
            \node[adam] at (1.5, -1.4)  (c) {$\nicefrac{1}{2}$ \nodepart{two} $\tF(\tb_1,\td_1)$};
            \node[adam] at (4.5, -1.4)  (d) {$\nicefrac{1}{2}$ \nodepart{two} $\tF(\tb_2,\td_1)$};
            \node[empty] at (1.5, -2.8)  (e222) {$\ldots$};
            
            \draw (a) edge[->] (b);
            \draw (b) edge[->] (c);
            \draw (b) edge[->] (d);
            \draw (c) edge[->] (e222);
        \end{tikzpicture}
    \end{center}
    The difference is that when narrowing the ADP,
    we have to decide how to rewrite $\te$ before we split the term 
    into different ones with a certain probability, i.e., before we
    rewrite $\ta$ to $\{ \nicefrac{1}{2}:\tb_1, \nicefrac{1}{2}:\tb_2 \}$.
\end{example}

Thus, in the probabilistic setting, we
can only transform the ADP by applying a rewrite rule 
if we can ensure the same conditions as for the rewriting processor. 
So in the probabilistic setting,
the narrowing processor can only instantiate the ADP by the narrowing
substitutions, but it must not perform any rewrite step. 
Instead, the rewrite steps have to be done via the rewriting processor afterwards.
Thus, instead of calling it \emph{narrowing processor}, we call it \emph{rule overlap
instantiation processor} as it only instantiates the ADPs but does not perform any rewrite
steps. 

\begin{restatable}[Rule Overlap Instantiation Processor]{theorem}{RuleOverlapInst} \label{theorem:rule-overlap-inst} 
    Let $\PP$ be an ADP problem with $\PP = \PP' \uplus \{\ell \ruleArr{}{}{} \{ p_1:r_{1}, \ldots, p_k: r_k\}^{m}\}$,
    let $1 \leq j \leq k$, and let $t \trianglelefteq_{\#} r_j$.
    Let $\delta_1,\ldots,\delta_d$ be all narrowing substitutions of
    $t$, where $d \geq 0$.   
    Then $\Proc_{\mathtt{roi}}\!=\!\{\PP' \cup N\}$ is sound and complete,
    where
    \[
        \begin{array}{rcl}
            N & = & \Bigl\{ \ell \delta_e \to \{p_1:r_1 \delta_e, \ldots,
            p_k:r_k \delta_e\}\ \Big| 1 \leq e \leq d \Bigr\}\\ 
            & \cup & \Bigl\{ 
                \begin{array}{rcrl}
                    \ell \ruleArr{}{}{} \{ p_1:\anno_{\capt_1(\delta_1,\ldots,\delta_d)}(r_{1}), \ldots, p_k: \anno_{\capt_k(\delta_1,\ldots,\delta_d)}(r_k)\}^{m}\! 
                \end{array}\ \Bigr\} \!
        \end{array}
    \]
\end{restatable}

\begin{example}
    Consider $\R$ with 

    \vspace*{-0.7cm}
    \begin{minipage}[t]{6cm}
        \vspace*{0.3cm}
        \begin{align*}
            \tf(\td(x)) &\to \{ \nicefrac{3}{4}: \te(\tf(\tg(x)),\tf(\th(x))), \nicefrac{1}{4}: \ta \}\!
        \end{align*}
    \end{minipage}
    \hspace*{.5cm}
    \begin{minipage}[t]{5cm}
        \begin{align*}
            \tg(\ta) &\to \{1:\td(\ta)\}\\
            \th(\tb) &\to \{1:\td(\tb)\}\!
        \end{align*}
    \end{minipage}
    
    \vspace*{.2cm}

    \noindent
    The usable rules and the usable terms processor transform the initial ADP problem 
    into $\PP$ with 

    \vspace*{-0.7cm}
    \begin{minipage}[t]{6cm}
        \vspace*{0.3cm}
        \begin{align*}
            \tf(\td(x)) &\to \{ \nicefrac{3}{4}: \te(\tF(\tg(x)),\tF(\th(x))), \nicefrac{1}{4}: \ta \}^{\tfalse}\!
        \end{align*}
    \end{minipage}
    \hspace*{.5cm}
    \begin{minipage}[t]{5cm}
        \begin{align*}
            \tg(\ta) &\to \{1:\td(\ta)\}^{\ttrue}\\
            \th(\tb) &\to \{1:\td(\tb)\}^{\ttrue}\!
        \end{align*}
    \end{minipage}
    \pagebreak

    \noindent
    The ADP problem $\PP$ (and thus also the original PTRS $\R$)
    is iAST because for every instantiation, at most one of the
    two ``recursive $\tF$-calls'' in the right-hand side of the $\tf$-ADP can be applied.
    The reason is that we can either use the $\tg$-rule if the variable $x$ is
    instantiated with $\ta$, or we can apply the $\th$-rule if the variable is
    instantiated with $\tb$, but not both.
    We apply $\Proc_{\mathtt{roi}}$ using the term $\tF(\tg(x))$,
 whose only narrowing substitution  is $\delta = \{x/\ta\}$.
 The other subterm $\tF(\th(x))$ with annotated root
 is not captured by this substitution, and hence, we have to generate an additional ADP
 where this second subterm is annotated. Thus, we replace the former $\tf$-ADP
 by the following two new ADPs.
   \begin{align*}
        \tf(\td(\ta)) &\to \{ \nicefrac{3}{4}: \te(\tF(\tg(\ta)),\tF(\th(\ta))), \nicefrac{1}{4}: \ta \}^{\tfalse}\\
        \tf(\td(x)) &\to \{ \nicefrac{3}{4}: \te(\tf(\tg(x)),\tF(\th(x))), \nicefrac{1}{4}: \ta \}^{\tfalse}\!
    \end{align*}
    Now one can remove the annotation of $\tF(\th(\ta))$ from the first ADP by the usable terms processor and
    then apply the reduction pair processor with the polynomial interpretation that maps $\tF$
    to 1 and all other symbols to 0 to remove all annotations, which proves
    iAST.
    Again, proving iAST with such a simple polynomial interpretation would not be possible
    without the rule overlap instantiation processor.
\end{example}

We now prove \Cref{theorem:rule-overlap-inst}, i.e., soundness and completeness of the rule overlap instantiation
processor.

\medskip

 \begin{myproof}
    \smallskip
   
    \noindent
    \underline{\emph{Soundness:}} We use the same idea as in the proof of soundness for \cref{theorem:forward-inst-proc} 
    but at a node $v$ that uses the position $\pi \in \posT(t_x)$ we do not look at the next 
    node that rewrites at an annotated position below or equal to $\pi$, but at all such nodes that can be
    either annotated or not.
    
    Let $\PP$ be not iAST.
    Then by \Cref{lemma:starting} there exists a $\PP$-CT $\F{T} = (V,E,L,P)$ that converges with probability
    $< 1$ that starts with $(1:t)$ and $t = s \theta \in \ANF_{\PP}$ for a
    substitution $\theta$ and an ADP $s \to \ldots \in \PP$, and $\posT(t) = \{\varepsilon\}$. 
    Let $\overline{\PP'} = \PP' \cup N$.
    We will now create a $\overline{\PP'}$-CT $\F{T}' = (V,E,L',P)$, with the same underlying tree structure, and an adjusted labeling such that $p_x^{\F{T}} = p_x^{\F{T}'}$ for all $x \in V$.
    Since the tree structure and the probabilities are the same, we then get $|\F{T}'|_{\ctleaf} = |\F{T}|_{\ctleaf} < 1$, and hence $\overline{\PP'}$ is not iAST either.
    
    The core idea of this construction is that every rewrite step with $\ell \ruleArr{}{}{} \{ p_1:r_{1}, \ldots, p_k: r_k\}^{m}$ can also be done with a rule from $N$.
    If we use $\ell \ruleArr{}{}{} \{ p_1:\anno_{\capt_1(\delta_1,\ldots,\delta_d)}(r_{1}), \ldots, p_k: \anno_{\capt_k(\delta_1,\ldots,\delta_d)}(r_k)\}^{m} \in N$,
    we may create fewer annotations than we did when using the old ADP $\ell \ruleArr{}{}{} \{ p_1:r_{1}, \ldots, p_k: r_k\}^{m}$.
    However, we will never rewrite at the position of the annotations that do not get created in the CT $\F{T}$, hence we can ignore them.
    We construct the new labeling $L'$ for the $\overline{\PP'}$-CT $\F{T}'$ inductively such that for all nodes $x \in V \setminus \ctleaf$ with $xE = \{y_1, \ldots, y_m\}$ we have $t'_x \itored{}{}{\overline{\PP'}} \{\tfrac{p_{y_1}}{p_x}:t'_{y_1}, \ldots, \tfrac{p_{y_m}}{p_x}:t'_{y_m}\}$.
    Let $X \subseteq V$ be the set of nodes $x$ where we have already defined the labeling $L'(x)$.
    During our construction, we ensure that the following property holds:
    \begin{equation} \label{soundness-roi-induction-hypothesis}
        \parbox{.9\textwidth}{For every $x \in X$ we have $t_x \doteq t'_x$ and $\posT(t_x) \setminus \Junk(t_x) \subseteq \posT(t'_x)$.}
    \end{equation}
    Here, we define $\Junk(t_x)$ as in the proof of \Cref{theorem:prob-UPP}.

    For the construction, we start with the same term at the root.
    Here, \eqref{soundness-roi-induction-hypothesis} obviously holds.
    As long as there is still an inner node $x \in X$ such that its successors are not contained in $X$, we do the following.
    Let $xE = \{y_1, \ldots, y_m\}$ be the set of its successors.
    We need to define the corresponding sets $t_{y_1}', \ldots, t_{y_m}'$ for the nodes $y_1, \ldots, y_m$.
    Since $x$ is not a leaf and $\F{T}$ is a $\PP$-CT, we have $t_x \itored{}{}{\PP} \{\tfrac{p_{y_1}}{p_x}:t_{y_1}, \ldots, \tfrac{p_{y_m}}{p_x}:t_{y_m}\}$.
    We have the following three cases:
    \begin{enumerate}
        \item[(A)] If it is a step with $\itored{}{}{\PP}$ using an ADP that is different from 
        $\ell \ruleArr{}{}{} \{ p_1:r_{1}, \ldots, p_k: r_k\}^{m}$ in $\F{T}$, 
          then we perform a rewrite step with the same ADP, the same redex, and the same substitution in $\F{T}'$.
        Analogous to Case (A) of the soundness proof for \Cref{theorem:inst-proc}, we can show \eqref{soundness-roi-induction-hypothesis} for the resulting terms.
        \item[(B)] If it is a step with $\itored{}{}{\PP}$ using the ADP 
        $\ell \ruleArr{}{}{} \{ p_1:r_{1}, \ldots, p_k: r_k\}^{m}$ at a position $\pi \notin \posT(t_x)$ in $\F{T}$,
        then we perform a rewrite step with $\ell \ruleArr{}{}{} \{ p_1:\anno_{\capt_1(\delta_1,\ldots,\delta_d)}(r_{1}), \ldots, p_k: \anno_{\capt_k(\delta_1,\ldots,\delta_d)}(r_k)\}^{m}$, same redex, same substitution, and same position in $\F{T}'$.
        Analogous to Case (B) of the soundness proof for \Cref{theorem:inst-proc}, we can show \eqref{soundness-roi-induction-hypothesis} for the resulting terms.
        Note that the rule that we use contains fewer annotations than the original rule, but since $\pi \notin \posT(t_x)$, we remove all annotations from the rule during the application of the rewrite step anyway.
        \item[(C)] If it is a step with $\itored{}{}{\PP}$ using the ADP 
        $\ell \ruleArr{}{}{} \{ p_1:r_{1}, \ldots, p_k: r_k\}^{m}$ at a position $\pi \in \posT(t_x)$ in $\F{T}$, 
        then we look at specific successors to find a substitution $\delta$ such that 
        $\ell \delta \ruleArr{}{}{} \{ p_1:r_{1}\delta, \ldots, p_k: r_k\delta\}^{m} \in N$ or we detect that we can use the ADP 
        $\ell \ruleArr{}{}{} \{ p_1:\anno_{\capt_1(\delta_1,\ldots,\delta_d)}(r_{1}), 
        \ldots, p_k: \anno_{\capt_k(\delta_1,\ldots,\delta_d)}(r_k)\}^{m}$
        and perform a rewrite step with this new ADP in $\F{T}'$.
    \end{enumerate}

    \noindent
    So it remains to consider Case (C) in detail. Here, we have
     $t_x \itored{}{}{\PP} \{\tfrac{p_{y_1}}{p_x}:t_{y_1}, \ldots, \tfrac{p_{y_k}}{p_x}:t_{y_k}\}$ using the ADP $\ell \ruleArr{}{}{} \{ p_1:r_{1}, \ldots, p_k: r_k\}^{m}$, the position $\pi \in \posT(t_x)$, and a substitution $\sigma$ such that $\flat(t_x|_{\pi}) = \ell \sigma \in \ANF_{\PP}$.
    
    We first consider the case where
    there is no successor $v$ of $x$ where an ADP is applied at an annotated position below or at $\pi$,
    or an ADP is applied on a position above $\pi$ before reaching such a node $v$.
    Then, we can use $\ell \ruleArr{}{}{} \{ p_1:\anno_{\capt_1(\delta_1,\ldots,\delta_d)}(r_{1}), \ldots, p_k: \anno_{\capt_k(\delta_1,\ldots,\delta_d)}(r_{k})\}^{m}$ instead, because the annotations will never be used, so they do not matter.

    Otherwise, there exists a successor $v$ of $x$ where an ADP is applied at an annotated position below or at $\pi$,
    and no ADP is applied on a position above $\pi$ before.
    Let $v_1, \ldots, v_n$ be all (not necessarily direct) successors that rewrite below position $\pi$, 
    or rewrite at position $\pi$, and on the path 
    from $x$ to $v$ there is no other node with this property,
    and no node that performs a rewrite step above $\pi$.
    Furthermore, let $t_1, \ldots, t_n$ be the used redexes 
    and $\rho_1, \ldots, \rho_n$ be the used substitutions.
    \begin{itemize}
        \item \textbf{(C1) If} none of the redexes $t_1, \ldots, t_n$ are captured by $t$, then we use the ADP $\ell \ruleArr{}{}{} \{ p_1:\anno_{\capt_1(\delta_1,\ldots,\delta_d)}(r_{1}), 
        \ldots, p_k: \anno_{\capt_k(\delta_1,\ldots,\delta_d)}(r_k)\}^{m}$ with the position
        $\pi \in \posT(t_{x}) \setminus \Junk(t_{x}) \subseteq_{(IH)} \posT(t'_{x})$ and the substitution $\sigma$.
        Once again, \eqref{soundness-roi-induction-hypothesis} is satisfied for our resulting terms.

        \item \textbf{(C2) If} $t = t_i$ for some $1 \leq i \leq n$, then we can find a narrowing substitution $\delta_e$ of $t$ that is more general than $\sigma$, i.e., we have $\delta_e \gamma = \sigma$.
        Now, we use the ADP $\ell \delta_e \ruleArr{}{}{} \{ p_1:r_{1}\delta_e, \ldots, p_k: r_k\delta_e\}^{m}$ with the position 
        $\pi \in \posT(t_{x}) \setminus \Junk(t_{x}) \subseteq_{(IH)} \posT(t'_{x})$ and the substitution $\gamma$ such that $\flat(t_x|_{\pi}) = \ell \delta_e \gamma = \ell \sigma \in \ANF_{\PP}$.
        Once again, \eqref{soundness-roi-induction-hypothesis} is satisfied for our resulting terms.

        \item \textbf{(C3) If} $t \neq t_i$ for all $1 \leq i \leq n$ but there is an $1 \leq i \leq n$ such that $t_i$ is captured, then, since $t_i$ is captured, there exists a narrowing substitution $\delta_e$ of $t$ that is more general than $\rho_i$, i.e., there exists a substitution $\kappa_1$ with $\delta_e \kappa_1 = \rho_i$,
        and since we use $\rho_i$ later on we additionally have that $\rho_i$ is more general than $\sigma$, i.e., there exists a substitution $\kappa_2$ with $\rho_i \kappa_2 = \sigma$.
        Now, we use the ADP $\ell \delta_e \ruleArr{}{}{} \{ p_1:r_{1}\delta_e, \ldots, p_k: r_k\delta_e\}^{m}$ with the position 
        $\pi \in \posT(t_{x}) \setminus \Junk(t_{x}) \subseteq_{(IH)} \posT(t'_{x})$ and the substitution $\kappa_1 \kappa_2$ such that $\flat(t_x|_{\pi}) = \ell \delta_e \kappa_1 \kappa_2 = \ell \sigma \in \ANF_{\PP}$.
        Once again, \eqref{soundness-roi-induction-hypothesis} is satisfied for our resulting terms.
    \end{itemize}

    \medskip
    
    \noindent
    \underline{\emph{Completeness:}} The proof is analogous to the completeness proof of \cref{theorem:inst-proc}.
    We can replace each ADP $\ell \delta_e \ruleArr{}{}{} \{ p_1:r_{1}\delta_e, \ldots, p_k: r_k\delta_e\}^{m}$ with the more general one $\ell \ruleArr{}{}{} \{ p_1:r_{1}, \ldots, p_k: r_k\}^{m}$, and each 
    ADP $\ell \ruleArr{}{}{} \{ p_1:\anno_{\capt_1(\delta_1,\ldots,\delta_d)}(r_{1}), 
    \ldots, p_k: \anno_{\capt_k(\delta_1,\ldots,\delta_d)}(r_k)\}^{m}$
    can be replaced by $\ell \ruleArr{}{}{} \{ p_1:r_{1}, \ldots, p_k: r_k\}^{m}$ as well, leading
    to more annotations than before.
\end{myproof}

    \section{Examples}\label{Examples}

In this section, we present several examples to
illustrate specific strengths and weaknesses of the new transformational
processors.

\vspace*{-.1cm}

\subsection{Probabilistic Quicksort}\label{Probabilistic Quicksort}

\vspace*{-.05cm}

In \Cref{Evaluation} we have already seen a version of the probabilistic
quicksort algorithm where iAST can only be proved
if we use the new transformational processors. 
The following PTRS represents the full implementation of this probabilistic quicksort algorithm.

\vspace*{-.5cm}

{\scriptsize
\begin{align*}
    \trotate(\tnil) & \!\to\! \{ 1 : \tnil \}\\
    \trotate(\tcons(x,\xs)) & \!\to\! \{ \nicefrac{1}{2} : \tcons(x,\xs), \nicefrac{1}{2} : \trotate(\tapp(\xs, \tcons(x,\tnil)))\}\\
    \tisempty(\tnil) & \!\to\! \{ 1 : \ttrue\}\\
    \tisempty(\tcons(x, \xs)) & \!\to\! \{ 1 : \tfalse\}\\
    \tqs(\xs) & \!\to\! \{ 1 : \tif(\tisempty(\xs), \tlow(\thd(\xs), \ttail(\xs)), \thd(\xs), \thigh(\thd(\xs), \ttail(\xs)))\}\\
    \tif(\ttrue, \xs, x, \ys) & \!\to\! \{ 1 : \tnil\}\\
    \tif(\tfalse, \xs, x, \ys) & \!\to\! \{ 1 : \tapp(\tqs(\trotate(\xs)), \tcons(x, \tqs(\trotate(\ys))))\}\\
    \thd(\tcons(x, \xs)) & \!\to\! \{ 1 : x\}\\
    \ttail(\tcons(x, \xs)) & \!\to\! \{ 1 : \xs\}\\
    \tlow(x,\tnil)  & \!\to\! \{ 1 : \tnil\}\\
    \tlow(x,\tcons(y,\ys)) & \!\to\! \{ 1 : \tiflow(\tleq(x,y),x,\tcons(y,\ys))\}\\
    \tiflow(\ttrue,x,\tcons(y,\ys)) & \!\to\! \{ 1 : \tlow(x,\ys)\}\\
    \tiflow(\tfalse,x,\tcons(y,\ys)) & \!\to\! \{ 1 : \tcons(y,\tlow(x,\ys))\}\\
    \thigh(x,\tnil) & \!\to\! \{ 1 : \tnil\}\\
    \thigh(x,\tcons(y,\ys)) & \!\to\! \{ 1 : \tifhigh(\tleq(x,y),x,\tcons(y,\ys))\}\\
    \tifhigh(\ttrue,x,\tcons(y,\ys)) & \!\to\! \{ 1 : \tcons(y,\thigh(x,\ys))\}\\
    \tifhigh(\tfalse,x,\tcons(y,\ys)) & \!\to\! \{ 1 : \thigh(x,\ys)\}\\
    \tleq(\tz,x) & \!\to\! \{ 1 : \ttrue\}\\
    \tleq(\ts(x),\tz) & \!\to\! \{ 1 : \tfalse\}\\
    \tleq(\ts(x),\ts(y)) & \!\to\! \{ 1 : \tleq(x,y)\}\\
    \tapp(\tnil,\ys) & \!\to\! \{ 1 : \ys\}\\
    \tapp(\tcons(x,\xs),\ys) & \!\to\! \{ 1 : \tcons(x,\tapp(\xs,\ys))\}\!
\end{align*}
}

This quicksort algorithm searches for a random pivot element using the first two $\trotate$ rules.
Here, we rotate the list and always move the head element to the end of the list.
With a chance of $\nicefrac{1}{2}$ we stop this iteration and use the current head element as the next pivot element.
The rest of the rules represents the classical quicksort algorithm without any probabilities.
Here, $\tapp$ computes list concatenation, 
$\tlow(x,\xs)$ returns all elements of the list $\xs$ that are smaller than $x$, and
$\thigh$ works analogously.
Furthermore, $\thd$ returns the head element of a list and $\ttail$ returns the rest of the list without the head.
Finally, $\tisempty$ checks whether the list is empty or not.

Using our new ADP framework with the new transformational processors,
\aprove can automatically prove that this PTRS $\R_{\tqs}$ is iAST.
In particular, we need the transformations (i.e., the rewriting processor) to evaluate the 
$\thd$, $\ttail$, and $\tisempty$ functions, and we need the rule overlap instantiation processor
to determine all possible terms that these functions can actually be applied on, e.g., 
we need to detect that if we have the term $\tisempty(x)$ for a variable $x$, then in order to
apply any rewrite step, the variable $x$ needs to be instantiated with either $\tnil$ or $\tcons(y,\ys)$
for some new variables $y$ and $\ys$.

\subsection{Moving Elements in Lists Probabilistically}

Another interesting probabilistic algorithm that deals with lists is the following:
We are given two lists $L_1$ and $L_2$.
If one of the two lists is empty, then the algorithm terminates.
Otherwise, we either move the head of list $L_1$ to $L_2$ or vice versa, both with a chance of $\nicefrac{1}{2}$, and then we repeat this procedure.
This algorithm is represented by the following PTRS.

{\scriptsize
\begin{align*}
    \tor(\tfalse,\tfalse) & \!\to\! \{ 1 : \tfalse\}\\
    \tor(\ttrue,x) & \!\to\! \{ 1 : \ttrue\}\\
    \tor(x,\ttrue) & \!\to\! \{ 1 : \ttrue\}\\
    \tmoveelements(\xs,\ys) & \!\to\!  \{ 1 : \tif(\tor(\tisempty(\xs),\tisempty(\ys)), \xs, \ys)\}\\
    \tif(\ttrue, \xs, \ys) & \!\to\! \{ 1 : \xs\}\\
    \tif(\tfalse, \xs, \ys) & \!\to\!
    \{ \nicefrac{1}{2} : \tmoveelements(\ttail(\xs),\tcons(\thd(\xs),\ys)),\\
    & \phantom{ \!\to\! \{\,} \nicefrac{1}{2} : \tmoveelements(\tcons(\thd(\ys),\xs), \ttail(\ys))\}\\
    \tisempty(\tnil) & \!\to\! \{ 1 : \ttrue\}\\
    \tisempty(\tcons(x, \xs)) & \!\to\! \{ 1 : \tfalse\}\\
    \thd(\tcons(x, \xs)) & \!\to\! \{ 1 : x\}\\
    \ttail(\tcons(x, \xs)) & \!\to\! \{ 1 : \xs\}\\
\end{align*}
}

This algorithm is iAST (and even AST) because we can view this as a classical random walk on the number of elements in the first list that is both bounded from below by $0$ and from above by the sum of the length of both lists.
In order to prove this automatically, we again have to use some kind of instantiation processor, e.g., the rule overlap instantiation processor, to find all possible terms that the functions $\thd$, $\ttail$, and $\tisempty$ can actually be applied on.
In addition, we also need the rewriting processor again to then evaluate these functions.
Once the $\tmoveelements$-ADP contains the symbols $\tcons$ and $\tnil$ in the left-hand side, we can detect the structure of the random walk using an application of the reduction pair processor that removes all annotations of this ADP.
After that, there is no SCC in the dependency graph left, and we have proven iAST.

\subsection{Conditions on Numbers}

Another important task of termination analysis is to be able to handle conditions on numbers.
These occur in nearly every program and often impact the termination behavior.
The same is true for probabilistic programs.
For a successful proof of iAST without transformations, we need that the rules of the PTRS
have these conditions integrated in their left-hand sides of the rules, i.e., if one wants
to check whether a number is zero or not, then one needs two different rules
$\tf(\tz) \to \ldots$ and $\tf(\ts(x)) \to \ldots$ to perform this case analysis.
However, in programs, conditions are mostly implemented by an $\tif$-construct, where one
could use, e.g., an
additional function $\tgt$ to check whether a number is greater than another.
The same is true for conditions on other data structures than numbers, as we have seen above. 
If one wants to check whether the list is empty or not, then without transformations, one needs two rules
(for $\tnil$ and $\tcons$), whereas with transformations, one can use conditions and auxiliary
functions like $\tisempty$.

The following PTRS depicts the classical random walk, but we check the condition $x > 0$
not directly in the left-hand side of the rule but with an additional function $\tgt(x,y)$
which checks whether $x > y$.
Additionally, in order to decrease the value of a number by one, we use the predecessor function $\tp(x) = x-1$.

{\scriptsize
\begin{align*}
    \tgt(\tz,\tz) & \!\to\! \{ 1 : \tfalse \}\\
    \tgt(\ts(x),\tz) & \!\to\! \{ 1 : \ttrue \}\\
    \tgt(\tz,\ts(y)) & \!\to\! \{ 1 : \tfalse \}\\
    \tgt(\ts(x),s(y)) & \!\to\! \{ 1 : \tgt(x,y) \}\\
    \tp(\tz) & \!\to\! \{ 1 : \tz \}\\
    \tp(\ts(x)) & \!\to\! \{ 1 : x \}\\
    \tloop(x) & \!\to\! \{ 1 : \tif(\tgt(x,\tz), x) \}\\
    \tif(\tfalse, x) & \!\to\! \{ 1 : \tstop \}\\
    \tif(\ttrue, x) & \!\to\! \{ \nicefrac{1}{2} : \tloop(\tp(x)), \nicefrac{1}{2} : \tloop(\ts(x)) \}\!
\end{align*}
}

In this case, we need the rewriting processor to evaluate the
functions $\tgt$ and $\tp$
and once again we need the rule overlap instantiation processor
to check for all possible terms that these functions can actually be applied on.

\subsection{Limits of the Instantiation Processors}

Whenever we have an ADP where the left-hand side of the rule is also contained in the support of the right-hand side, 
then instantiations become useless, because we will always have at least the same ADP again after applying the processor.
For example, we need to apply one of the instantiation processors in order to prove 
termination for the TRS with the rules $\tf(x,y,z) \to \tg(x,y,z)$ and $\tg(\ta,\tb,z) \to \tf(z,z,z)$.
If we make the rules probabilistic by adding the possibility to do nothing in a rewrite step with the probability $\nicefrac{1}{2}$,
then we result in the following PTRS.

\vspace*{-.2cm}

{\small
\begin{align*}
    \tf(x,y,z) & \!\to\! \{\nicefrac{1}{2} : \tg(x,y,z), \nicefrac{1}{2} : \tf(x,y,z)\}\\
    \tg(\ta,\tb,z) & \!\to\! \{\nicefrac{1}{2} : \tf(z,z,z), \nicefrac{1}{2} : \tg(\ta,\tb,z)\}\!
\end{align*}
}

\noindent
This PTRS is iAST (since the original TRS was innermost terminating) but we are unable to show this
using the instantiation processor, because if one tries to instantiate any of the rules, this will result in at least
the same rule after the processor. 
In contrast, in \cref{example:inst} we had nearly the same PTRS but the first rule remained non-probabilistic.
There, we were able to apply the instantiation processor and prove iAST using the ADP
Framework. (However, while instantiation does not help in our example above, we can prove
iAST using the rule overlap instantiation processor.)

\subsection{Transformations do not Suffice for Inductive Reasoning}

Transformational processors are useful to perform a case analysis, but they do not suffice
for PTRSs where one needs inductive reasoning for the termination analysis.
For example, we cannot show iAST of the following PTRS even though the 
$\tevenif$-structure seems similar to the one of the
probabilistic quicksort example. This example was proposed to us by Johannes Niederhauser at the 0-th probabilistic termination competition in August 2023.

\vspace*{-.5cm}

{\small
\begin{align*}
    \teven(\tz) & \!\to\! \{ 1 : \ttrue \} & \teven(\ts(\tz)) & \!\to\! \{ 1 : \tfalse \}\\
    \teven(\ts(\ts(x))) & \!\to\! \{ 1 : \teven(x) \} & \tloop(x) & \!\to\! \{ 1 : \tevenif(\teven(x), x) \}\\
    \tevenif(\tfalse, x) & \!\to\! \{ 1 : \tstop \} & \tevenif(\ttrue, x) & \!\to\! \{ \nicefrac{1}{2} : \tloop(x), \nicefrac{1}{2} : \tloop(\ts(x)) \}\!
\end{align*}
}

\noindent
The idea here is that the recursion of $\tloop$ stops if its argument contains an even number.
If it is not even, then we either increase the value by 1 or use the same value again.
Here, a simple case analysis does not suffice, but we actually have to show (inductively), 
that if a number $x$ is odd, then $x+1$ is even.
This is not possible with the new transformational processors but needs other types of
processors for inductive reasoning (e.g., as in \cite{DPInductionJAR2011} for the non-probabilistic DP framework).

}
\end{document}